\documentclass[12pt]{iopart}
\usepackage{hyperref}
\usepackage{iopams}          \usepackage{graphicx}        \usepackage{cite}            \usepackage{setstack}
\usepackage{hyperref}
\usepackage{caption}
\usepackage[makeroom]{cancel}
\usepackage{indentfirst}
\usepackage{enumerate}
\usepackage{scalerel}
\usepackage{stmaryrd}
\usepackage{xcolor}
\usepackage{mdframed}
\usepackage{comment}
\usepackage[normalem]{ulem}
\usepackage{tikz}
\usetikzlibrary{shapes.geometric}

\usepackage{fancyhdr}
\usepackage{ulem}
\newcommand{\comments}[1]{}

\def\Xint#1{\mathchoice
    {\XXint\displaystyle\textstyle{#1}}{\XXint\textstyle\scriptstyle{#1}}{\XXint\scriptstyle\scriptscriptstyle{#1}}{\XXint\scriptscriptstyle\scriptscriptstyle{#1}}\!\int}
\def\XXint#1#2#3{{\setbox0=\hbox{$#1{#2#3}{\int}$}
    \vcenter{\hbox{$#2#3$}}\kern-.5\wd0}}
\def\dashint{\Xint-}
\definecolor{mygreen}{rgb}{.0,.5,.1}

\newenvironment{packed_enum}{
\begin{itemize}
  \setlength{\itemsep}{0pt}
  \setlength{\parskip}{0pt}
  \setlength{\parsep}{0pt}
}{\end{itemize}}

\begin{document}
\setlength{\headheight}{16.0pt}

\title[Langevin dynamics towards GGEs]{Langevin dynamics towards Generalized Gibbs Ensembles}

\author{Rafael Gonz\'alez-Albaladejo
\footnote{Corresponding author.
Present address: Euneiz University, Vitoria-Gasteiz, Spain.}}
\address{Sorbonne Universit\'e, Laboratoire de Physique Th\'eorique et Hautes Energies, CNRS UMR 7589, 4 Place Jussieu, 75252 Paris Cedex 05, France}
\ead{rafael.gonzalez@euneiz.com}
\author{Leticia F. Cugliandolo}
\address{Sorbonne Universit\'e, Laboratoire de Physique Th\'eorique et Hautes Energies, CNRS UMR 7589, 4 Place Jussieu, 75252 Paris Cedex 05, France}
\author{Gustavo S. Lozano}
\address{Departamento de F\'{\i}sica, Universidad de Buenos Aires and IFIBA CONICET, Argentina}
\author{Daniel A. Stariolo}
\address{Universidade Federal Fluminense, Departamento de Física,
Av. Gal. Milton Tavares de Souza s/n, Campus da Praia Vermelha, 24210-346 Niterói, RJ, Brazil}

\begin{abstract}
In recent years, statistical physics tools have been used to study the
dynamics of macroscopic quantum and classical integrable models. 
These studies have shown that, in the thermodynamic limit,
 long term  time-averages of local observables equal statistical averages calculated with a Generalized
Gibbs Ensemble (GGE). In the latter, the weight in phase space involves all constants of motion 
multiplied by Lagrange multipliers which are fixed by their initial values. We argue that the GGE can be interpreted as a proper Boltzmann measure of a 
different (non-standard) mechanical model. This measure can therefore be sampled
with stochastic methods implemented, for example, as a Langevin process.
We tailor the deterministic force in the Langevin equations so that at long times the 
stochastic trajectories sample the desired measure. 
{We use the integrable Neumann Model -- the
model of a classical particle constrained to move
on a sphere under anisotropic harmonic potentials --
as a benchmark. 
We test the method in a setup where the Lagrange  multipliers are fixed by 
the GGE governing the asymptotic dynamics of the isolated Neumann model following a quench.
This problem has different phases depending on the initial conditions and quench protocol and,}
in all cases, 
with the stochastic evolution of an adequately tuned Langevin process 
we reproduce the asymptotic values of relevant observables.
\end{abstract}

\enlargethispage{2\baselineskip}
\noindent{\it Keywords\/}: Classical integrability, Generalised Gibbs Ensembles, stochastic processes

\maketitle

\newpage
\tableofcontents

\section{Introduction}
\label{sec:introduction}

There is renewed interest in the dynamics of classical closed macroscopic systems, where questions regarding the eventual equilibration 
to a stationary measure naturally arise~\cite{DeLuca16,Doyon17,Doyon18,Bastianello18,Goldfriend18,CuLoNePiTa18,Cao19,Goldfriend19,Doyon19,Spohn19,BaCuLoNePiTa19,BaCuLoNe20,Myers20,Koch22,BaCuLoNe22,Baldovin21,Kethepalli23,Roy23,Pandey24,Singh24,Baldovin25}. 
Concretely, consider the mechanical motion 
of a particle, with position $\vec x$ and momentum $\vec p$, 
in an $N$ dimensional space. The coordinates and momentum vectors
components are $x_\mu$ and $p_\mu$, with 
$\mu=1,\dots, N$. Define a time-independent
Hamiltonian ${\mathcal H}(\{ x_\mu, p_\mu\})$ which determines the 
Newtonian particle's dynamics. 
The issue is to know 
whether after some characteristic time, $t_{\rm eq}$, the ergodic property
\begin{equation}
\overline{A(\{x_\mu(t), p_\mu(t) \})} = \langle A(\{ x_\mu, p_\mu \}) \rangle
\label{eq:ergodic}
\end{equation}
applies to any non-pathological observable $A$, in some sense local. 
The overline in the left-hand-side represents a long time average performed after $t_{\rm eq}$ 
and the angular brackets in the  right-hand-side denote an average over the relevant 
stationary measure.

For non-integrable models, the long time evolution (possibly after a time-scale
$t_{\rm eq}$ 
diverging with system size) are expected to reach Gibbs-Boltzmann equilibrium 
\begin{eqnarray}
&& \ \ P_{\rm GB}(\{ x_\mu, p_\mu \}) = {\mathcal Z}_{\rm GB}^{-1} \; e^{-\beta {\mathcal H}(\{ x_\mu, p_\mu \})}
\; , 
\label{eq:GB-intro}
\end{eqnarray}
with the same Hamiltonian ${\mathcal H}$ that rules the dynamics appearing in the 
Boltzmann weight. The inverse temperature $\beta=1/(k_BT)$, with $k_B$ the Boltzmann constant 
and $T$ a temperature, is a Lagrange multiplier
which controls the allowed energy fluctuations in this canonical ensemble. 

Instead, for integrable models with as many constants of motion, $I_\nu(\{ x_\mu, p_\mu \})$ with $\nu=1, \dots, N$, 
as degrees of freedom, the asymptotic measure (\ref{eq:GB-intro}) is expected to be replaced by the one of 
a Generalized Gibbs Ensemble
\begin{eqnarray}
&& P_{\rm GGE}(\{ x_\mu, p_\mu \}) = {\mathcal Z}_{\rm GGE}^{-1} \; e^{- \beta\sum_\nu \bar\gamma_\nu I_\nu(\{ x_\mu, p_\mu \})}
\; .
\label{eq:GGE}
\end{eqnarray}
The integrals of motion $I_\nu(\{ x_\mu, p_\mu \})$ are functions of the phase space variables $\{ x_\mu, p_\mu \}$ 
and are multiplied by Lagrange multipliers, $\beta \bar\gamma_\nu$, in the exponential.
(The dimensioned pre-factor $\beta$ was introduced for 
convenience, as also done in the grand-canonical ensemble in which $\beta \mu$, with $\mu$ the chemical potential, 
multiplies the particle number $N$. 
The full expression in the exponential is dimensionless, as it should).
In the dynamical process, the parameter dependence of all 
$\beta \bar\gamma_\nu$ is 
determined by the initial conditions and the Newtonian evolution. 

In tests of the ``ergodic hypothesis'' in Eq.~(\ref{eq:ergodic}), for systems expected to be ruled by the 
GGE measure (\ref{eq:GGE}), the  dynamical process 
is chosen to be of the kind performed in quantum mechanical studies, often called  ``quantum quenches''. 
This means that the system, prepared in some set of initial 
conditions which know about a given initial Hamiltonian ${\mathcal H}_0$, are let evolve 
with Newtonian dynamics, using another Hamiltonian ${\mathcal H}$. The parameters 
$\beta \bar\gamma_\nu$ are then fixed by the conditions
\begin{equation}
I_\nu(\{x_\mu(0^+), p_\mu(0^+)\}) 
= \overline{I_\nu(\{x_\mu(t), p_\mu(t) \})} = \langle I_\nu(\{ x_\mu, p_\mu \}) \rangle_{\rm GGE}
\; , 
\label{eq:Imu-conditions}
\end{equation}
$\forall \; \nu=1, \dots, N$.

The main question we want to address in this paper is:

\qquad ``How to sample efficiently a GGE measure like the one in Eq.~(\ref{eq:GGE})?'' 
\\
In other words, how to generate independent phase space variables
$\{ x_\mu, p_\mu \}$ distributed according to Eq.~(\ref{eq:GGE}).

Interestingly enough, the distribution (\ref{eq:GGE}) can be alternatively interpreted as the conventional 
Gibbs-Boltzmann distribution at inverse temperature $\beta$ of a model with Hamiltonian
 ${\mathcal H}_{\rm GGE} = \sum_\nu \bar\gamma_\nu I_\nu$, which is not necessarily of the conventional 
 kinetic + potential energy form. That is, 
\begin{eqnarray}
&& \ \ P_{\rm GB}^{({\mathcal H_{\rm GGE}})}(\{ x_\mu, p_\mu \}) = {\mathcal Z}_{\rm GB}^{-1} \; e^{-\beta {\mathcal H}_{\rm GGE}(\{ x_\mu, p_\mu \})}
\nonumber\\
[10pt]
&& \qquad\qquad\quad{\rm with}\qquad
{\mathcal H}_{\rm GGE}(\{ x_\mu, p_\mu \})=\sum_\nu \bar\gamma_\nu I_\nu(\{ x_\mu, p_\mu \})
\; .
\label{eq:GB-GGE0}
\end{eqnarray}
Interpreted in this way, one can ask questions about the conventional equilibrium behaviour of the 
model defined by ${\mathcal H}_{\rm GGE}$: whether there is a phase transition as a function 
of $\beta$ times the parameters in ${\mathcal H}_{\rm GGE}$, 
which is the nature of the phases, which is the critical behaviour, etc. {Two properties of the GGE
Hamiltonian should be noted:
\begin{packed_enum}
\item
${\mathcal H}_{\rm GGE}$  is integrable under classical Newtonian dynamics 
for any choice of the parameters
$\{ \bar\gamma_\nu \}$, so the conventional canonical equilibrium analysis of the model ${\mathcal H}_{\rm GGE}$ is 
already a non-standard one, which we will carry out in great detail for some special choices of the set
$\{\bar\gamma_\nu\}$. 
\item 
The position and momentum components can be coupled in a non-trivial way, so that the momentum does not enter in the usual 
kinetic energy form. 
\end{packed_enum}
}

Taking advantage of the interpretation of the measure in Eq.~(\ref{eq:GGE})
as a  conventional Gibbs-Boltzmann measure~(\ref{eq:GB-GGE0}), the sampling  can be achieved with 
the help of a stochastic process representing the coupling of the system to an equilibrium 
heat bath at inverse temperature $\beta$. 
The latter  induces dissipation and noise and, if the stochastic process respects detailed balance, it 
should converge, in the long time run, 
to the conventional equilibrium distribution in Eq.~(\ref{eq:GB-GGE0}). 
In particular, a Markov process of Langevin kind~\cite{Gardiner2009, Coffey2004,Kloeden2011,LFC25}
achieves this goal. Importantly enough, in certain slow dynamic cases it might be necessary to scale times with system 
size and only then take the thermodynamic limit to reach the goal. 

The idea that we propose in this paper is to use a Langevin process (for simplicity memory-less) to sample 
 the GGE measure~(\ref{eq:GGE}) which is identical to the Gibbs-Boltzmann measure~(\ref{eq:GB-GGE0}). 
 We apply this method to a case in which the 
constants of motion $I_\nu$ are those of the 
Neumann model of classical mechanics~\cite{Neumann}, 
that is, the model of a particle constrained to move on a sphere and under the effect of 
anisotropic harmonic potentials. 
The potential energy of the Neumann model is equivalent to the one of the 
Spherical Sherrington-Kirkpatrick (SSK) model of statistical physics, where the spring 
constants, $\lambda_\mu$, drawn from a Wigner semi-circle  distribution, are the eigenvalues of a 
GOE random matrix~\cite{KoThJo76}. We consider two choices for the parameters
$\bar\gamma_\nu$. They are either fixed to the values determined by the quenches of the 
isolated ${\mathcal H}$, in which case we sample the GGE measure, or else we take them 
to be free parameters, defining a set of new integrable models, which we call 
Generalized Neumann Models (GNM).

\begin{center}
\begin{table}[h!]
\begin{center}
\begin{tabular}{|c|c|c|}
\hline
{\small Newton dynamics}  
& 
{\small  Ergodicity with} 
& 
{\small Langevin relaxation}
\\
{\small after a quench of}  
& 
$P_{\rm GGE} \propto e^{-\beta {\mathcal H}_{\rm GGE}} \propto P^{({\mathcal H}_{\rm GGE})}_{\rm GB}$ 
& 
{\small at} $k_BT = \beta^{-1}$ 
\\
${\mathcal H}$, 
{\small the integrable model}
&
${\mathcal H}_{\rm GGE}=\sum_\mu \bar\gamma_\mu I_\mu$ &
of ${\mathcal H}_{\rm GGE}$ 
\\
\hline
& & 
\\
[-15pt]
\hline
& & 
\\
$\overline{\langle I_\nu(t>0)\rangle_{i.c.}}$
& 
$ \langle I_\nu\rangle_{\rm GGE} $ 
&
$\lim\limits_{t\to\infty} \langle I_\nu(t) \rangle_{\xi, i.c.} $
\\
[15pt]
{\small Fixed by initial conditions}
& 
{\small = average with} $P_{\rm GGE}$
&
{\small = asymptotic limit}
\\
[7pt]
& {\small fixes}  $\{\beta \bar\gamma_\mu\}$ 
&
\\
[2pt]
& {\small to represent the quench of}  
&
\\
[2pt]
& 
{\small the isolated} ${\mathcal H}$
&
\\
[15pt]
$\overline{\langle p^2_\mu(t>0)\rangle_{i.c.}}$
& 
$ \langle p^2_\mu\rangle_{\rm GGE} = \langle p^2_\mu\rangle^{({\mathcal H}_{\rm GGE})}_{\rm GB} $ 
&
$\lim\limits_{t\to\infty} \langle p^2_\mu(t) \rangle_{\xi, i.c.} $
\\
[15pt]
$\overline{\langle x^2_\mu(t>0)\rangle_{i.c.}}$
& 
$ \langle x^2_\mu\rangle_{\rm GGE} = \langle x^2_\mu\rangle^{({\mathcal H}_{\rm GGE})}_{\rm GB} $ 
&
$\lim\limits_{t\to\infty} \langle x^2_\mu(t) \rangle_{\xi, i.c.} $
\\
[15pt]
\hline
\end{tabular}
\end{center}
\caption{\small
In the first column, results from the Newton's dynamics of the isolated integrable model
${\mathcal H}$ (in the application, it will be the Generalized Neumann Model) 
after an instantaneous quench, as derived in~\cite{BaCuLoNePiTa19,BaCuLoNe20,BaCuLoNe22}. Second column, 
static averages calculated with the GGE measure, also calculated in~\cite{BaCuLoNe22}. Third column, the Langevin 
dynamics with deterministic force derived from a Generalized Neumann Model in which the
Lagrange multipliers are fixed to be the ones determined with the protocol of the second column
and which we call
${\mathcal H}_{\rm GGE}$. The angular brackets denote 
here an average over the thermal noise 
and initial conditions. If the $\{\beta\bar\gamma_\mu\}$ are fixed to be the ones of a quench of the isolated 
integrable model evolving with Newton's dynamics, then the asymptotic stochastic dynamics samples 
the corresponding GGE measure. Otherwise, one can choose the $\{\bar\gamma_\mu\}$ 
at will to study the stochastic evolution of the corresponding ${\mathcal H}_{\rm GNM}$, generalisations of the Neumann model. 
In Sec.~\ref{subsec:quadratic-gamma} and Sec.~\ref{sec:numerics} we will use a simple 
quadratic form $\bar\gamma_\mu = a/J \, \lambda^2_\mu - \lambda_\mu/2$ which allows us to identify various 
phases with similar qualitative properties 
to the ones of the Newtonian dynamics of the isolated model.
}
\label{table:uno}
\end{table}
\end{center}

The Table~\ref{table:uno} summarises the three settings in which the original model is studied.
In the first column, we place the Newtonian dynamics of an isolated integrable model,  
${\mathcal H}$, 
for which, in the thermodynamic limit and 
asymptotically, local observables are described by a GGE measure with parameters $\{\beta\bar\gamma_\nu\}$ determined
by the initial conditions (characterised by the parameter $T_0/J_0$ in the application) and quench (characterised by the parameter $J/J_0$ in the same 
application). 
The GGE measure, with its alternative interpretation as the GB measure of another model,  
is described in the  second column. 
The new aspect discussed in this paper appears
in the third column, with the stochastic evolution leading, asymptotically, to the GGE/GB measure of the second column. 

To illustrate how the method works, we first fix the $\bar\gamma_\nu$ to 
be simple quadratic functions of the $\lambda_\nu$. In doing this, we go
beyond the SSK/Neumann Hamiltonian and we use a Generalized Neumann Model, ${\mathcal H}_{\rm GNM}$,
which mixes non trivially the positions and momenta.
In the process of studying the static averages, and 
solving the stochastic process and its asymptotic properties, 
we exhibit a rich phase diagram with interesting properties which extend the 
ones of the standard SSK model. We note that 
generalizations of the Neumann model of this kind have been 
discussed in the mathematical-physics  literature~\cite{Lund80,Lund80b}.

Finally, in order to check compatibility with the results in ~\cite{BaCuLoNe22}, we consider the case 
which represents the asymptotic behaviour of the quenched isolated SSK/Neumann model
by tuning the $\bar\gamma_\mu(\lambda_\mu)$ to be the ones
 fixed by the requirements~(\ref{eq:Imu-conditions}).
 {While performing this analysis, we identified a mismatch that revealed an error in the determination of $\bar\gamma_\mu(\lambda_\mu)$ in Ref.~\cite{BaCuLoNe22}. We provide the correct derivation in App.~\ref{app:new-derivation}. Upon incorporating the correct $\bar\gamma_\mu(\lambda_\mu)$, the asymptotic predictions obtained from the stochastic evolution show excellent agreement with the static results.}

{We close this introduction with a comment
on previous work.} With the purpose of putting the predictions of Generalized Hydrodynamics 
in classical integrable models and field theories to the test, GGE initial conditions were sampled with the Metropolis rule, and they were further evolved with conservative dynamics~\cite{Bastianello18,Koch22}. 
Slightly earlier, the use of Monte Carlo methods to sample the Hilbert space of a quantum spin chain was proposed in~\cite{Alba15}. 
The results in our paper 
show that performing such sampling efficiently can be tricky, 
even in cases in  which the stochastic dynamics of a finite size system should converge to 
the GGE measure in the long time limit, {due to long relaxation times.}

The paper is organized as follows. 
In Sec.~\ref{sec:definition-SSK} we 
introduce the Neumann model of classical mechanics, or Spherical Sherrington-Kirkpatrick model of statistical physics with the addition of a kinetic energy, ${\mathcal H}_{\rm SSK}$.
In Sec.~\ref{sec:GGE-Model} we introduce the Generalized Neumann model (GNM) in which the 
Hamiltonian, ${\mathcal H}_{\rm GNM}$, is a generic linear combination of Uhlenbeck's 
constants of motion conserved by the Newton's dynamics of the SSK/Neumann model. 
In this Section we discuss the Gibbs-Boltzmann equilibrium  of this model.
Next, in Sec.~\ref{sec:evol-GGE}, we introduce the Langevin equations for spherically constrained 
generic models, and we specialize them to the SSK/Neumann and GNM cases.
We also explain in this section why these equations are expected to 
take the systems to equilibrium at the Gibbs-Boltzmann measure with 
${\mathcal H}_{\rm SSK/NM}$ and ${\mathcal H}_{\rm GNM}$, respectively.
The numerical tests are presented and discussed in Sec.~\ref{sec:numerics} and Sec.~\ref{sec:sampling}. 
In Sec.~\ref{sec:numerics} we use quadratic parameters $\bar\gamma_\mu= (a/J) \, \lambda_\mu^2 - \lambda_\mu/2$.
In Sec.~\ref{sec:sampling} we use the
spectra of Lagrange multipliers which correspond to different quenches of the isolated 
SSK/Neumann model. 
{We show that the stochastic dynamics succeed in sampling the GGE in all phases, 
for the system sizes that we analyze. Slow dynamics can hinder convergence in some of the 
phases for larger system sizes, as common in models with slow dynamics.}
We conclude in Sec.~\ref{sec:conclusions}. 

The paper is complemented by several Appendices.
In~App.~\ref{app:Kramers}  we derive the Kramers equation for the spherically constrained model and 
we show that it admits a 
stationary solution of usual canonical equilibrium form.  Appendix~\ref{app:GGE-to-SSK}
recovers the SSK/Neumann model from the GNM one for $\bar\gamma_\mu = -\lambda_\mu/2$ and presents
some numerical results for the Langevin dynamics of the SSK/Neumann model. The following 
App.~\ref{app:N2} consider the 
cases $N=2$ and its generalization to finite $N$ in detail. In App.~\ref{app:numerics}
we give some details on the code used to integrate the Langevin equations.
Appendix~\ref{app:momentum-vs-velocity} compares the 
behaviour of the momentum and velocity in the GNM. In App.~\ref{app:details-of-calculations}
we  detail how the results in Sec.~\ref{sec:GGE-Model} are derived and, in particular, the phase diagram of the 
model with quadratic $\bar\gamma_\mu= (a/J) \, \lambda_\mu^2 - \lambda_\mu/2$. In App.~\ref{app:temperature}
we investigate numerically the role played by the temperature of the Langevin bath. 
In App.~\ref{sec:stationary-states}  
we analyse the various stationary states of the GNM.
{Finally, in App.~\ref{app:new-derivation} we present the 
correct derivation of the $\bar\gamma_\mu(\lambda_\mu)$ needed in the GGE description of the 
quenched dynamics of the Neumann model.}

\section{The Spherical Sherrington-Kirkpatrick and the Neumann Models }
\label{sec:definition-SSK}

This section presents the Spherical Sherrington-Kirkpatrick (SSK) model
of the disordered-systems literature, 
and its dynamical systems counterpart, the Neumann model. 
We start with a general introduction of Hamilton's equations for a 
particle moving on a sphere in Sec.~\ref{sec:Hamilton}.
In Sec.~\ref{sec:model} we introduce the SSK model and we connect it 
to the Neumann model after introducing kinetic energy.
Finally, in Sec.~\ref{sec:charges}
we recall the expression of Uhlenbeck's integrals of motion.

\subsection{Hamilton's equations on the sphere}
\label{sec:Hamilton}

Let us use a generic Hamiltonian ${\mathcal H}(\vec x, \vec p )$ and impose the constraints 
\begin{equation}
\phi_1: \sum_\mu x_\mu^2 =N 
\; , 
\qquad\qquad
\phi_2: \sum_\mu x_\mu p_\mu =0
\; ,
\label{eq:constraints-GGE}
\end{equation}
by supplementing ${\mathcal H}$ with two Lagrange multipliers, $z_1$ and $z_2$,
\begin{equation}
 {\mathcal H}^z= 
 {\mathcal H}
+\frac{z_1}{2}  \left( \sum_\mu x_\mu^2 - N\right)
 + z_2 \sum_\mu x_\mu p_\mu  
\end{equation}
which depend on $\vec x$ and $\vec p$.
Secondary constraints derivable from Eq.~(\ref{eq:constraints-GGE}) are 
\begin{equation}
\phi'_1: \sum_\mu x_\mu \dot x_\mu =0 \qquad\qquad
\phi'_2: \sum_\mu \left( \dot x_\mu p_\mu + x_\mu \dot p_\mu \right) =0
\; . 
\label{eq:phi-prime}
\end{equation}

Hamilton's equations read
\begin{equation}
\dot x_\mu = \frac{\partial {\mathcal H}}{\partial p_\mu} + z_2 x_\mu
\qquad\qquad
\dot p_\mu = -\frac{\partial {\mathcal H}}{\partial x_\mu} - z_1 x_\mu - z_2 p_\mu
\; . 
\label{eq:Hamilton-eqs-GGE}
\end{equation}

In order to fix the expressions of the Lagrange multiplier $z_2$ 
we multiply the first equation by $x_\mu$ and sum over $\mu$:
\begin{equation}
0 = \sum_\mu x_\mu \dot x_\mu = \sum_\mu x_\mu \frac{\partial {\mathcal H}}{\partial p_\mu} + z_2 \sum_\mu x_\mu x_\mu
=
\sum_\mu x_\mu \frac{\partial {\mathcal H}}{\partial p_\mu} + z_2 N
\; .
\end{equation}
We obtain
\begin{equation}
z_2 = 
- \frac{1}{N} \sum_\mu x_\mu \frac{\partial {\mathcal H}}{\partial p_\mu} 
\; . 
\label{eq:z2}
\end{equation}
This expression for $z_2$ is independent of the particular form of $ {\mathcal H}$. Note that if ${\mathcal H}$ is quadratic in $p_\mu$, then $z_2=- \frac{1}{N} \sum_\mu x_\mu  p_\mu /m = 0$, thanks to 
$\phi_2$. Because of this reason, 
we will not include this secondary constraint in the discussion of the SSK model below. Next, we 
fix $z_1$ by multiplying the second equation also by $x_\mu$ and summing over $\mu$:
\begin{equation}
\sum_\mu x_\mu \dot p_\mu = - \sum_\mu x_\mu \frac{\partial {\mathcal H}}{\partial x_\mu} - z_1 N
\; . 
\end{equation}
We can now rewrite the left-hand side using (\ref{eq:phi-prime})  and derive
\begin{equation}
z_1 =  \frac{1}{N} \sum_\mu \dot x_\mu p_\mu -  \frac{1}{N} \sum_\mu x_\mu \frac{\partial {\mathcal H}}{\partial x_\mu} 
\; , 
\end{equation}
which, using the first equation of motion in (\ref{eq:Hamilton-eqs-GGE}) and the orthogonality between $\vec x$ and 
$\vec p$ becomes
\begin{equation}
z_1 =  \frac{1}{N} \sum_\mu  \frac{\partial {\mathcal H}}{\partial p_\mu}  p_\mu -  \frac{1}{N} \sum_\mu x_\mu \frac{\partial {\mathcal H}}{\partial x_\mu} 
\; .
\label{eq:z1}
\end{equation}
In the case of a Hamiltonian ${\mathcal H} = E_{\rm kin}(p)+V(\vec x)$, 
like in the SSK/Neumann model defined in Eq.~(\ref{eq:Hssk}) below, 
$z_1$ becomes  $N z_1=2E_{\rm kin}(p) - 2V(\vec x)$. 
With the expressions of $z_1$ and $z_2$ in Eqs.~(\ref{eq:z1}) and (\ref{eq:z2}), 
Hamilton's equations read
\begin{eqnarray}
&& 
\dot x_\mu = 
\frac{\partial {\mathcal H}}{\partial p_\mu} - \Big( \frac{1}{N} \sum_\nu x_\nu \frac{\partial {\mathcal H}}{\partial p_\nu} \Big) \;  x_\mu 
\; , 
\label{eq:Hamiltoni1}
\\
 &&
 \dot p_\mu = -\frac{\partial {\mathcal H}}{\partial x_\mu} - 
\Big( \frac{1}{N} \sum_\nu  p_\nu \frac{\partial {\mathcal H}}{\partial p_\nu}  -  \frac{1}{N} \sum_\nu x_\nu \frac{\partial {\mathcal H}}{\partial x_\nu} \Big)
 x_\mu  
 \nonumber\\
 && 
 \qquad
 +
 \Big(  \frac{1}{N} \sum_\nu x_\nu \frac{\partial {\mathcal H}}{\partial p_\nu} \Big) \; p_\mu
 \; . 
 \ \ \ \ \ \
 \label{eq:Hamiltoni2}
\end{eqnarray}

\subsection{The SSK/Neumann model}
\label{sec:model}

The Spherical Sherrington--Kirkpatrick (SSK) model -- also called the spherical $p=2$ model --
was introduced by Kosterlitz, Thouless and Jones~\cite{KoThJo76}. 
It was formulated  
as a classical continuous spin model with two-body interactions controlled by couplings $J_{ij}$ drawn from a Gaussian probability distribution
and a global spherical constraint on the spins. 
It has become a well-known and, because of the continuous and quadratic
character of the variables, a relatively simple and fully solvable model with quenched randomness~\cite{CuDe95a,CuDe95b,BenAr01,ChCuYo06,CuDeYo07,vanDu10,FyPeSc15,BaFrCuSt21,Lahoche22,Freitas23,DiTa24,Lahoche24a,Lahoche24b,Niedda24,Tarabolo25,Zhenpeng26}. Beyond the relation to Bose-Einstein condensation already pointed out in~\cite{KoThJo76},
the model has found applications in a large variety of fields including neural networks~\cite{dAscoli2022,Fournier23,montanari25,Janssen25,Fournier26b},
ecological and biological problems where interactions are not necessarily symmetric~\cite{Lorenzana25,StarioloMetz2026}, synchronization in disordered
oscillator systems~\cite{Ikeda23}
and even polariton chemistry~\cite{Rubio}.

Here, 
we adopt a related problem which arises when we add intrinsic dynamics to the SSK model~\cite{CL1,CL2}. In this case, the model corresponds to that of a classical particle 
under harmonic potentials, 
placed in an $N$-dimensional space but constrained to move on the $(N-1)$-dimensional sphere with 
radius $\sqrt{N}$. 
Concretely, we call $\vec x = (x_1, \dots, x_N)$ the particle's position and
$x_\mu = \vec x \cdot \vec v_\mu$ the projections onto the orthogonal directions $\vec v_\mu$,
with $\mu = 1, \dots, N$, which are the eigenvectors of the symmetric real
matrix with elements $J_{ij}$. The particle interpretation allows us to make contact with the Neumann model~\cite{Neumann} of 
classical integrability~\cite{Moser81,Uhlenbeck,Lund80,Lund80b,AvTa90,BaTa92,Dullin01,Dullin12,BaBeTa09,BaCuLoNePiTa19,BaCuLoNe20,BaCuLoNe22,Baron24}, as we explain below.

In the particle interpretation, the particle is subject to
an anisotropic quadratic potential,
\begin{equation}
V(\vec x) = - \frac{1}{2} \sum\limits_{\mu=1}^N \lambda_\mu x_\mu^2
\label{eq:harmonic-potential-SSK}
\end{equation}
with harmonic constants, $\lambda_\mu$, which we 
order according to $\lambda_1 < \lambda_2 < \dots < \lambda_{N-1} < \lambda_N$.
These parameters can be chosen freely. 
In previous studies of this problem, performed in the thermodynamic $N\to\infty$ limit,
the $\lambda_\mu$ were 
drawn from 
the Wigner semi-circle law~\cite{BaCuLoNePiTa19,BaCuLoNe20,BaCuLoNe22},
\begin{equation}
\rho(\lambda_\mu) = \frac{1}{2\pi J^2} \sqrt{(2J)^2-\lambda_\mu^2}
\qquad \mbox{for} \qquad \lambda_\mu \in [-2J, 2J]
\label{eq:semi-circle-SSK}
\end{equation}
and zero otherwise. Apart from finite $N$ fluctuations, 
there are as many positive as negative $\lambda_\mu$s
but the motion is not unstable since the particle is constrained to remain
(strictly or on average) on the sphere.
The motivation for making this choice is that with the $\lambda_\mu$s distributed
as in Eq.~(\ref{eq:semi-circle-SSK}), the potential in Eq.~(\ref{eq:harmonic-potential-SSK}) 
is the one of the SSK model~\cite{KoThJo76}.
Most of the qualitative
results are recovered in very similar form
for other distributions $\rho$ with finite support and no repeated values of the~$\lambda_\mu$.
In finite $N$ studies, the $\lambda_\mu$ are selected to be the eigenvalues of an $N\times N$
random matrix of the GOE ensemble, whose averaged density approaches~(\ref{eq:semi-circle-SSK}) in the $N\to\infty$ limit.

In order to induce dynamics to this model we add a kinetic energy 
with the standard quadratic form
\begin{equation}
E_{\rm kin}(p)=\frac{1}{2m} \sum_\mu p_\mu^2
\; .
\label{eq:Ekin-SSK}
\end{equation}
The total energy reads
\begin{equation}
{\mathcal H}_{\rm SSK}(\vec x, \vec p ) = E_{\rm kin}(p) + V(\vec x)
\label{eq:Hssk}
\end{equation}
with $E_{\rm kin}(p)$ in Eq.~(\ref{eq:Ekin-SSK}) and 
$V(\vec x)$ the harmonic potential energy in Eq.~(\ref{eq:harmonic-potential-SSK}).

As already discussed in the previous subsection, 
the second Lagrange multiplier vanishes, $z_2=0$. 
We then simplify the notation and just call $z_{\rm SSK}$ the remaining one, $z_1=z_{\rm SSK}$. 
Equations~(\ref{eq:Hamiltoni1}) and 
(\ref{eq:Hamiltoni2}) become
\begin{eqnarray} 
\dot x_\mu & = & p_\mu/m \; ,\\
\dot p_\mu & = & -[z_{\rm SSK}(\vec x,\vec p) - \lambda_{\mu}]\; x_{\mu} \label{eq:Newton} 
\nonumber\\
& = & \lambda_\mu x_\mu - \frac{2}{N} \Big[E_{\rm kin}(p) - V(\vec x)\Big] x_\mu
\; . 
\end{eqnarray}
We wrote the explicit dependence of $z_{\rm SSK}$ on $\vec x$ and $\vec p$ to stress the fact that 
Hamilton's equations are not linear in the phase space variables.
We note that, in this model, the velocity is proportional to the momentum, as usual.
Below, we will also use the notation
\begin{equation}
  {\mathcal H}^z_{\rm SSK}(\vec x, \vec p ) = E_{\rm kin}(p) + V(\vec x) + \frac{z_{\rm SSK}(\vec x, \vec p)}{2}
  \left( x^2-N \right)
\; , 
\end{equation}
where $x^2 \equiv \sum_{\mu=1}^N x_{\mu}^2$.
The extension of the SSK model which includes a standard kinetic energy is identical to the 
Neumann model of classical integrability.
Beyond its obvious mathematical interest, the Neumann model appears in 
different areas of physics. It is the structural dual to Jacobi's problem of free geodesic motion on an ellipsoid: the trajectories of a particle on a sphere under a quadratic potential are geometrically equivalent to free particles gliding on an ellipsoidal manifold~\cite{Moser80,Moser81}. In high-energy physics, the semiclassical behaviour of strings rotating in curved background spaces 
can be mapped to the classical Neumann model 
or its extended variant, the Neumann-Rosochatius system~\cite{Russo04}.

\subsection{The conserved quantities}
\label{sec:charges}

The $N$ phase-space functions 
\begin{eqnarray}
\label{eq:imu}
I_{\mu}(\vec x, \vec p) =
x^2_\mu
+ \frac{1}{mN} \sum_{\nu (\neq \mu)}
\frac{  x^2_\mu p^2_\nu+  p^2_\mu x^2_\nu
-  2 x_\mu p_\mu x_\nu p_\nu}{\lambda_\nu-\lambda_\mu}
\end{eqnarray}
are conserved under Hamilton's dynamics on each trajectory~\cite{Moser81,Uhlenbeck}.  
This is proven as follows: take the time derivative of $I_\mu$  and 
use the relation $m\dot x_\mu = p_\mu$, 
\begin{equation}
\frac{m}{2}
\frac{d I_{\mu}}{d t}=
x_{\mu}p_{\mu}
+ x^2_{\mu} \left(\frac{1}{N}\sum_{\nu(\neq\mu)}x_{\nu}p_{\nu}\right)- x_{\mu}p_{\mu} \left(\frac{1}{N}\sum_{\nu(\neq\mu)}x^2_{\nu}\right)
\; .
\end{equation}
One can now complete the two sums adding the missing 
$\nu=\mu$ terms which cancel each other. For strict spherical constraints, valid on each trajectory, 
the  first factor between parenthesis vanishes, the second 
equals one, and the first and last terms cancel out.  
All charges $I_\mu$ are then conserved and satisfy 
\begin{equation}
\sum_\mu I_\mu = N
\qquad \mbox{and} \qquad
\sum_\mu \lambda_\mu I_\mu = 
-2(E_{\rm kin}+V) = -2 {\mathcal H}_{\rm SSK} 
\; .
\end{equation}

\section{The Generalized Neumann Model}
\label{sec:GGE-Model}

We next consider the Generalized Neumann Model (GNM), described by the Hamiltonian
\begin{equation}
{\mathcal H}_{\rm GNM}^z =
\sum_\mu \bar{\gamma}_\mu I_\mu
+ \frac{z_{\rm GNM}}{2}
\left(\sum_\mu x_\mu^2 - N\right)
+ z_2 \sum_\mu x_\mu p_\mu 
\; ,
\label{eq:HGGE}
\end{equation}
where the coefficients $\bar{\gamma}_\mu(\lambda_\mu)$ are arbitrary functions of the parameters $\lambda_\mu$. The particular choice $\bar{\gamma}_\mu=-\lambda_\mu/2$ reduces the GNM to the standard Neumann model. Models of this type have been studied in the mathematical literature in connection with constrained harmonic systems and nonlinear wave equations \cite{Lund80,Lund80b}. Moreover, as discussed above, Hamiltonians of this form arise naturally in the Generalized Gibbs Ensemble (GGE) description of the post-quench dynamics of the Neumann model.

\subsection{Hamilton's equations of motion} 
\label{subsec:Hamilton-GGE}

We now specify Hamilton's equations (\ref{eq:Hamiltoni1}) and  (\ref{eq:Hamiltoni2}) for the GNM model~(\ref{eq:HGGE}).
From Eqs.~(\ref{eq:imu}) and (\ref{eq:HGGE}) the variation of ${\mathcal  H}_{\rm GNM}$ with respect to $p_\mu$ reads
\begin{eqnarray}
\frac{mN}{2} \frac{\partial {\mathcal H}_{\rm GNM}}{\partial p_\mu} 
&=& 
 - \sum_{\eta(\neq \mu)}
\left( x_\eta^2 p_\mu - x_\mu p_\eta x_\eta \right)
\frac{\bar\gamma_\mu-\bar\gamma_\eta}{\lambda_\mu -\lambda_\eta} 
\; 
\label{eq:xdot}
\end{eqnarray}
and we note that 
\begin{eqnarray}
z_2 = \sum\limits_\mu x_\mu \frac{\partial {\mathcal  H}_{\rm GNM}}{\partial p_\mu} =
- \frac{2}{mN} \sum_{\eta\neq \mu}
\left( x_\eta^2 x_\mu p_\mu - x^2_\mu p_\eta x_\eta \right)
\frac{\bar\gamma_\mu-\bar\gamma_\eta}{\lambda_\mu -\lambda_\eta} 
=
0
\label{eq:z2-equal-0}
\end{eqnarray}
for all $N$. 
The equations  (\ref{eq:Hamiltoni1}) and  (\ref{eq:Hamiltoni2}) then simplify to 
\begin{eqnarray}
&&
\dot x_\mu = \frac{\partial {\mathcal  H}_{\rm GNM}}{\partial p_\mu}  
\; ,
 \label{eq:Hamilton-eqs-GGE-final1}
\\
&&
 \dot p_\mu =  -\frac{\partial  {\mathcal  H}_{\rm GNM}}{\partial x_\mu} - 
\left( \frac{1}{N} \sum_\nu \frac{\partial {\mathcal H}_{\rm GNM}}{\partial p_\nu}   p_\nu -  \frac{1}{N} \sum_\nu x_\nu \frac{\partial  {\mathcal H}_{\rm GNM}}{\partial x_\nu} \right)
 x_\mu  
 \; , 
 \ \ \ \ \ \
 \label{eq:Hamilton-eqs-GGE-final2}
\end{eqnarray}
and they are compatible with the constraint $\phi_2$, that is $\vec x$ and $\vec p$  being
 orthogonal. Indeed, if $\vec x \cdot \vec p =0 $, then $\dot{\vec x} \cdot \vec p + \vec x \cdot \dot{\vec p} =0 $
 and this form is simple to check. Using the equations of motion in their general form:
 \begin{eqnarray}
&& 
\dot{\vec x} \cdot \vec p + \vec x \cdot \dot{\vec p} 
\nonumber\\
&&
\qquad
 =
\sum_\mu 
\frac{\partial {\mathcal H}_{\rm GNM}}{\partial p_\mu}  p_\mu
\nonumber\\
&&
\qquad\quad
+ 
\sum_\mu 
\left[ -\frac{\partial  {\mathcal H}_{\rm GNM}}{\partial x_\mu} - 
\left( \frac{1}{N} \sum_\nu \dot x_\nu p_\nu -  \frac{1}{N} \sum_\nu x_\nu \frac{\partial  {\mathcal H}_{\rm GNM}}{\partial x_\nu} \right)
 x_\mu \right] x_\mu
 \nonumber\\
 && 
 \quad\quad = 0
 \; . 
 \end{eqnarray}
 Therefore, $\vec x \cdot \vec p \ $ is a constant and, if we set this constant  to zero initially,  
 the identity is henceforth conserved.
The dynamics are such that $\dot{\vec x}$ and $\vec p$ are both perpendicular to $\vec x$
 but not necessarily parallel to each other. 

We can verify that all $I_\mu$ are constants of motion under these dynamics 
for any $\bar\gamma_\mu$:
\begin{equation}
\frac{dI_\mu}{dt} = \{ I_\mu, {\mathcal H}_{\rm GNM} \} = \{ I_\mu, \sum_\nu \bar\gamma_\nu I_\nu  \} 
=
\sum_\nu \, \bar\gamma_\nu \, \{ I_\mu,  I_\nu  \}  =0
\end{equation} 
since the $I_\mu$ are in involution 
\begin{eqnarray}
\{ I_\mu,  I_\nu  \} = 
\sum_\rho 
\left( 
\frac{\partial I_\mu}{\partial x_\rho} \frac{\partial I_\nu}{\partial p_\rho} 
- \frac{\partial I_\nu}{\partial x_\rho}\frac{\partial I_\mu}{\partial p_\rho}
\right) = 0
\end{eqnarray}
(the curly brackets represent the Poisson brackets).

\subsection{Gibbs-Boltzmann equilibrium for generic $\overline \gamma_\mu$}
\label{subsec:generic-gamma}

The standard statistical physics setting is one in which the model is considered to be in Gibbs--Boltzmann equilibrium with a thermal bath at temperature $T$. 
The configurations are then sampled from the canonical distribution
\begin{equation}
P_{\rm GB}(\vec x, \vec p) = 
{\mathcal Z}_{\rm GB}^{-1} \; e^{-\beta {\mathcal H}^z_{\rm GNM}(\vec x, \vec p)}
\; , 
\label{eq:GB}
\end{equation}
where $\beta = (k_B T)^{-1}$, and we set $k_B = 1$ henceforth.
The spherical constraint is imposed on average via an integration over $z_{\rm GNM}$.

We perform the evaluation of the measure via a saddle-point calculation in the limit $N \rightarrow \infty$. The calculation is formally equivalent to the saddle-point calculation of the GGE measure
\begin{equation}
P_{\rm GGE} \propto e^{-\sum_\mu \gamma_\mu I_\mu} \, e^{-z_{\rm GGE} (\sum_\mu x_\mu^2-N)}
\end{equation}
performed in Ref.~\cite{BaCuLoNe22}. Two important remarks are in order. First, in Ref.~\cite{BaCuLoNe22} there was no temperature, and therefore 
one should keep in mind the identification $\beta \bar\gamma_\mu \equiv \gamma_\mu$
and $\beta z_{\rm GNM}=z_{\rm GGE}$. Second, 
{in the formal saddle-point calculation $\bar\gamma_\mu$ 
can be considered to be arbitrary; a canonical probability interpretation additionally requires normalizability.}
In Ref.~\cite{BaCuLoNe22} they had to be determined {\it a posteriori} via a consistency condition (namely, the conservation of the charges after the quench). In the latter setting, 
the $\bar\gamma_{\mu}$ turned out to be complicated functions of the 
$\lambda_{\mu}$. We refer the interested reader to Ref.~\cite{BaCuLoNe22} for the details of the calculations and quote here the most relevant results which we will use in the rest of the article.

Assuming a harmonic {\it Ansatz}, the expectation values of the square powers of the position and momentum of the particle on the sphere, 
under the measure~(\ref{eq:GB}) read
{
\begin{eqnarray}
\langle x_\mu^2 \rangle_{\rm GNM} &=& 
\left\{
\begin{array}{ll}
\displaystyle \frac{T_\mu}{z_{\rm GNM} - \lambda_\mu}  & \quad q=0
\\
[10pt]
\displaystyle \frac{T_\mu}{z_{\rm GNM} - \lambda_\mu} (1 - \delta_{\mu N}) + q N \delta_{\mu N} & \quad q \neq 0
\end{array}
\right.
\label{eq:xmu2GGE}
\end{eqnarray}
}
and
\begin{eqnarray}
\langle p_\mu^2 \rangle_{\rm GNM} &=& m T_\mu
\; , 
\label{eq:pmu2GGE}
\\
\langle x_\mu p_\mu \rangle_{\rm GNM} &=& 0
\; ,
\label{eq:xmupmuGGE}
\end{eqnarray}
where we included the possibility of there being a condensation on the last mode, 
captured by a non-vanishing Edwards--Anderson parameter $q = \langle x_N^2 \rangle_{\rm GNM} / N$, 
as in the standard SSK case. {These expressions belong to the large-$N$ harmonic Ansatz.}
The saddle point equations used to evaluate these averages with the measure (\ref{eq:GB}) 
and arbitrary $\bar\gamma_\mu$
reduce to $N$ coupled equations for the
mode ``temperatures'' $T_\mu$, the Lagrange multiplier $z_{\rm GNM}$ and $q$ 
{when there is condensation. In cases with $q=0$,
\begin{equation}
\frac{2T_\mu}{T}  = - \left(  \frac{1}{N} \sum_{\nu(\neq \mu)}  \frac{T_\nu}{z_{\rm GNM}-\lambda_\nu}
\frac{\bar\gamma_\mu-\bar\gamma_\nu}{\lambda_\mu-\lambda_\nu}  \right)^{-1} 
\; ,
\label{eq:Tmu-q0}
\end{equation}
together with the spherical constraint
\begin{equation}
\sum_{\mu} \frac{T_\mu}{z_{\rm GNM} - \lambda_\mu}  = N \, .
\end{equation}
In cases with $q\neq 0$
\begin{equation}
\frac{2T_\mu}{T}  = - \left(  \frac{1}{N} \sum_{\nu(\neq \mu, N)}  \frac{T_\nu}{z_{\rm GNM}-\lambda_\nu}
\frac{\bar\gamma_\mu-\bar\gamma_\nu}{\lambda_\mu-\lambda_\nu} 
+ q \;  \frac{\bar\gamma_N-\bar\gamma_\mu}{\lambda_N -\lambda_\mu} \right)^{-1} 
,
\label{eq:Tmu}
\end{equation}
together with 
\begin{equation}
\sum_{\mu (\neq N)} \frac{T_\mu}{z_{\rm GNM} - \lambda_\mu} + q N = N \, .
\end{equation}
The sums refer to the ordinary modes; in the condensed case the $N$th mode is separated explicitly. Coincident indices are 
excluded before taking the continuum limit.}
The parameter $q$ vanishes when there is no condensation, while it takes values between 0 and 1 in condensed phases in which the particle's position aligns along the direction~$\vec v_N$. 

For future reference, we signal that the dimensions of variables and parameters are
$[x_\mu]=[I_\mu]=1$, $[\lambda_\mu]=[\bar\gamma_\mu]= [z_{\rm GGE}]= [k_BT] = [J]$, $[t]=[p]/[J]$, $[m]=[p^2]/[J]$.
In the following we set $m=1$ unless otherwise stated.

\subsection{Gibbs-Boltzmann equilibrium for quadratic  $\bar\gamma_{\mu}(\lambda_{\mu})$}
\label{subsec:quadratic-gamma}

In order to proceed, we focus on the simplest generalization of the Neumann model, obtained by choosing a quadratic form for $\bar{\gamma}_\mu$. As we will show, despite its simplicity, this model exhibits a rich and nontrivial phase diagram.

We take parameters $\{\bar\gamma_\mu\}$ which depend
quadratically on the $\{ \lambda_\mu \}$:
\begin{equation}
\bar\gamma_\mu =  \frac{a}{J} \lambda_\mu^2 - b \lambda_\mu = 
2b \left(  \frac{a}{2bJ} \lambda_\mu^2 - \frac{1}{2} \lambda_\mu  \right)
\; ,   
\label{eq:choice-of-gamma}
\end{equation}
with $a$ and $b$ two dimensionless parameters.
For $a=0$ and $b=1/2$ the $\bar\gamma_\mu$ decrease linearly with $\lambda_\mu$ and 
the model boils down to the SSK one~(\ref{eq:harmonic-potential-SSK}), 
see App.~\ref{app:GGE-to-SSK}. 

The extrema of the quadratic function are
\begin{equation}
0 = \frac{2a}{J} \lambda^*_\mu   - b \quad \Rightarrow \quad \lambda^*_\mu = \frac{b}{2a} \, J
\end{equation}
and their stability is decided by  the sign of $2a/b$ ($J$ is positive).
Taking $a>0$ and $b>0$, for
\begin{equation}
b > 4a \; , 
\label{eq:condition-on-b}
\end{equation}
we respect the ordering
$\bar\gamma_1 > \dots > \bar\gamma_N$ ($|\lambda_\mu| < 2J$). 
The function (\ref{eq:choice-of-gamma}) is convex, 
its minimum 
is placed at the right of $2J$, and it reaches $2J$ for $a=b/4$. 
Taking, instead, $a<0$ and still $b>0$, the function (\ref{eq:choice-of-gamma}) is concave 
and as long as $a>-b/4$ the maximum is placed to the left of $-2J$, reaching $-2J$ for $a=-b/4$.

The limiting values $a=\pm 1/8$ for $b=1/2$, are shown with dashed vertical lines in the phase diagram in Fig.~\ref{fig:phase-diagram}. 
Representative $\bar\gamma_\mu(\lambda_\mu)$ for various values of $a$ are 
plotted in the lower left and right panels of the same figure for $a$ negative and 
positive, respectively.
Note that factorizing $b$ we can derive the results for 
generic $b$ from the ones for $b=1/2$ by renaming $a \mapsto a/(2b)$ and 
$\beta \mapsto \beta 2b$. 
We will evaluate the averages (\ref{eq:xmu2GGE})-(\ref{eq:xmupmuGGE}), 
and from them we will 
derive the static phase diagram of Fig.~\ref{fig:phase-diagram}, 
that will give us a framework to rationalize the stochastic dynamics.

In the rest of this section we derive a number of exact results for the statistical properties of this model 
in the $N\to\infty$ limit, which allow us to build the phase diagram in Fig.~\ref{fig:phase-diagram}. 
We confront them against the asymptotic stochastic evolution  in Sec.~\ref{sec:numerics}. 

\subsubsection{Extended states.}
\label{subsec:extended-quadratic}

For the quadratic $\bar\gamma_\mu$ in Eq.~(\ref{eq:choice-of-gamma}), and no condensation $q=0$, 
the equation (\ref{eq:Tmu-q0}) for the mode-resolved
temperatures reads
\begin{eqnarray}
- \frac{T_\mu}{T}  &=&  \left\{  \frac{2}{N} \sum_\nu  \frac{T_\nu}{z_{\rm GNM}-\lambda_\nu} 
\left[ \frac{a}{J}  (\lambda_\mu+\lambda_\nu) - b \right] \right\}^{-1}
 \label{eq:Tmu1}
\end{eqnarray}
which can also be written as
\begin{equation}
     \frac{2a}{J}  \,  \frac{T_\mu}{T}  = \frac{1}{A -  \lambda_\mu} 
     \label{eq:Tmu-biss}
\end{equation}
with 
\begin{eqnarray}
&& A = \frac{b}{a} J-  z_{\rm GNM}  + 2e_{\rm kin}
\; , 
\\
&& 2e_{\rm kin} \equiv 
\frac{1}{N} \sum_\mu T_\mu =   \frac{1}{mN}   \sum_\mu \langle p^2_\mu\rangle_{\rm GNM}
\; .
\label{eq:A}
\end{eqnarray} 
We have decided to call $2e_{\rm kin} $ the (normalised) sum of all averaged square momenta. In the model 
we are dealing with, momenta and velocities are not necessarily proportional. A study of how they compare
is presented in App.~\ref{app:momentum-vs-velocity}. In the following we will always refer to $e_{\rm kin}$ in this way.

All mode temperatures must be semi-positive definite. Therefore,  Eq.~(\ref{eq:Tmu-biss}) implies
\begin{eqnarray}
A-\lambda_\mu > 0 \;\; \forall \mu \;\;
\Rightarrow \;\; 
A>\lambda_N = 2J & &  \qquad {\rm for} \; a> 0 \; , 
\\
A-\lambda_\mu < 0 \;\; \forall \mu \;\;
\Rightarrow \;\; 
A<\lambda_1 \ = -2J & & \qquad {\rm for} \; a<  0 \; . 
    \end{eqnarray}
Equation~(\ref{eq:Tmu-biss}) also implies 
\begin{eqnarray}
&& {\rm sign} \displaystyle{ \frac{d\langle p_\mu^2\rangle_{\rm GNM}}{d\lambda_\mu} }  =
 {\rm sign} \; a \qquad \mbox{for all} \; \mu \; , 
 \\ && {\rm sign} \displaystyle{ \frac{d^2\langle p_\mu^2\rangle_{\rm GNM}}{d\lambda^2_\mu} } =
+ \qquad\quad\;\, \mbox{for all} \; \mu \; .
\end{eqnarray}
Thus,  a positive (negative) $a$ implies that $\langle p_\mu^2\rangle_{\rm GNM}$ is a growing (decreasing) 
function of $\lambda_\mu$ in the interval. The curvature of $ \langle p_\mu^2\rangle_{\rm GNM}$ 
as a function of $\lambda_\mu$ is positive for all $a$, since $a(A-\lambda_\mu)^3 >0$ 
irrespectively of the sign of $a$ once $T_\mu>0$ has been ensured. 

Some simple manipulations of Eq.~(\ref{eq:Tmu-biss}), explained in App.~\ref{app:details-of-calculations}, lead to 
                 \begin{eqnarray}
                 && e_{\rm kin} 
=   \frac{bT}{16 a^2} \left[ 1 - \sqrt{ 1 -  \frac{8  a^2}{b^2} \left( \frac{ 4 a J+T }{2 a J+T} \right) } \right] 
                \; . 
                 \label{eq:ekin-analytic-GGE0}
                 \end{eqnarray}
                          This expression  restricts the  parameters such that $e_{\rm kin}$ is real and positive,  to
                            \begin{equation}
0 \ \leq \ \frac{4aJ+T}{2aJ+T} \ \leq \  \frac{b^2}{8a^2}
\; .
\label{eq:one-condition0}
                            \end{equation}
                            The second condition ensures that $e_{\rm kin}$ is real.
                            Once the second condition enforced, the first inequality 
                            arises from the requirement $e_{\rm kin} \geq 0$ 
                            and imposes $a\geq -T/(4J)$ if $2aJ+T\geq 0$, or $a
                            \leq -T/(2J)$.
                             Focusing now on the right 
                            inequalities, one has
                            \begin{eqnarray}
                            \frac{T}{J}  \gtrless  2a \, \displaystyle{\frac{16a^2-b^2}{b^2-8a^2}} 
                                   \;\;
                                   {\mbox{for}}
                                   \;\;
                                   \begin{array}{l}
                                       \left\{
                                            \begin{array}{l}
                                                  2aJ+T>0 \;\; \& \;\; b^2 - 8a^2>0 
                                                  \\
                                                  2aJ+T<0 \;\; \& \;\; b^2 - 8a^2<0
                                             \end{array}
                                       \right.
                                       \;\; \mbox{or}
                                    \\
                                   [20pt]
                                      \left\{
                                          \begin{array}{l}
                                             2aJ+T>0 \;\; \& \;\; b^2 - 8a^2<0
                                             \\
                                             2aJ+T<0 \;\; \& \;\; b^2 - 8a^2>0
                                           \end{array}
                                     \right.
                                     \;\; \mbox{or}
                                 \end{array}
                                 \label{eq:conditions-yellow0}
                           \end{eqnarray}
                            These upper and lower bounds for $T/J$ vanish at $|a|=|b|/4$ and diverge 
                            at $|a|=|b|/\sqrt{8}$. In the numerical application, $b=1/2$, and these two values are
                            $a=\pm 1/8$ and $a=\pm 1/(4\sqrt{2})$. In the  interval $a\in [|b|/4, |b|/\sqrt{8}]$,  
                            the relevant parameters are the ones in the first line in Eq.~(\ref{eq:conditions-yellow0}), 
                            and the condition on $T/J$ 
                            boils down to 
                            $T/J>2a \, (16a^2-b^2)/(b^2-8a^2)$ which is displayed with a yellow solid line 
                            in the phase diagram in Fig.~\ref{fig:phase-diagram}. This line marks a transition between 
                            the extended phase (E) with $\langle x^2_\mu\rangle ={\mathcal O}(1)$ and 
                            $\langle p^2_\mu\rangle ={\mathcal O}(1)$ for all $\mu$, and a moving phase (M$_R$) in which some 
                            components of the momentum are not bounded and diverge. For negative $a$, the conditions 
                            above impose $T/J<2a \, (16a^2-b^2)/(b^2-8a^2)$ for $a<-|b|/\sqrt{8}$, and the bound 
                            is also displayed with a green solid line in Fig.~\ref{fig:phase-diagram}, which also 
                            separates the extended (E) phase from a moving phase (M$_L$).
                    
                    One can also derive the limits 
                 \begin{eqnarray}
e_{\rm kin} \to 
\left\{
\begin{array}{ll}
\displaystyle{ \frac{T}{4b} }+  \frac{a}{2b} J+  {\mathcal O}(a^2) & \quad a\sim 0 \;\; (\sim {\rm SSK}) 
\\ 
[10pt]
\displaystyle{ \frac{1}{16 a^2 } \left(b-\sqrt{b^2-8 a^2}\right) \; T}
+{\mathcal O}\left(1\right)
 & \quad T\to \infty \;\; \& \;\;  a \neq 0 \label{eq:ekin-Tinf}
\end{array}
 \right.
 \end{eqnarray}

The expression for the Lagrange multiplier reads (see App.~\ref{app:details-of-calculations})
\begin{eqnarray}
&& 
z_{\rm GNM} =
 \frac{b [ 4(8aJ+7T) (aJ)^2 + (8aJ+T) T^2]}
{8 a^2 (2aJ+T) (4 a J + T)}
\nonumber\\
&& 
\qquad\quad
- \frac{bT }{8a^2} \frac{6aJ+T}{4aJ+T} 
\sqrt{1 - \frac{8a^2}{b^2} \left( \frac{4aJ+T}{2aJ+T} \right)}
\label{eq:zGGE-aboveTc0}
\end{eqnarray}
      with limits
       \begin{eqnarray}
z_{\rm GNM}
 \to 
\left\{
\begin{array}{ll}
\displaystyle{ \frac{4 b^2 J^2+T^2}{2 b T}}
+  {\mathcal O}(a)  
\qquad\quad\qquad\quad  
a\sim 0  \;\; (\sim {\rm SSK})&
\\
[10pt]
\displaystyle{ \frac{T}{8a^2} \left(
b-\sqrt{b^2-8 a^2}\right)}
+{\mathcal O}\left(1\right)
  \quad\;\;\;
    T\to \infty \;\; \& \;\;  a \neq 0 \label{eq:zext-Tinf}&
\end{array}
\right.
\end{eqnarray}
The slope of $z_{\rm GNM}$ at large $T$ is twice the one of $e_{\rm kin}$.
The second condition in Eq.~(\ref{eq:one-condition0}) ensures that the square root yields a real result.

We determine the critical temperature towards the condensed phase from the condition $z_{\rm GNM}=2J$, 
see App.~\ref{app:details-of-calculations}. It reads
\begin{equation}
T_c = b  J \Big(1 -  \frac{3a}{b}  \pm \sqrt{1- \frac{2a}{b} - \frac{7a^2}{b^2}} \Big)
\; . 
\label{eq:Tc0}
\end{equation}
This expression approaches $T_c =2bJ$ (the critical temperature of the SSK model)
for $a\to 0$ if we choose the $+$ sign. This sign selection 
corresponds to the blue solid line in the phase diagram of Fig.~\ref{fig:phase-diagram}, and yields
a real critical temperature for $a$ in between 
$a^-=-b(1+2\sqrt{2})/7$ and  $a^+=b(-1+2\sqrt{2})/7 $. 
For $b=1/2$, these two extremes are $a^-=-0.273$ and $a^+\sim 0.130$
and the corresponding temperatures are
$T_c^-/J =  (5+3\sqrt{2})/7 \simeq 1.320$ and $T_c^+/J =  (5-3\sqrt{2})/7 \simeq 0.108$. 
For positive $a$ and below $T_c^+$ the critical line is continued 
by the branch of Eq.~(\ref{eq:Tc0}) with the 
negative sign,  drawn as the red curve in Fig.~\ref{fig:phase-diagram}, which is easier to visualise in the 
enlargement of the right upper panel. It
 joins $(a^+\sim 0.130, T_c^+/J =  0.108)$ and  
$(T_c/J=0, a=1/8)$. Interestingly enough, there is a re-entrance  of the extended phase below
the condensed one as temperature decreases at fixed $a \in [b/4, a^+]$.
The kinetic energy on the transition towards the condensed phase {equals 
\begin{equation}
e_{\rm kin}^c = \frac{JT_c}{2(T_c+2aJ)}
\end{equation}
both on the blue and red branches of $T_c$ in Eq.~(\ref{eq:Tc0}), and
it  correctly yields $e_{\rm kin}^c=J/2$ for $a=0$.}

\subsubsection{Condensed states.}

We will now assume that there is a low $T$ phase in which the $N$th mode condenses, in the sense that $\langle x_N^2\rangle = qN$.
The mode temperatures $T_\mu$ and the parameter $q>0$  are determined by the constraint and Eqs.~(\ref{eq:Tmu}):
\begin{eqnarray}
\quad \;\; 1 & = & \frac{1}{N} \sum_{\mu (\neq N)} \frac{T_\mu}{2J-\lambda_\mu} + q 
\; , 
\\
 - \frac{T_\mu}{T}  &  = &   \left\{  \frac{2}{N} \sum_{\nu\neq N}  \frac{T_\nu}{2J-\lambda_\nu} 
\left[ \frac{a}{J}  (\lambda_\mu+\lambda_\nu) - b \right]  +   2q \left( \frac{\bar\gamma_\mu-\bar\gamma_N}{\lambda_\mu -\lambda_N} \right)
\right\}^{-1}
\!\!\!\! , 
\end{eqnarray}
where we used $z_{\rm GNM}=2J$. In the same way as for the conventional SSK model in canonical 
equilibrium, one needs to give a macroscopic weight to the last Cartesian component to satisfy the 
spherical constraint and the Lagrange multiplier freezes at the minimal value it can take~\cite{KoThJo76}.

Following similar steps as the ones used in the analysis of the extended phase, 
which are also detailed in App.~\ref{app:details-of-calculations}, 
one derives
\begin{eqnarray}
- \frac{T_\mu}{T} &=&
 \Big\{ \frac{2a}{J} (1-q) \lambda_\mu - 2b (1-q) 
-   \frac{4a}{J} \, e_{\rm kin}
+ 
 4a (1-q)
 \nonumber\\
 &&
 \quad
 + \; 2q 
 \left[ \frac{a}{J}  (\lambda_\mu+2J) - b \right]
\Big\}^{-1}
 \; . 
 \label{eq:Tmu-condensed0}
\end{eqnarray}
In the $T\to 0$ limit, 
\begin{equation}
\lim\limits_{T\to 0} T_\mu = -  \displaystyle{\frac{T}{2}} \displaystyle{\frac{1}{\left[ \displaystyle{\frac{a}{J}}  (\lambda_\mu+2J) - b \right]}}
=
-  \displaystyle{\frac{T}{2}} \displaystyle{\frac{1}{\left[ \displaystyle{\frac{a}{J}}  \lambda_\mu+ (2a - b) \right]}}
\end{equation}
since $q=1- c T$, as we will show below.
As a way of check, we confirm that in the SSK/Neumann case $a=0$ and $b=1/2$, 
\begin{eqnarray}
T_\mu &=& T
 \left\{   2b (1-q) 
 + \; 2b \, q \right\}^{-1}  = T/(2b) 
 = T
 \; , 
 \label{eq:Tmu-condensed-a0=appF}
\end{eqnarray}
where we used $q=1-T/J$, which is self-consistently derived from the spherical constraint
setting $T_\mu=T$. In the $T\to 0$ limit, this solution can only exist for $a<b/4$, which 
ensures $T_N>0$.

For general $a$ and $b$, the kinetic energy density reads
\begin{eqnarray}
&& 
\frac{2e_{\rm kin}}{T}  
=
\frac{b-2a 
-
\sqrt{(b-2a)^2 - \frac{2a}{J} ( 2a J- T ) }
}
{\frac{2a}{J}  (2a J - T )}
\; . 
\label{eq:ekin-belowTc0}
\end{eqnarray}
We chose the sign in front of the square root to get the correct $a\to 0$, $b=1/2$ limit, 
$2e_{\rm kin} = T$ at all $T$. 
At $T\to 0$:
\begin{eqnarray}
&& e_{\rm kin} =\frac{T}{2}  \left[
\frac{b-2a - \sqrt{b^2-4ab}}{(2a)^2}
\right]
\; . 
\end{eqnarray}
We note that the functional 
form that we got for $e_{\rm kin}$ in this condensed phase is not the same as the one we had 
in the extended phase. In Fig.~\ref{fig:ekin-zGGE} in Sec.~\ref{sec:numerics} we show that the full expression, with Eq.~(\ref{eq:ekin-analytic-GGE0}) in the extended phase and 
Eq.~(\ref{eq:ekin-belowTc0}) in the condensed phase, describes the asymptotic kinetic energy at all temperatures
in these two phases.
 
 Evaluating Eq.~(\ref{eq:Tmu-condensed0}) at $\mu=N$, and setting $\lambda_N=2J$, one has
\begin{eqnarray}
- \frac{T_N}{T} 
&=&
  \Big( 8a  -2b 
-   \frac{4a}{J} \, e_{\rm kin}
\Big)^{-1}
\end{eqnarray}
which does not explicitly depend on $q$.
Using now the expression for $e_{\rm kin}$ in Eq.~(\ref{eq:ekin-belowTc0})
\begin{eqnarray}
-\frac{T_N}{T} 
&=& 
 \left( 8a  -2b 
-   \, T
\frac{b-2a 
-
\sqrt{(b-2a)^2 - \frac{2a}{J} ( 2a J- T ) }
}
{ (2a J - T )}
\right)^{-1}
 \label{eq:TN-lowT0}
\end{eqnarray}
which for $T\to 0$ approaches
\begin{equation}
T_N \ \  \mathop{\longrightarrow}\limits_{T\to 0} \  \
\frac{T}{2}  (b-4a)^{-1}
\; . 
\end{equation}
If we set $4a=b$, before taking $T\to 0$, 
\begin{equation}
T_N
= 
\left( 
\frac{2a 
-
\sqrt{ \frac{2a}{J} T }
}
{ 2a J - T }
\right)^{-1} 
 = 
 \frac{2aJ-T}{2a - \sqrt{ \frac{2a}{J} T} }  \ \ \mathop{\longrightarrow}\limits_{T\to 0} \  \ J
\end{equation}

The expression for $q$ in terms of $a$, $b$, $T$ and $J$ is
\begin{eqnarray}
&&
q =
1+ 
\frac{ T - 
 T \sqrt{
 \displaystyle{
 \frac{-4 a b J + 2 a T + b T + T \sqrt{-4 a b + b^2 + \frac{2 a T}{J}}}{
  16 a^2 J - 4 a b J - 6 a T + b T + 
   T \sqrt{-4 a b + b^2 + \frac{2 a T}{J}}}}}}{4 a J}
   \label{eq:qEA0}
\end{eqnarray}
and it is plotted with solid lines for several values of $a$ in Fig.~\ref{fig:xNpN-highT}.
 For $a\to 0$ one recovers the correct linear law $q=1-T/(2bJ)$.  We can also verify that 
$q=0$ on the critical line towards the extended phase, on both (blue and red in Fig.~\ref{fig:phase-diagram}) 
branches of it. 
  For $T\to 0$ and any $a<b/4$, 
\begin{eqnarray}
q = 1 - \frac{T}{4aJ} \left( - 1 + \sqrt{\frac{b}{b-4a}} \right)
\label{eq:qEA-lowT}
\end{eqnarray}
which, further taking $a\to 0$ recovers $q=1-T/(2bJ)$.  Another interesting limit is $q\to 1+ T/(4aJ)$ for $a\to -\infty$. For $0<a<b/4$ the concavity of $q(T)$ is positive while for $a<0$ it is negative. Interestingly enough, for $a>b/4$ in the re-entrant region, $q(T)$ is non monotonic and varies from zero to zero at the two ends of the critical line (see the gray curve in Fig.~\ref{fig:xNpN-highT}).

We can derive a limit of validity of the condensed phase from the condition that $q$ must be real. 
By requiring that the internal square root be positive or zero one has
\begin{eqnarray}
\frac{T}{J} \lessgtr \displaystyle{\frac{4ab-b^2}{2a}}   \qquad a  \lessgtr 0
\label{eq:limit-real-q0}
\end{eqnarray}
 For $a<0$, $q$ is real below the critical line 
 \begin{equation}
 T_c(a)/J = (4ab-b^2)/(2a)
\label{eq:Tc-ab}    
 \end{equation}
 drawn with violet colour 
 in Fig.~\ref{fig:phase-diagram}. The condensed phase extends all the way to $a\to -\infty$.
 
 We can note that the three limiting curves  obtained for $a<0$ meet at a single point.
Indeed, equating $T_c/J$ from the real condition on the kinetic energy (yellow line in Fig.~\ref{fig:phase-diagram}), 
$T_c/J$ from the real condition on $q$ (violet line), and  $T_c/J$ in Eq.~(\ref{eq:Tc0}), one finds
\begin{equation}
64 a^4 - 12 a^2 b^2 - 4 ab^3 + 32 a^3 b + b^4 = 0
\end{equation}
which, for the particular case $b=1/2$, yields $a^*=-1/4$ and 
$T_c^*/J=3/2$.

Instead, for $a>0$ the condition (\ref{eq:limit-real-q0}) becomes a lower limit and the condensed phase can 
only exist above the curve $(4ab-b^2)(2a)$ which touches $T=0$ at $a=b/4$ and tends to 
$2b$ for $a\to\infty$. The condensed phase is, however, limited by a more restrictive 
condition which was already derived from the analysis of $T_c$ in Eq.~(\ref{eq:Tc0})
for $a>0$.

\subsubsection{The phase diagram}

\begin{figure}[h!]
\centerline{
\includegraphics[scale=0.45]{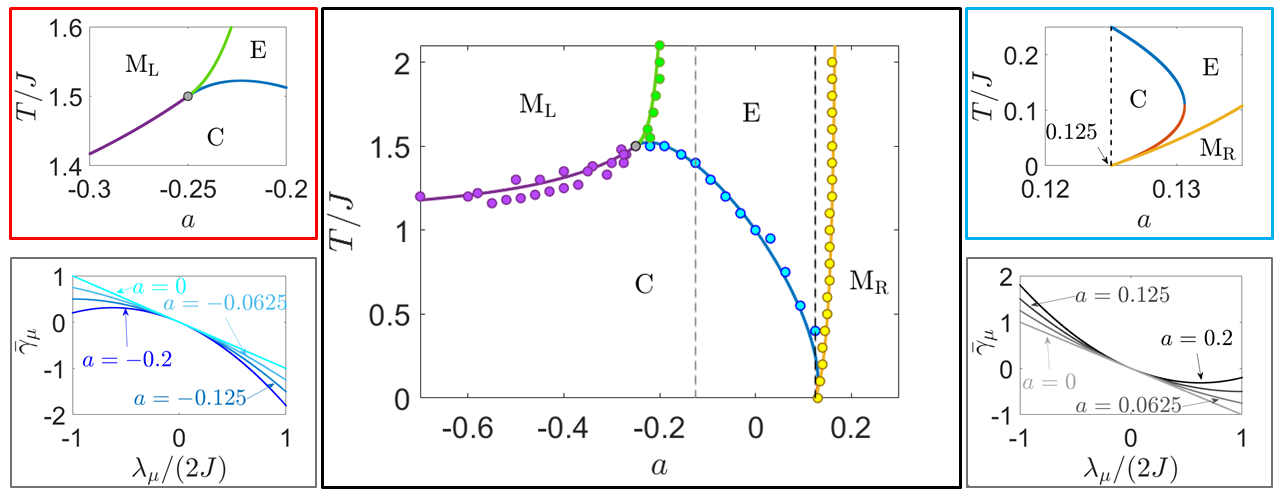}
}
\caption{\small
The full phase diagram {in the ($a, T/J$) plane} of the Generalized Neumann model ${\mathcal H}_{\rm GNM} =
\sum_\mu \bar\gamma_\mu I_\mu$ with  
quadratic $\bar\gamma_\mu =a\lambda_\mu^2/J - b\lambda_\mu$
and $b=1/2$. The vertical dashed lines are situated at $a=\pm 1/8$: 
for $a<-1/8$ ($a>1/8$), 
$\bar\gamma_\mu(\lambda_\mu)$ has a maximum (minimum) within the interval $[-2J, 2J]$.
There are three phases: condensed (C, with  
$\langle x_N^2 \rangle/N = q\neq 0$, $\langle x_{\mu\neq N}^2 \rangle = {\mathcal O}(1)$
and $\langle p_\mu^2 \rangle = {\mathcal O}(1)$), 
extended (E, with $\langle x_\mu^2 \rangle = {\mathcal O}(1)$
and $\langle p_\mu^2 \rangle = {\mathcal O}(1)$) and 
two phases in motion (M$_R$ and M$_L$, in which modes close to the right of the spectrum in 
M$_R$ and modes close to the left of the spectrum in M$_L$ increase their momentum indefinitely).
The transition between the extended and condensed phase  is given by Eq.~(\ref{eq:Tc0}) with the $+$ sign 
in front of the square root (blue solid line) and the reentrance (red solid line, zoom in the  upper right panel) 
represents Eq.~(\ref{eq:Tc0}) with the $-$ sign in front of it. This transition is continuous, with $q=0$. 
The yellow and green  lines are the transition between the extended and moving regions M$_R$ and M$_L$, 
respectively, 
and are the limiting parameters for which $e_{\rm kin}$ in Eq.~(\ref{eq:ekin-analytic-GGE0}) is real.
The asymptotic limit of the yellow line is $a\sim 1/\sqrt{32}\sim 0.177$ for $T\to \infty$.
The parameter $q$ depends on $T$ and $a$, according to Eq.~(\ref{eq:qEA0}).  When it ceases being real there
is a phase transition between the condensed and M$_L$ phase (violet line, see the upper left panel, and Eq.~(\ref{eq:Tc-ab})). $q$ is finite 
at this transition.
The data points are estimates from the numerical analysis of the stochastic dynamics. 
The two small lower panels show how the 
parameters $\bar\gamma_\mu$ change when varying $a$: negative values induce a negative curvature and a maximum
within the interval $[-2J,2J]$ for $a<-0.125$ while positive values lead to a positive curvature and a minimum within the 
interval $[-2J,2J]$ for $a>0.125$. The two small upper panels show zooms over the regions where three
critical lines merge. {Outside the domain of a normalizable canonical measure, 
the large-$N$ saddle-points predict dynamical regimes, rather than global equilibrium phases.} 
}
\label{fig:phase-diagram}
\end{figure}

The results derived in the previous subsections and in 
App.~\ref{app:details-of-calculations}
are summarised in the phase diagram of Fig.~\ref{fig:phase-diagram}. 
The critical curves are drawn with solid lines. Three types of 
phases were identified: extended (E), condensed (C) and two 
moving ones (M$_L$ and M$_R$). Further details are given in the 
caption.

The moving phases are specific to the quadratic model and do not exist in the 
SSK/Neumann ($a=0$) one. In these phases, some components of the momentum are unbounded and
their $\langle p_\mu^2\rangle$ diverge while the corresponding coordinates remain bounded because of the spherical 
constraint. The motion with unbounded momentum is confined to subspaces of the sphere. 
This occurs for modes $\mu$ which lie to the 
right of the minimum of $\bar\gamma_\mu(\lambda_\mu)$ in M$_R$ or to the left of 
the maximum of $\bar\gamma_\mu(\lambda_\mu)$ in M$_L$. However, the limits of existence 
of the extended phase depend on temperature and do not coincide with $a=\pm 1/8$, where the minimum 
and maximum of $\bar\gamma_\mu(\lambda_\mu)$ enter the interval $[-2J, 2J]$ from the 
right and left ends, respectively. 
Interestingly enough, the extended phase 
shows a reentrance driven by temperature, as it extends beyond $a=1/8$. Similarly, on the other side of the 
phase diagram, the extended phase subsists for  $a<-1/8$. Therefore, a simple mechanical stability 
analysis performed for finite $N$, as the one in App.~\ref{app:N2} is not enough to derive the 
macroscopic collective behaviour of the system in the thermodynamic limit.

\section{Langevin Dynamics}
\label{sec:evol-GGE}

In this section we describe the stochastic evolution of spherically constrained models. 
We start by 
introducing noise and dissipation in the equations of motion for the SSK/Neumann model 
in Sec.~\ref{subsec:Langevin-GGE1}. We then specialize to the model of interest,
${\mathcal H}_{\rm GNM}$, in Sec.~\ref{subsec:Langevin-GGE}. 
Finally, in Sec.~\ref{subsec:approach-to-equil} we prove
that the stochastic process approaches the desired asymptote by using the Kramers equation.

\subsection{Stochastic equations of the SSK/Neumann model}
\label{subsec:Langevin-GGE1}

The relaxational noisy dynamics of the SSK model was studied in detail in~\cite{CuDe95a,CuDe95b,CuDeYo07} 
in the over-damped limit. In the underdamped case, the Langevin equations of motion are 
\begin{eqnarray}
&& m\dot x_\mu = p_\mu
\; , 
\\
&&
 \dot p_\mu + \gamma \dot x_\mu = 
  - \frac{\partial {\mathcal H}^z_{\rm SSK}}{\partial x_\mu}  + \xi_\mu 
  = 
  - [\bar z_{\rm SSK}(\vec x, \vec p, \vec \xi) - \lambda_\mu] x_\mu +\xi_\mu
\; , \label{eq:langevin-ssk}
\end{eqnarray}
with $\xi_\mu$ independent white noises with zero mean, $\langle \xi_\mu(t) \rangle = 0$, and delta correlations
$\langle \xi_\mu(t) \xi_\nu(t') \rangle = 2\gamma T \delta_{\mu\nu} \delta(t-t')$. Note that the Lagrange
multiplier $\bar z_{\rm SSK}(\vec x, \vec p, \vec \xi)$ depends on the phase space variables and the noise (compare with
expression (\ref{eq:Newton}) for the noiseless equations of motion).
The friction parameter $\gamma$ represents the strength of the coupling to the bath. The ratio  $m/\gamma$
is a time-scale that typically controls the equilibration of the momenta to the Maxwell probability density.
(One can show that for second order time evolution $\langle \dot x(t) x(t^-\to t)\rangle = \langle \dot x(t^-\to t) x(t)\rangle \to 0$, contrary to what happens in the strict over-damped case.) 

As we will show below, in order to keep the particle on the sphere, the Lagrange multiplier 
$\bar z_{\rm SSK}$ has to depend
explicitly on the noise realisation. Indeed, a projection of the random force on a direction tangential to the 
sphere is needed to keep the particle on the sphere. We can anticipate here that after fixing the 
expression of $\bar z_{\rm SSK}$ in terms of $\vec x, \vec p$ and $\vec \xi$, the noise term will end up being of the form
\begin{equation}
\sum_{\nu=1}^N {\rm P}^\mu_\nu \, \xi_\nu , 
\qquad\mbox{with}\qquad
{\rm P}^\mu_{\, \nu} \equiv \delta^{\mu}_{\nu} - \frac{1}{N} x^\mu x_\nu \; ,  
\label{eq:projector}
\end{equation}
a projector that keeps the random force tangent to the sphere and satisfies
\begin{equation}
P^\mu_\eta P^\eta_\nu = \Big(\delta^\mu_\eta - \frac{1}{N} x^\mu x_\eta \Big)  \Big(\delta^\eta_\nu - \frac{1}{N} x^\eta x_\nu \Big) = 
\delta^\mu_\nu -\frac{1}{N}  x^\mu x_\nu = P^\mu_\nu
\; . 
\end{equation}
The noise has become multiplicative.

The time derivative of functions of stochastic processes with multiplicative noise 
can be tricky and depend on the discretisation scheme used. 
For the multiplicative noise in Eq.~(\ref{eq:projector}) and the linear discretisation rule
\begin{equation}
\overline x_t^\mu = x^\mu_t + \alpha \Delta x_t^\mu
\qquad\mbox{and}\qquad
\overline p_t^\mu = p^\mu_t + \alpha \Delta p_t^\mu
\; , 
\label{eq:linear-discretization}
\end{equation} 
one has
\begin{equation}
\frac{dF}{dt} = \frac{\partial F}{\partial x^\mu} \dot x^\mu + \frac{\partial F}{\partial p^\mu} \dot p^\mu 
+ (1-2\alpha) \, \gamma k_BT \, {\rm P}^{\mu\eta} {\rm P}^{\nu\eta} \frac{\partial^2 F}{\partial p^\mu \partial p^\nu}
\; , 
\end{equation}
for a generic function $F$ of $\vec x$ and $\vec p$. For an $F$ which depends only on $\vec x$, as for the spherical constraint, 
the last term vanishes identically and 
it is licit to take a conventional  time derivative of it.

Now, multiplying  the second Langevin equation by $x_\mu$
\begin{eqnarray}
&& \dot p_\mu x_\mu + \gamma \,  \dot x_\mu x_\mu=  - [\bar z_{\rm SSK}(\vec x, \vec p, \vec \xi) - \lambda_\mu] \, x_\mu x_\mu +\xi_\mu x_\mu
\; ,
\end{eqnarray}
summing over $\mu$ and using the relation~(\ref{eq:phi-prime}) due to the
orthogonality of $\vec p$ and $\vec x$, we obtain an expression for the Lagrange multiplier
\begin{eqnarray}
N \bar z_{\rm SSK}(\vec x, \vec p, \vec \xi) =2 E_{\rm kin}(p)  -2  V (\vec x) + \sum_{\mu=1}^N \xi_\mu x_\mu
\; .
\end{eqnarray}
(Note that on average over the noise the last term vanishes since it is related to the linear response at equal times. 
With no average,  it should still be 
sub-leading with respect to the other ones in the large $N$ limit.)
Back in the Langevin equation (\ref{eq:langevin-ssk})
\begin{eqnarray}
&& \dot p_\mu  
= \lambda_\mu x_\mu - \frac{2}{N} \Big[ E_{\rm kin}(p)   - V (\vec x)   \Big] x_\mu 
 - \gamma   \dot x_\mu + \sum\limits_{\nu=1}^N {\rm P}^\mu_{\, \nu}(x_\mu x_\nu) \, \xi_\nu 
 \; , 
 \label{eq:Langevin-pmu}
\end{eqnarray}
with the projector ${\rm P}^\mu_{\, \nu} $ acting on the noise as announced above.

The equation (\ref{eq:Langevin-pmu}) does not have a balanced form for the coupling to the bath: the noise is acted upon with the
projector, while the friction term is not. However, under the spherical constraint,
\begin{equation}
\dot x^\mu = {\rm P}^{\mu}_{\, \nu} \, \dot x^\nu \qquad \mbox{using} \qquad \sum_\nu x_\nu \dot x_\nu = 0
\; . 
\end{equation}
Thus, 
\begin{eqnarray}
\dot p_\mu   
 &  =   &  \underbrace{\left[ \lambda_\mu - z_{\rm SSK}(\vec x, \vec p) \right] 
 x_\mu }_{ f_\mu^{(p)}(\vec x, \vec p)} 
 + \sum\limits_{\nu=1}^N  {\rm P}_\mu^{\, \nu}(x_\mu x_\nu) \left(
- \gamma   \dot x_\nu +   \xi_\nu \right)
 \label{eq:Langevin-pmu-final}
\end{eqnarray}
with
\begin{eqnarray}
 z_{\rm SSK}(\vec x, \vec p) \equiv  \frac{2}{N} \Big[ E_{\rm kin}(p)   - V (\vec x)   \Big] 
\; . 
\label{eq:z-Langevin}
\end{eqnarray}
Note that $z_{\rm SSK}$ is now the same as in the Hamiltonian case.
The first term in the right-hand side of Eq.~(\ref{eq:Langevin-pmu-final}) 
is the $\mu$th component of a ``deterministic'' force $\vec f^{(p)}(\vec x, \vec p)$. 

In order to make contact with other studies of stochastic dynamics on constrained surfaces, 
we notice that we can rewrite Eq.~(\ref{eq:Langevin-pmu-final}) in the form of Eq.~(8) in Ref.~\cite{Schwartz}. 

We remark here that for the numerical implementation, Eq.~(\ref{eq:Langevin-pmu}) is efficient in keeping the 
particle on the sphere while Eq.~(\ref{eq:Langevin-pmu-final}) enhances the numerical error that violates the constraint 
and it is not convenient to use. In the numerical checks 
we therefore use Eq.~(\ref{eq:Langevin-pmu}). 

The relation between the Langevin process and the Kramers formulation
for a model with constraints is derived in App.~\ref{app:Kramers}.

\subsection{Stochastic equations of the GNM
}
\label{subsec:Langevin-GGE}

The Langevin equations 
for the model ${\mathcal H}_{\rm GNM}= \sum_\nu \bar\gamma_\nu I_\nu$ with the constraints
$\phi_1$ and $\phi_2$ are
\begin{eqnarray}
&& \dot x_\mu = \frac{\partial {\mathcal H}_{\rm GNM}}{\partial p_\mu} + z_2 x_\mu \, , 
\label{eq:Langevin-eqs-GGE1}
\\
&& 
\dot p_\mu = -\frac{\partial  {\mathcal H}_{\rm GNM}}{\partial x_\mu} - \! z_1 x_\mu - \! z_2 p_\mu \! + \!
(-\gamma \dot x_\mu + \xi_\mu) 
\; .
\label{eq:Langevin-eqs-GGE2}
\end{eqnarray}
We fix $z_2$ just as we did above, 
\begin{equation}
z_2 = 
- \frac{1}{N} \sum_\mu x_\mu \frac{\partial {\mathcal H}_{\rm GNM}}{\partial p_\mu} = 0
\; , 
\end{equation}
thanks to Eq.~(\ref{eq:z2-equal-0}). We note that this identity follows from symmetry reasons and as such it will not be 
affected by imprecisions  in the numerical solution that we will implement later.
Next, we fix $z_1$ by multiplying the second equation by $x_\mu$ and summing over $\mu$:
\begin{equation}
\sum_\mu x_\mu \dot p_\mu = - \sum_\mu x_\mu \frac{\partial  {\mathcal H}_{\rm GNM}}{\partial x_\mu} - z_1 \sum_\mu x_\mu^2
+ \sum_\mu (-\gamma \dot x_\mu + \xi_\mu) x_\mu
\; . 
\end{equation}
The last two terms come from the coupling to the bath. The term proportional to $\gamma$ vanishes and is 
not affected by numerical errors.
Now, we exchange the term in the left-hand side using Eq.~(\ref{eq:phi-prime})  (an operation that may lead to numerical 
errors) and
\begin{equation}
z_1 =   \frac{1}{\sum\limits_\nu x_\nu^2} 
\left( \sum_\mu \dot x_\mu p_\mu -   \sum_\mu x_\mu \frac{\partial  {\mathcal H}_{\rm GNM}}{\partial x_\mu} 
+ \sum_\mu  \xi_\mu x_\mu 
\right)
\; . 
\end{equation}
Analytically, the  denominator equals $N$. However, we find that the numerical  stability of the  
code that we will use is greatly improved by keeping this full expression there.
In the case of a Hamiltonian $ E_{\rm kin}(p)+V(\vec x)$, with $E_{\rm kin}(p)$ quadratic 
in $p_\mu$ the two first terms in the right-hand side are  $2E_{\rm kin}(p) - 2V(\vec x)$, but this is not the 
case for the Hamiltonian of our interest here. 
The complete equations then read
\begin{eqnarray}
 \dot x_\mu & = & \frac{\partial {\mathcal H}_{\rm GNM}}{\partial p_\mu} \ , \nonumber \\
\dot p_\mu & = & -\frac{\partial  {\mathcal H}_{\rm GNM}}{\partial x_\mu} 
-
 \frac{1}{\sum\limits_\eta x_\eta^2} 
\left(  \sum_\nu \frac{\partial {\mathcal H}_{\rm GNM}}{\partial p_\nu}  p_\nu -   \sum_\nu x_\nu \frac{\partial  {\mathcal H}_{\rm GNM}}{\partial x_\nu} 
+ \sum_\nu  \xi_\nu x_\nu
 \right) x_\mu 
 \nonumber\\ 
&&  -\gamma \dot x_\mu + \xi_\mu \; , 
 \ \ \ \ \ \ \ \ 
\label{eq:Langvein-eqs-GGE-1}
\end{eqnarray}
where we replaced $\dot x_\nu$ by $\partial {\mathcal H}_{\rm GNM}/\partial p_\nu$ in the first line of the right-hand side.
Note that we can rewrite the second equation with the projector acting on the terms
that correspond to the bath
\begin{eqnarray}
&& 
\!\!\!\!\!\!\!\!\!\!
\dot x_\mu = \frac{\partial {\mathcal H}_{\rm GNM}}{\partial p_\mu} 
\; , 
\label{eq:Langevin-eqs-GGE-2}
\\
&&
\!\!\!\!\!\!\!\!\!\!
\dot p_\mu = 
\underbrace{ 
-\frac{\partial  {\mathcal H}_{\rm GNM}}{\partial x_\mu} 
-
\underbrace{   \frac{1}{\sum\limits_\eta x_\eta^2} \left(   \sum_\nu \frac{\partial {\mathcal H}_{\rm GNM}}{\partial p_\nu} p_\nu -  \sum_\nu x_\nu \frac{\partial  {\mathcal H}_{\rm GNM}}{\partial x_\nu} 
 \right) }_{z_{\rm GNM}(\vec x, \vec p)} x_\mu 
 }_{f_\mu^{(p)}}
 \nonumber\\
 &&
 +
\sum_\nu {\rm P}^{\mu\nu} (-\gamma \dot x_\nu + \xi_\nu)
\; .
 \ \ \ \ \ \ \ \ 
\label{eq:Langevin-eqs-GGE-3}
\end{eqnarray}
One can easily verify that $\dot{\vec x} \cdot \vec p + \vec x \cdot  \dot{\vec p}=0$ 
also under the stochastic evolution. The only new terms arise from the 
coupling to the bath but they vanish identically thanks to $\sum_\mu {\rm P}^{\mu\nu} x_\mu = 0$.
Therefore, if we start from an initial condition that satisfies $\vec x\cdot \vec p =0$, the orthogonality is 
conserved by the Langevin process. This set of equations should then 
be apt to be integrated numerically.

A comment is in order here. The final equation (\ref{eq:Langevin-eqs-GGE-3}) has multiplicative noise, 
which stems from the evaluation of the Lagrange multiplier $z_1$, where we used $d(\vec x \cdot \vec p)/dt=0$
to replace $\sum_\mu x_\mu \dot p_\mu$ by $-\sum_\mu \dot x_\mu p_\mu$. With multiplicative noise, 
as already said, one needs to be specially careful about the discretization scheme used in the definition of the stochastic process and, in particular, 
worry about the validity of the chain rule applied to time derivatives. It turns out that because of the structure of  
the Eqs.~(\ref{eq:Langevin-eqs-GGE-2})-(\ref{eq:Langevin-eqs-GGE-3}), and the simple form of the 
function to be derived which is just $\vec x \cdot \vec p$, the standard chain rule applies for any discretization scheme.
Its use in the derivation of $z_1$ above is justified {\it a posteriori}.

Even more explicitly,  $\partial {\mathcal H}_{\rm GNM}/\partial x_\mu$ is
\begin{eqnarray}
&&
\frac{\partial {\mathcal H}_{\rm GNM}}{\partial x_\mu}
=
 2 \bar\gamma_\mu x_\mu + \frac{2}{mN} 
 \sum_{\nu (\neq \mu)}  \, p_\nu  \, ({p_\mu x_\nu - x_\mu p_\nu}) \, 
 \frac{\bar\gamma_\mu - \bar\gamma_\nu}{\lambda_\mu - \lambda_\nu}
\label{eq:primera}
\end{eqnarray}
and  $\partial {\mathcal H}_{\rm GNM}/\partial p_\mu$ is
\begin{eqnarray}
&&
\frac{\partial {\mathcal H}_{\rm GNM}}{\partial p_\mu}
=
  \frac{2}{mN} 
 \sum_{\nu (\neq \mu)}  \, x_\nu  \, ({x_\mu p_\nu - p_\mu x_\nu}) \, 
 \frac{\bar\gamma_\mu - \bar\gamma_\nu}{\lambda_\mu - \lambda_\nu}
 \; . 
\end{eqnarray}
 This  last equation can be checked by noting that it should be equal to the second (sum) term in Eq.~(\ref{eq:primera})
 changing $x_\rho \leftrightarrow p_\rho$ for all $\rho$.
We then deduce a very simple expression for the Lagrange multiplier  $z_{\rm GNM}(\vec x, \vec p)$:
\begin{eqnarray}
&& 
N z_{\rm GNM}(\vec x, \vec p) =
N
\left( \frac{1}{N} \sum_\nu  \frac{\partial  {\mathcal H}_{\rm GNM}}{\partial p_\nu} p_\nu -  \frac{1}{N} \sum_\nu x_\nu \frac{\partial  {\mathcal H}_{\rm GNM}}{\partial x_\nu} \right) 
 \nonumber\\
 &&
  \qquad\quad\ \  
 =
 - 2
 \sum_{\nu}  
 \bar\gamma_\nu x^2_\nu 
 \nonumber\\
 &&
  \qquad\qquad  \ \ 
 +
 \frac{2}{mN} 
 \sum_{\nu \neq \eta}  \,  \underbrace{(p_\nu \, x_\eta + p_\eta x_\nu)}_{{\rm symm} \; \eta \leftrightarrow \nu}\, 
 \underbrace{(x_\nu p_\eta - p_\nu x_\eta)}_{\rm anti symm} \, 
 \underbrace{\frac{\bar\gamma_\nu - \bar\gamma_\eta}{\lambda_\nu - \lambda_\eta}}_{\rm symm}
 \end{eqnarray}
 and then 
 \begin{eqnarray}
 N z_{\rm GNM}(\vec x, \vec p) 
 =
 - 2
 \sum_{\nu}  
 \bar\gamma_\nu x^2_\nu 
 \; . 
 \label{eq:z-in-the-end}
 \end{eqnarray}
Putting all these details together, the Langevin equations become
\begin{eqnarray}
&&
\dot x_\mu  =
\underbrace{
 \frac{2}{mN} 
 \sum_{\nu (\neq \mu)}  \, x_\nu  \, (x_\mu p_\nu - p_\mu x_\nu) \, 
 \frac{\bar\gamma_\mu - \bar\gamma_\nu}{\lambda_\mu - \lambda_\nu}}_{f^{(x)}_\mu 
 \ = \  \partial {\mathcal H}_{\rm GNM}/\partial p_\mu}
 \label{eq:explicit1}
\\
&&
\dot p_\mu = 
\underbrace{- 
2 \bar\gamma_\mu x_\mu 
+ \frac{2}{mN} 
 \sum_{\nu (\neq \mu)}  \, p_\nu  \, (x_\mu p_\nu - p_\mu x_\nu) \, 
 \frac{\bar\gamma_\mu - \bar\gamma_\nu}{\lambda_\mu - \lambda_\nu}
 }_{\textcolor{black}{-\partial {\mathcal H}_{\rm GNM}/{\partial x_\mu}}}
 \underbrace{+ \frac{2}{N} \sum_\nu \bar\gamma_\nu x_\nu^2 }_{\textcolor{black}{- z_{\rm GNM}}} x_\mu
 \nonumber\\
 &&
 \qquad 
 +
\sum_\nu {\rm P}^{\mu\nu} (-\gamma \dot x_\nu + \xi_\nu)
\nonumber\\
&& =
 \underbrace{
 \underbrace{- 2 \bar\gamma_\mu x_\mu 
\underbrace{+ \frac{2}{mN} 
 \sum_{\nu (\neq \mu)}  \, p_\nu  \, (x_\mu p_\nu - p_\mu x_\nu) \, 
 \frac{\bar\gamma_\mu - \bar\gamma_\nu}{\lambda_\mu - \lambda_\nu}}_{\textcolor{black}{{\rm if} \;\; \bar\gamma_\mu=-\lambda_\mu/2 \;\; \rightarrow \;\; -2 E_{\rm kin} x_\mu/N}}
 }_{-\partial {\mathcal H}_{\rm GNM}/\partial x_\mu}
\ \
 +  \underbrace{ \frac{2}{N}  \sum_\nu \bar\gamma_\nu x_\nu^2}_{\textcolor{black}{{\rm if} \; \bar\gamma_\mu=-\lambda_\mu/2 \;\; \rightarrow \;\;
 2V_{\rm SSK}/N}} \!\!\!\! x_\mu}_{f^{(p)}_\mu \ \ = \ \ \textcolor{black}{\lambda_\mu x_\mu \ -\  (2E_{\rm kin}-2V_{\rm SSK})/N \, x_\mu}}
\nonumber\\
 &&
 +
\sum_\nu {\rm P}^{\mu\nu} \left(-\gamma \frac{\partial {\mathcal H}_{\rm GNM}}{\partial p_\nu} + \xi_\nu \right)
\label{eq:Langevin-eqs-GGE-4}
\end{eqnarray}
where we underlined the origin of each term,
and we highlighted the limits  for ${\mathcal H}_{\rm GNM} \mapsto {\mathcal H}_{\rm SSK}$.

As a way of test, we now show that the usual expression 
$m\dot x_\mu = p_\mu$ appears for $\bar\gamma_\mu = - \lambda_\mu/2$. Indeed, 
fixing in this way the $\bar\gamma_\mu$, Eq.~(\ref{eq:explicit1}) reduces to 
\begin{eqnarray}
&&
m \dot x_\mu =\frac{1}{N} \left(
 p_\mu \sum_{\eta}  x_\eta^2 - x_\mu \sum_{\eta}  x_\eta p_\eta  \right)
 \; . 
\end{eqnarray}
We can safely use the primary spherical constraint to set the sum in the first 
term of the right-hand side to $N$ and get
\begin{eqnarray}
&&
m \dot x_\mu =
 p_\mu - x_\mu \frac{1}{N} \sum_{\eta}  x_\eta p_\eta 
 \; . 
\end{eqnarray}
From here we see that $m\dot x_\mu = p_\mu$ is a solution to this equation, since in that case the second 
sum vanishes identically. 

Setting $\bar\gamma_\mu=-\lambda_\mu/2$ now in Eq.~(\ref{eq:Langevin-eqs-GGE-3}),
 one can simply see that the
Lagrange multiplier takes the form $2(E_{\rm kin}-V_{\rm SSK})/N$ and the usual equation is derived.
Indeed, the kinetic term is absorbed into the definition of the Lagrange multiplier, yet it reappears outside this definition 
through the term $-{\partial {\mathcal H}_{\rm GNM}}/{\partial x_\mu}$. 
The 
term $\sum_\nu  x_\nu \xi_\nu$ (from the last line), would contribute to the original $z_1$ but for large $N$ or on average 
over the noise it vanishes (an assumption also usually made in the analysis of the SSK model). 

\subsection{Kramers equation.}
\label{subsec:approach-to-equil}

We proceed here in a completely general way, for generic ${\mathcal H}$.
The generic $N$ dimensional Fokker-Planck equation for a set of Langevin equations with first 
order derivatives and multiplicative noise is~\cite{Arnoulx23}
\begin{equation}
\partial_t P(y^{\bar \mu},t) = -\partial_{\bar\mu} [({\hat f}^{\bar\mu} + 2 g^{\bar\nu i} (\partial_{\bar\nu} g^{\bar\mu i})) P(y^{\bar \mu},t) ] + 
{\gamma k_BT} \partial_{\bar\mu} \partial_{\bar\nu} [ \omega^{\bar\mu\bar\nu} P(y^{\bar \mu},t) ]
\end{equation}
We read from the Langevin equations the vectors
\begin{eqnarray}
y^{\bar\mu} = (x^\mu, p^\mu) \qquad\qquad \bar\mu=1,\dots, 2N
\; , 
\end{eqnarray}
the factor that multiplies the noise
\begin{equation}
g^{\bar\mu i} = 
\left\{
\begin{array}{ll}
0 & \bar\mu = 1, \dots, N
\\
{\rm P}^{\mu i } & \bar\mu = N+1, \dots, 2N
\end{array}
\right.
\end{equation}
and the metric
\begin{equation}
\omega^{\bar\mu \bar\nu} = g^{\bar\mu i} g^{\bar \nu i} = 
\left(
\begin{array}{cc}
0 & 0 \\
0 &  {\rm P}^{\mu\nu}
\end{array}
\right)
\; . 
\end{equation}
The two ``forces'' in {${\hat f}^{\bar \mu}$} include
{the terms which have been highlighted with the underbracing in Eqs.~(\ref{eq:explicit1})
and (\ref{eq:Langevin-eqs-GGE-4}) and also the friction one:}
\begin{eqnarray}
{\hat f}_\mu^{(x)} = \frac{\partial {\mathcal H}}{\partial p_\mu}
\; , 
\qquad\qquad
{\hat f}_\mu^{(p)} = - \frac{\partial {\mathcal H}}{\partial x_\mu} -z(\vec x, \vec p) x_\mu - \gamma {\rm P}_\mu^\nu \dot x_\nu
\; .
\label{eq:forces}
\end{eqnarray}
{For the model at hand, $g^{\bar\nu i}\partial_{\bar\nu}g^{\bar\mu i}=0$: the noise acts only on momenta, 
whereas its amplitude depends only on positions. Thus the It\^o and Stratonovich drifts coincide in these coordinates.}

Putting all this together, the Kramers equation for stochastic motion on the sphere of a generic 
model ${\mathcal H}$ reads
\begin{eqnarray}
\frac{\partial P}{\partial t} 
&=&  - \frac{\partial }{\partial x_\mu} ({\hat f}^{(x)}_\mu P)
 -  \frac{\partial }{\partial p_\mu} ( {\hat f}^{(p)}_\mu P 
)
 + \gamma k_BT \frac{\partial}{\partial p_\mu}   \frac{\partial}{\partial p_\nu}  ( {\rm P}^{\mu\nu} P)
 \nonumber\\
&=&
-
\frac{\partial }{\partial p_\mu}
\left[
 \left( 
 -\frac{\partial {\mathcal H}}{\partial x^\mu} - z x^\mu 
- \gamma {\rm P}^\nu_{\, \mu} \frac{\partial {\mathcal H}}{\partial p^\nu}
\right) P
\right]
\nonumber\\
&&
-
\frac{\partial }{\partial x_\mu} 
\left( 
\frac{\partial {\mathcal H}}{\partial p^\mu} P
\right)
+ \, \gamma k_BT \; {\rm P}^\mu_{\,\nu} \;  \frac{\partial^2 P }{\partial p^\mu\partial p^\nu} 
\; .
\label{eq:kramers}
\end{eqnarray}

 The expectation is that  the conventional Gibbs-Boltzmann measure, 
 \begin{equation}
 P_{\rm GB} \propto e^{-\beta {\mathcal H} - \beta \frac{z}{2} (x^2-N)} 
\label{eq:GB-GGE}
 \end{equation}
  will set the right-hand side of the 
 Kramers equation (\ref{eq:kramers}) to zero, under the spherical constraint. This statement can 
 be proven without difficulty. Indeed the distribution (\ref{eq:GB-GGE}) sets to zero the 
 terms originating from the coupling to the bath on the one hand, and the deterministic terms
 on the other hand, see App.~\ref{app:Kramers}. Thus, the expression in~(\ref{eq:GB-GGE})  
 is a {formal stationary weight. It defines a stationary probability measure
 only when it is normalizable}.
 
\section{Dynamics of the  GNM with quadratic $\bar\gamma_\mu (\lambda_\mu)$}
\label{sec:numerics}

In this section we study the stochastic relaxation dynamics of the Generalized Neumann model with 
\begin{equation}
\bar\gamma_\mu =   \frac{a}{J} \lambda_\mu^2 - b \lambda_\mu 
\label{eq:quadratic-gamma}
\end{equation}
and $b=1/2$. This is a simple choice for which we derived the static phase diagram,
shown in Fig.~\ref{fig:phase-diagram},
and many properties of the order parameter and other observables in Sec.~\ref{subsec:quadratic-gamma}. 
The Langevin equations were introduced and carefully discussed in Sec.~\ref{subsec:Langevin-GGE},
and here we simply use the $\bar\gamma_\mu$ in Eq.~(\ref{eq:quadratic-gamma}).

We recall that for  $a=0$ we recover the inertial SSK model, while for $a\neq 0$, either positive or negative, 
we get a generalised model.
The quadratic  form can be concave or convex depending on the sign of $a$, and it can have a minimum or 
a maximum outside or inside the interval $[-2J, 2J]$.  
To get a monotonically decreasing function $\bar\gamma_\mu(\lambda_\mu)$ we need:
\begin{equation}
 \frac{d\bar\gamma_\mu}{d\lambda_\mu} =  \frac{2 a}{J}  \lambda_\mu - b <0 
\; . 
\end{equation}
The unique extremum of $\bar\gamma_\mu(\lambda_\mu)$ -- minimum or maximum -- sits at 
 $\lambda_\mu^* = bJ/(2a)$ and it will be outside the interval $[-2J,2J]$ if
\begin{eqnarray}
4 |a| \leq b
\; .
\end{eqnarray}

\subsection{Numerical procedure}
\label{subsec:numerical-procedure}

We initialize the system with  (a set of) random initial condition(s) satisfying the spherical constraint
strictly. We distinguish  ``extended'' initial conditions with high kinetic energy and no macroscopic
spatial component, 
and ``condensed'' initial conditions with low kinetic energy and a macroscopic alignment of the 
particle's position along the North/South direction.

A concrete way to generate such typical extended configurations is: 
\begin{itemize}
\item[(i)] Choose a vector $\vec x$  with i.i.d. random components of order one and normalise it so that 
$x^2 =N$.
\item[(ii)] Choose another  vector $\vec t$ with i.i.d. random components 
and construct  $\vec p = \vec t - \vec x (\vec x \cdot \vec t)/N$,
with components  $p^\mu = {\rm P}^\mu_{\, \nu} t^\nu$ where 
${\rm P}^\mu_{\, \nu} = \delta^\mu_{\, \nu} - x^\mu x_\nu/N$.
\item[ (iii)] Normalize $\vec p$  so that $p^2 =1$ and then multiply it by the parameter $\sqrt{2 m E_{\rm kin}} = \sqrt{2m (T_0/2) \, N}$.
 This fixes the kinetic energy $E_{\rm kin}$ of the initial condition $p^2 =2mE_{\rm kin}$. Note that $E_{\rm kin}$
 must be O($N$).
\end{itemize}

 Typical condensed configurations are generated as follows:
\begin{packed_enum}
 \item[(i)] Set $x_N = \Xi [(1-T_0/J) N]^{1/2}$ with $\Xi$ a random variable O(1).
  \item[(ii)] Choose the other $\mu=1, \dots N-1$ components $x_\mu$ to be i.i.d. random O(1) values 
  and normalise them so that $\sum_{\mu=1}^{N-1} x_\mu^2 = [1-\Xi^2 (1-T_0/J)] N$. Then $x^2 =N$.
  \item[(iii)] Choose another random vector $\vec t$ with i.i.d. random components 
and construct  $\vec p = \vec t - \vec x (\vec x \cdot \vec t)/N$,
the components of which read $p^\mu = {\rm P}^\mu_{\, \nu} \, t^\nu$.
\item[ (iv)] Normalize $\vec p$  to $1$ and then multiply all components by $\sqrt{2 m E_{\rm kin}}=  \sqrt{2m (T_0/2) \, N}$
so as to fix the kinetic energy of the initial condition $p^2 =2mE_{\rm kin}=$ O($N$).
\end{packed_enum}

We notice that these initial conditions are not the ones of Boltzmann equilibrium of the model 
${\mathcal H}_{\rm GNM}^z$ at any temperature. In particular, the variances of all components of the position vector
in the extended case, and all components but the $N$th one in the condensed case, are identical here while they are
not expected to be so in equilibrium.  A similar difference applies to the components of the momentum vector.
$T_0$ is a parameter that characterises the initial conditions and its numerical values will be specified below.

Next, we separately evolved these two kinds of initial conditions with the Langevin process 
(\ref{eq:explicit1})-(\ref{eq:Langevin-eqs-GGE-4}) at temperatures $T$. 

Depending on the choice of the parameters in the  quadratic $\bar\gamma_\mu$ different dynamic behaviours are observed. 
For example, the stochastic dynamics at $T$ and $a$ in the extended phase 
will be typical of a disordered  phase, while the ones for quenches to $T$ and $a$ in the condensed phase
will tend to condense.  
For choices of $T$ and $a$ in the moving phases, more exotic behaviours appear.
We will discuss all these in the rest of this section.

The basic features of the numerical code are explained in App.~\ref{app:numerics}.
In the numerics, we set $J=1$, the mass to $m=1$, the friction coefficient to $\gamma=5$ or $\gamma=10$, and 
we averaged over five independent initial configurations throughout. We will discuss the dependence on the 
size of the system. In the Langevin code we use a very tiny time-step to avoid annoying numerical errors, 
see App.~\ref{app:numerics} for more details. Problems 
with Runge-Kutta 2 using $\Delta t = 0.001$ were solved with $\Delta t = 0.0001$. In all plots, time 
refers to the  number of iterations times $\Delta t$.

\subsection{Asymptotic regimes}

Depending on the parameter $a$ and temperature $T/J$, the system has 
very different dynamic behaviour, in correspondence with the 
various phases in the phase diagram derived with the static calculation 
using the measure $\propto  \exp(-\beta\sum_\mu \bar\gamma_\mu I_\mu) $
in Sec.~\ref{sec:GGE-Model}
and App.~\ref{app:details-of-calculations} and displayed in Fig.~\ref{fig:phase-diagram}.
We now check whether the stochastic dynamics of large but finite size systems 
approach asymptotically the static predictions of Sec.~\ref{sec:GGE-Model}.

\begin{figure}[h!]
\centerline{
\includegraphics[scale=0.45]{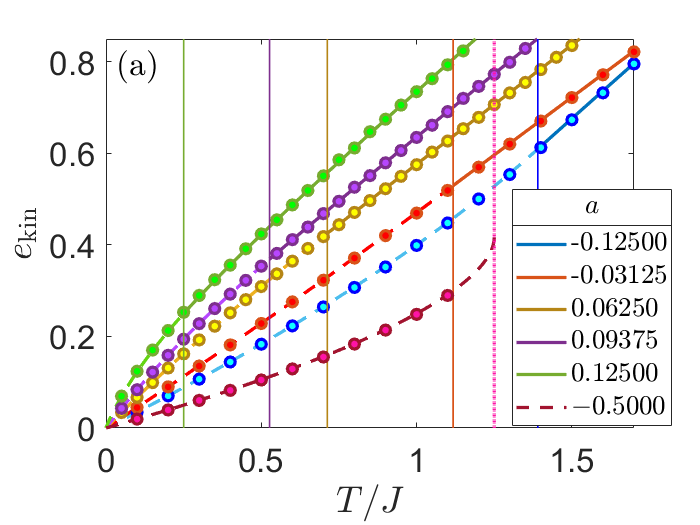}
\hspace{-0.5cm}
\includegraphics[scale=0.45]{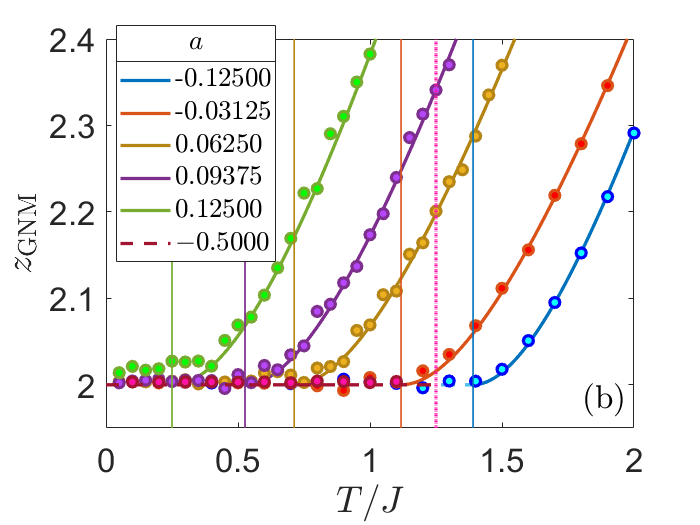}
}
\caption{\small The (asymptotic) kinetic energy density, $e_{\rm kin}= (2mN)^{-1} \sum_\mu \langle p_\mu^2\rangle$ (a), 
and Lagrange multiplier, 
$z_{\rm GNM}$ (b), as functions of temperature, for $b=1/2$ and different values of $a$ given in the keys.
The data points stem from the numerical integration of the Langevin process, for $N=500$, and are averaged over 
a few runs.
The thin vertical lines correspond to $T_c(a)$ towards the extended phase, given in Eq.~(\ref{eq:Tc0}) and drawn in the main central panel of 
Fig.~\ref{fig:phase-diagram} as a solid blue line with some numerical data points with the same blue colour.
The dotted pink vertical line represents $T_c(a=-0.5) = 1.25 J$ on the violet critical line towards the moving phase M$_L$ of the phase diagram of Fig.~\ref{fig:phase-diagram}.
The solid and dashed lines across the data points are the analytic predictions from the static calculation, 
above  and below the phase transition between the extended and localised phases.  
The data for $a=-0.5$  stop since the system enters the moving phase at higher temperatures.
}
\label{fig:ekin-zGGE}
\end{figure}

\subsubsection{Intensive quantities.}
We first analyse intensive quantities, like the kinetic energy density 
and the Lagrange multiplier, which are plotted in Fig.~\ref{fig:ekin-zGGE} against the temperature at which the
process is run. $b$ has been fixed to $1/2$ without 
loss of generality and various values of $a$, all yielding a monotonically decreasing $\bar\gamma_\mu(\lambda_\mu)$, 
are considered. The agreement 
between the numerical data, obtained in the asymptotic limit of the stochastic evolution, 
and the analytic predictions from the static evaluation is excellent, at all temperatures 
and for all values of the parameter $a$ (apart for the limiting one, $a=0.125$, at low $T$ in the right panel for which we do not have enough numerical precision). 
The static analytic predictions, with solid and dashed lines, 
are those of Eq.~(\ref{eq:ekin-analytic-GGE0}) at $T>T_c$ and Eq.~(\ref{eq:ekin-belowTc0}) 
at $T<T_c$ for $e_{\rm kin}$, and Eq.~(\ref{eq:zGGE-aboveTc0}) at $T>T_c$ and $2J$
at $T<T_c$ for $z_{\rm GNM}$. The crossover between high and low temperature behaviours occurs at $T_c(a)$ 
given in Eq.~(\ref{eq:Tc0}) 
and these values are reported with thin coloured vertical  lines following the code of the data points. 

\subsubsection{High temperature - extended phase.}

We now examine the stochastic dynamics  in the extended phase in more detail. 

\begin{figure}[h!]
\centerline{
\includegraphics[scale=0.4]{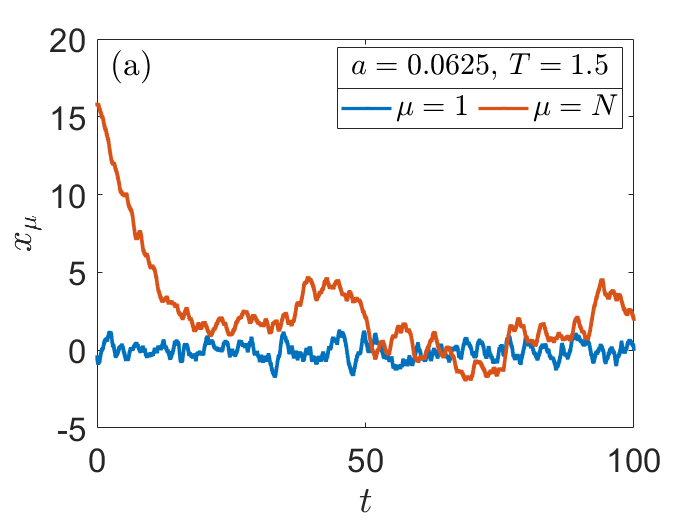}
\hspace{-0.5cm}
\includegraphics[scale=0.4]{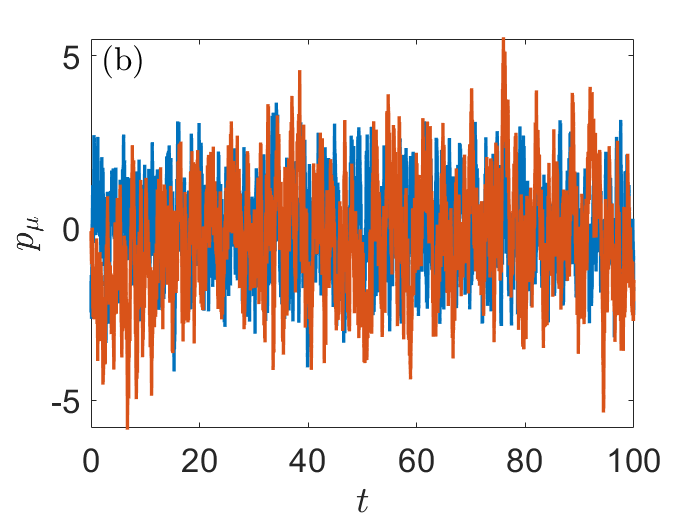}
}
\caption{\small Typical time dependence of two coordinates (a) and two components of the momentum (b) 
after a quench to  the extended phase. 
}
\label{fig:trajectories}
\end{figure}

Two representative components of the trajectories,  $x_1(t)$ and $x_N(t)$ (a), and
the momentum, $p_1(t)$ and $p_N(t)$ (b), are shown in Fig.~\ref{fig:trajectories}.
The comparison between the same two components of the momentum and 
the instantaneous velocity are discussed in App.~\ref{app:momentum-vs-velocity}.
These trajectories were evolved at $(T/J, a)$  in the extended phase starting  from a condensed initial condition at $t=0$, 
which is very quickly forgotten. All these quantities average -- over time or noise -- to zero. 
We then focus on their variances.

\begin{figure}[h!]
\centerline{
\includegraphics[scale=0.4]{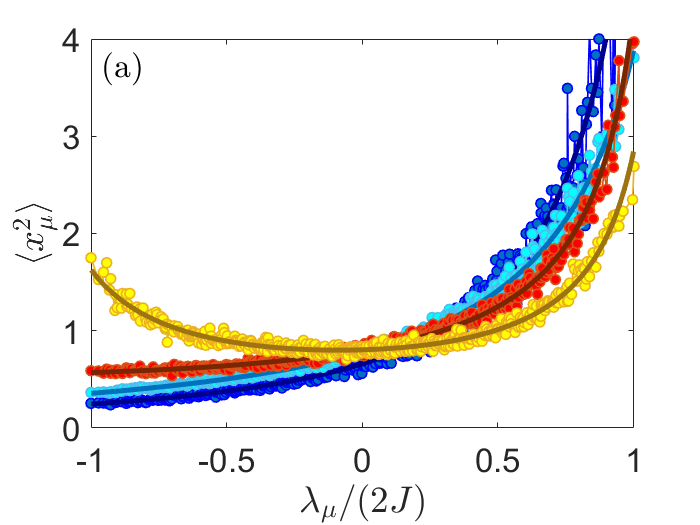}
\hspace{-0.5cm}
\includegraphics[scale=0.4]{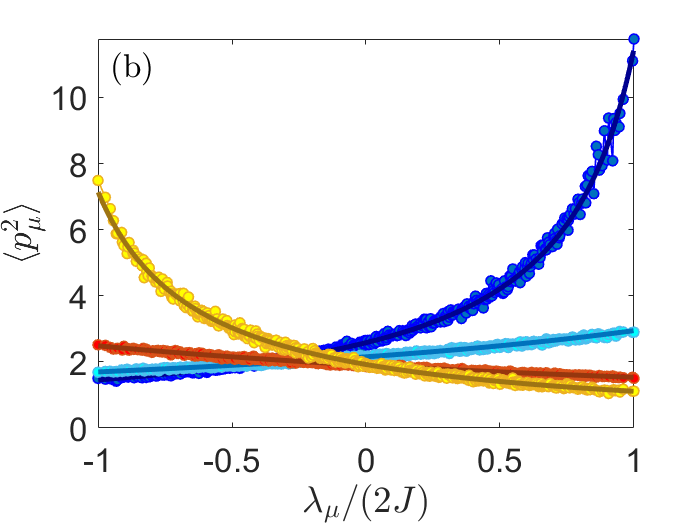}
}
\caption{\small
Mode-resolved squared positions and momenta in the extended phase,
$a=0.15$ (dark blue), $a=0.0625$ (light blue), $a=-0.0625$ (red), and $a=-0.19$ (orange), all at $T=2$. 
(a) Spectrum of  square coordinate values, $\langle x_\mu^2\rangle$, limited to  $\langle x_\mu^2\rangle < 4$
to improve visualisation
and (b) squared momenta components, $\langle p_\mu^2\rangle$.
The asymptotic Langevin numerical data points for $N=500$ are compared to the static predictions for $N\to\infty$, 
 $\langle x_\mu^2\rangle = \langle p^2_\mu\rangle/(z_{\rm GNM}-\lambda_\mu)$ and $\langle p_\mu^2\rangle$, 
with $\langle p^2_\mu\rangle$ given in Eq.~(\ref{eq:Tmu-biss}), and 
$e_{\rm kin}$ and $z_{\rm GNM}$ replaced by their analytic expressions in 
Eqs.~(\ref{eq:ekin-analytic-GGE0}) and (\ref{eq:zGGE-aboveTc0}), respectively.
}
\label{fig:x2p2-highT}
\end{figure}

In Fig.~\ref{fig:x2p2-highT} we compare the asymptotic values of dynamic mode-resolved observables 
to the static analytic ones, see Eqs.~(\ref{eq:Tmu-biss})-(\ref{eq:A}) and Eqs.~(\ref{eq:ekin-analytic-GGE0}),~(\ref{eq:zGGE-aboveTc0}). 
The notation $\langle \dots \rangle$ represents either the GNM
static average $\langle \dots \rangle_{\rm GNM}$ 
or the stochastic one (averaged over initial 
conditions $\langle \dots \rangle_{i.c.}$ and over time window of length $\tau$ by the
end of the simulation run, $\overline{\langle \dots \rangle_{i.c.}}$.
At high temperatures we do not need to use a very large system size to see convergence. 
In this case, we use $N=500$, and we choose fully random initial conditions respecting the spherical constraint. 
We evolve at $T=2$ and we use four values of $a$ (data points) ranging from close to the transition 
towards the moving phase M$_L$ for $a=-0.19$,  well within the extended phase for $a=\pm 0.0625$, 
and close to the transition to the moving phase M$_R$ for $a=0.15$,  
see the phase diagram in Fig.~\ref{fig:phase-diagram}. Note that for $a=-0.19$ and $a=0.15$, $\bar\gamma_\mu(\lambda_\mu)$ is not monotonic.
In the former case this function has 
a maximum and in the latter it has a minimum in the interval $[-2J, 2J]$, 
see the lower-left and lower-right panels in Fig.~\ref{fig:phase-diagram}. 
Still, the model is stable for the temperature considered.

We plot the long time noise-averages $\langle x_\mu^2\rangle$ and $\langle p_\mu^2\rangle$ 
together with the static analytic evaluation of the same quantities performed at $N\to\infty$ (solid lines). The latter are Eq.~(\ref{eq:Tmu-biss}) 
for $\langle p^2_\mu\rangle$, with the corresponding $z_{\rm GNM}$ and $e_{\rm kin}$
in Eq.~(\ref{eq:ekin-analytic-GGE0}) and Eq.~(\ref{eq:zGGE-aboveTc0}), respectively, and 
$\langle x_\mu^2 \rangle = \langle p_\mu^2 \rangle/(z_{\rm GNM}-\lambda_\mu)$. 
We find very good agreement between the numerical and analytic data,
even for this modest system size.  
{The simulated profiles rapidly become consistent with the large-$N$ saddle-point predictions 
over the time interval studied.}
From Eq.~(\ref{eq:Tmu-biss}) one easily derives that 
$d\langle p^2_\mu\rangle/d\lambda_\mu \propto 1/a$ justifying the different trends for $a>0$ and $a<0$. 
Plotting the spectra of $\langle p_\mu^2\rangle$ 
for other values of $a$, one confirms that the curves cross the working temperature 
$T$ at an $a$-dependent $\lambda_{\mu^*}$. This feature can be easily understood at $T\to \infty$ and 
turns out to hold also at any finite $T$ in the extended phase. Indeed, to have  $T_\mu=T$, one needs $\lambda_{\mu^*} +z_{\rm GNM} - 2e_{\rm kin}=0$.
{At high temperature, $z_{\rm GNM}$ and $2e_{\rm kin}$ have the same leading slope, but their difference need not vanish. For $a\ne0$, on the branch considered here,
\begin{equation}
\lambda_{\mu^*}\to-4aJ \bar\eta,\qquad
\bar\eta=\lim_{T\to\infty}\frac{e_{\rm kin}}T
=\frac1{2(b+\sqrt{b^2-8a^2})} \; .
\end{equation}
At $a=0$, the SSK spectrum is flat and every mode has $T_\mu=T$.
For $a=0.0625$, $b=1/2$ and $J=1$, $\lambda_{\mu^*} = 2 e_{\rm kin}-z_{\rm GNM} \sim -0.129$.}
A similar mechanism is at work at all temperatures, with the $\lambda_{\mu^*}$ arranging for $T_{\mu^*}=T$.

\begin{figure}[b!]
\centerline{
\includegraphics[scale=0.45]{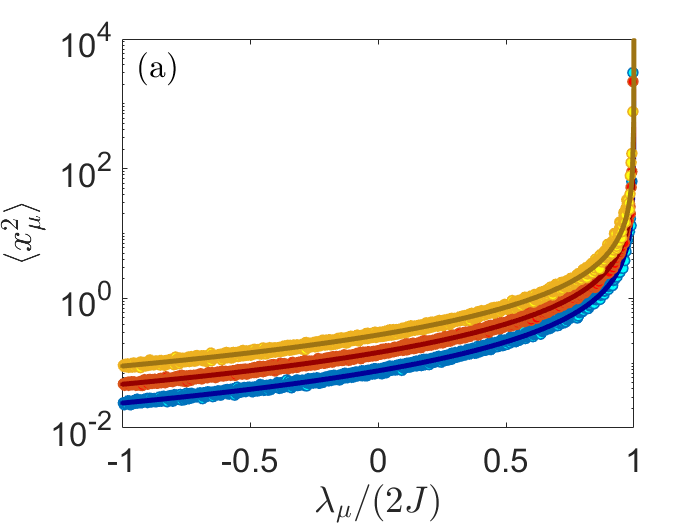}
\hspace{-0.5cm}
\includegraphics[scale=0.45]{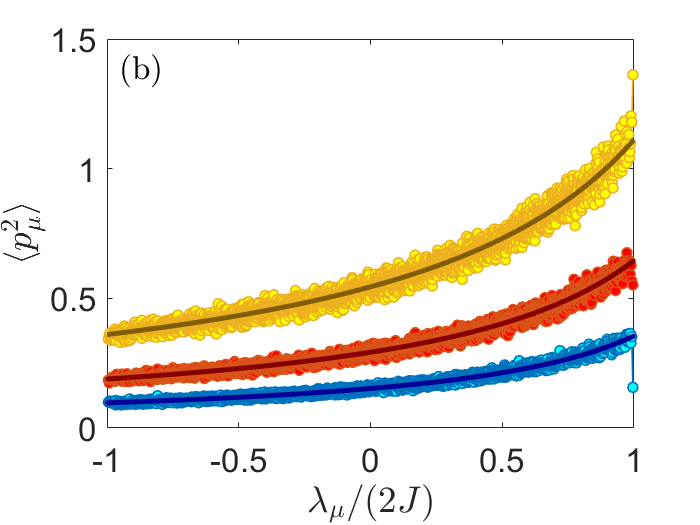}
}
\caption{\small The  spectrum of $\langle x_\mu^2\rangle$ (a) and $\langle p_\mu^2\rangle$ (b) 
for system with $a=0.09375$ and $N=4000$ at different temperatures, $T=0.1, \, 0.2, \, 0.4$, all 
in the condensed phase. The data points correspond to the asymptotic (long-time limit) 
stochastic values obtained with condensed initial conditions at $T_0=0.5$. The solid  lines are the 
analytic predictions in Sec.~\ref{sec:GGE-Model}.
}
\label{fig:x2p2-lowT}
\end{figure}

The self-correlation, $C(t,t_w) = N^{-1} \sum_\mu \langle x_\mu(t) x_\mu(t_w)\rangle_{\xi, i.c.}$, 
in the extended phase rapidly becomes stationary, $C(t,t_w) \mapsto C(t-t_w)$,  
and approaches zero exponentially in the time delay $t-t_w$ (not shown).
 
\subsubsection{Low temperature - condensed phase.}

The full spectra of $\langle x_\mu^2 \rangle$ and $\langle p_\mu^2 \rangle$
at three low temperatures and the same $a$ value, $a=0.09375$ are shown in  Fig.~\ref{fig:x2p2-lowT} (a) and (b).
The agreement between the numerical data with the analytic predictions is excellent, with 
the exception of $\langle p_N^2\rangle$ which we analyse next in more detail. 

\begin{figure}[t!]
\hspace{0.75cm}
\centerline{
\hspace{-1cm}
\includegraphics[scale=0.42]{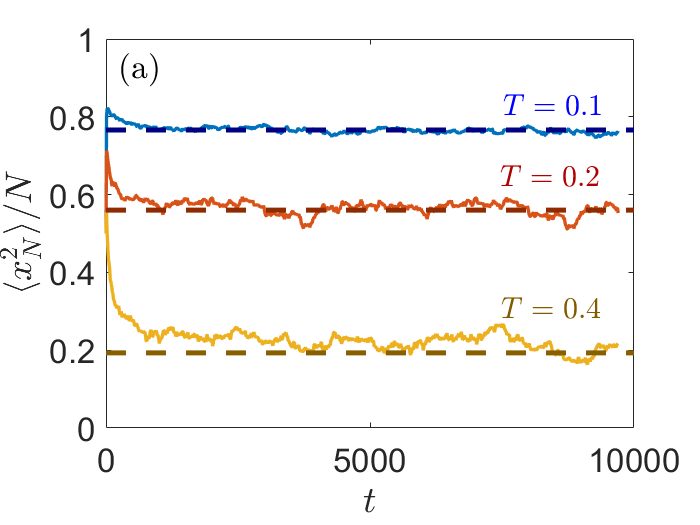}
\includegraphics[scale=0.42]{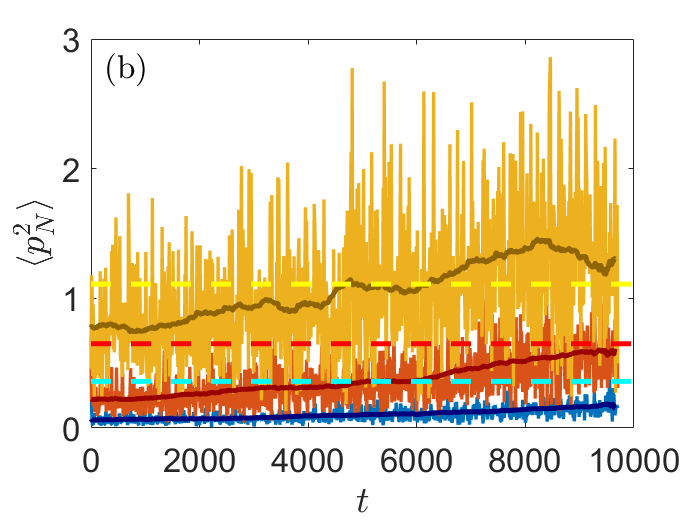}
}
\caption{\small 
(a) Time dependence of $\langle x_N^2\rangle/N$ at $T=0.1$ (blue, above), $T=0.2$ (red, intermediate) and 
$T=0.4$ (orange, below), all in the condensed phase for $a=0.09375$, 
$N=4000, \, \Delta t=0.001$. 
The horizontal dashed lines are the analytic predictions for $q$ in Eq.~(\ref{eq:qEA0}). 
(b)  Time dependence of $\langle p^2_N\rangle $ for the same parameters.
These data are much noisier and for the lower temperature they  have not reached their asymptotic value, Eq.~(\ref{eq:TN-lowT0}), 
shown with the horizontal dashed lines. 
Data were averaged over 10 initial conditions of a condensed phase at $T_0 = 0.5$. 
The solid lines in (b) correspond to time averages of the data over a running time window of width $100$ time-steps.
}
\label{fig:time-dep-lowT}
\end{figure}

In the condensed phase the $N$th mode should condense, $\langle x_N^2\rangle =qN$,  while keeping a finite 
variance of the corresponding momentum component. In Fig.~\ref{fig:time-dep-lowT} we show the time evolution
of  $\langle x_N^2\rangle/N$ and $\langle p_N^2\rangle$, at three representative 
low temperatures, starting from condensed  initial conditions 
$T_0 = 0.5$, $N=4000$ and $a=0.09375$.
As expected, $\langle p_N^2\rangle$ does not scale with $N$.
While the square of the $N$th coordinate rapidly reaches its asymptotic value, the square of the momentum does not. This difference is accentuated
at low $T$. The time dependence of $\langle p_N^2\rangle$ is much noisier than the one of $\langle x_N^2\rangle$, 
and although the  data get close to their
expected asymptotic  values, given in Eq.~(\ref{eq:TN-lowT0}), they also fluctuate strongly around
these values. (The instantaneous data were averaged over ten initial conditions and the 
solid lines were further averaged over time windows of 100 time step duration.) 
The difficulty with getting good results for the momentum 
seems to be related to the fact that the numerical error in the orthogonality with the coordinate and the spherical constraint
increase at higher temperature, see App.~\ref{app:numerics} for more details. 

\begin{figure}[t!]
(a) \hspace{7cm} (b)
\\
\centerline{
\includegraphics[scale=0.42]{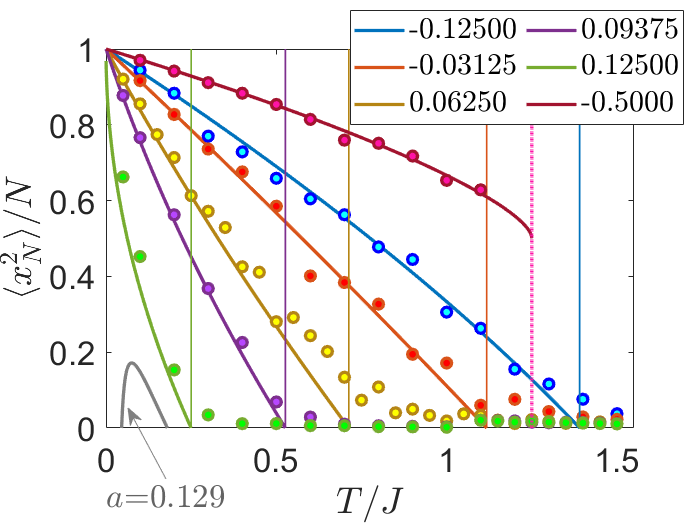}
\hspace{-0.45cm}
\includegraphics[scale=0.42]{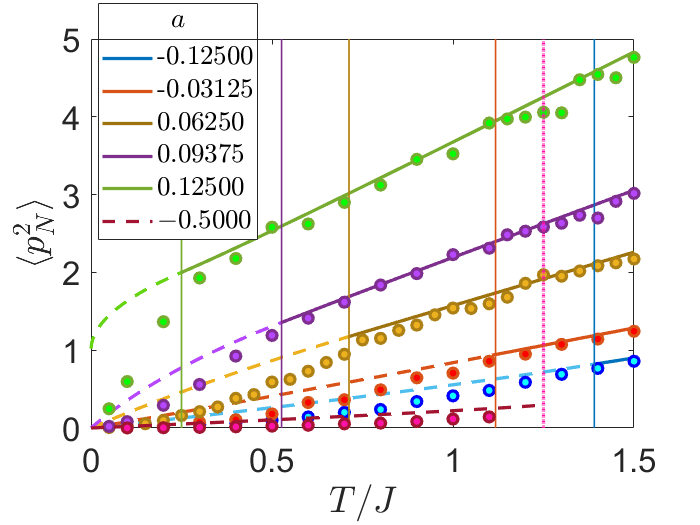}
}
\caption{\small 
The largest mode  $\langle x_N^2\rangle/N$ (a) and  $\langle p_N^2\rangle$ (b). 
The data points for $a = 0.125$ and $a = 0.09375$ are for $N = 4000$ and the rest for
 $N = 500$. The initial conditions were drawn from equilibrium at 
$T_0=0.5$ with $ x_N \approx 0.5\sqrt{N}$ and $p_\mu$ orthogonal to $x_\mu$.
Data points from the numerical evaluation and solid and dashed lines from the 
analytic expressions, Eq.~(\ref{eq:qEA0}) for $\langle x_N^2\rangle/N$ and 
 Eq.~(\ref{eq:TN-lowT0}) for $\langle p_N^2\rangle/N$.  The mismatch between
 data points and analytic result in (b) at low $T$ is discussed in the text.
 The vertical thin lines mark the transition between the extended and condensed phases 
 while the vertical dotted line signals the transition between the condensed and moving 
 M$_L$ phase.
}
\label{fig:xNpN-highT}
\end{figure}

We now study, in Fig.~\ref{fig:xNpN-highT},  the dependence of the asymptotic values of  $\langle x_N^2\rangle/N$ and $\langle p_N^2\rangle$
as a function of $T$ for different $a$, both in the condensed and extended phases. 
There is good agreement between the numerical measurement and the 
analytic prediction in the high temperature regime, where $\langle x_N^2\rangle/N \to 0$ and $\langle p_N^2\rangle$
is given by a non-trivial function shown with solid lines and going through the numerical data. In the low $T$ regime, 
$\langle x_N^2\rangle/N$ is favourably compared to $q$. However, as mentioned in the discussion of 
Fig.~\ref{fig:time-dep-lowT} below, the dynamic data for $\langle p_N^2\rangle$ have not saturated and they are still very far from the analytic one in the condensed phase. The larger the $a$, the worse the 
agreement with the analytics so,  most probably, the farther away from equilibrium.

\begin{figure}[h!]
\hspace*{0.5cm} (a)  \hspace{4.75cm} (b) \hspace{4.25cm} (c)
    \begin{center}
    \includegraphics[width=0.34\textwidth]{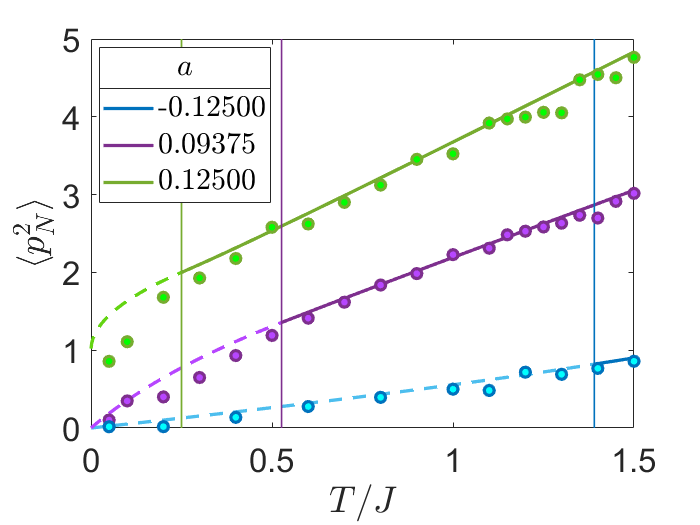}
    \hspace{-0.4cm}
    \includegraphics[width=0.34\textwidth]{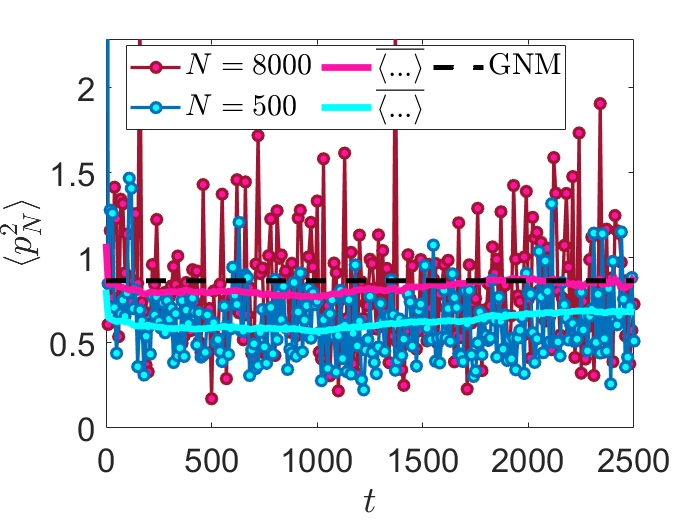}
        \hspace{-0.4cm}
    \includegraphics[width=0.34\textwidth]{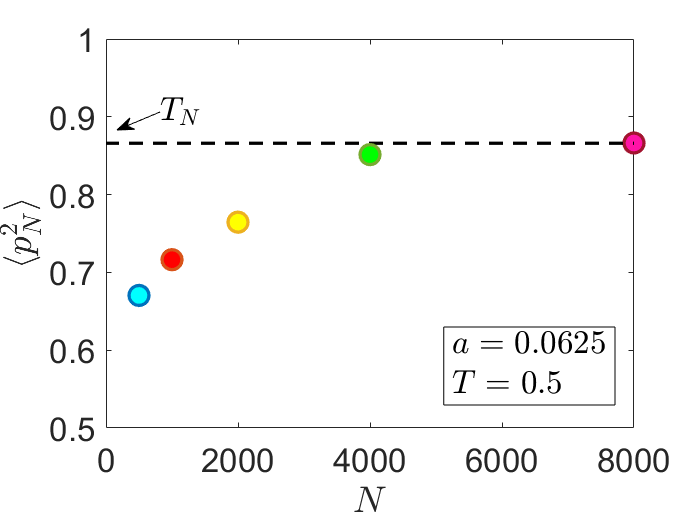}
    \end{center}
    \caption{\small Difficulty to  reach the equilibrium value of $\langle p_N^2 \rangle$ within the condensed phase. $N=4000$.
(a) Dependence of $\langle p_N^2 \rangle$ on temperature $T$ for different values of $a$ at time $t=5000$.
In the condensed region (dashed lines), the numerical points do not collapse onto the theoretical prediction, due 
to finite size effects.
(b) Time evolution of $\langle p_N^2 \rangle$ for $a=0.0625$ and $T=0.5$. The points represent instantaneous values, the solid lines are time averages over windows of length 100. The dashed horizontal line corresponds to the static expectations for $N\to\infty$. 
Two system sizes are shown, with the colour code also used in (c), 
where the dependence of the asymptotic $\langle p_N^2 \rangle$ on system size $N$ for the same parameters is used. The dashed horizontal line marks its value for $N\to\infty$. 
}
    \label{fig:hardpN}
\end{figure}

The difficulty to reach the asymptotic value of $\langle p_N^2\rangle$ is due to both 
finite size and finite time effects. This is explained in Fig.~\ref{fig:hardpN}.
Finite-size effects increase as the temperature decreases and as $a$ approaches $0.125$, making it very costly to reach the thermodynamic limit. However,  the trend converges to the GNM equilibrium
when calculations are pushed to the computational limit.
It can be seen that for $N=8000$ (pink line) the numerical value saturates at the GNM equilibrium, whereas for $N=500$ (blue line) it stabilizes at a clearly smaller value. These numerical limitations pose an additional challenge, illustrated in the following figures.

The condensed phase presents different dynamics for $a$ relatively close to 0 and $a$ negative and far from zero. The former 
is the conventional one with slow relaxation. The latter, has a different 
and very rapid dynamics.  We discuss them next.

\subsection{Pre-asymptotic evolutions}

\subsubsection{Aging in the condensed phase.}
\label{sec:531}

The out of equilibrium relaxation of the SSK model follows a pattern similar to the one of coarsening 
systems~\cite{Bray94,Onuki04,Puri09,KrReBe10,HePl10,LesHouches,Cugliandolo15,Cugliandolo24}. For subcritical quenches, 
from initial conditions which are typical of the high temperature phase, the details of the initial conditions are
soon forgotten and the self-correlation decays in two steps~\cite{CuDe95a}:
\begin{eqnarray}
C(t,t_w) = \lim_{N\to\infty} \frac{1}{N} \sum_\mu \langle x_\mu(t) x_\mu(t_w)\rangle_{\xi, i.c.} =
C_{\rm st}(t-t_w) + C_{\rm ag}(t,t_w)
\end{eqnarray}
and
\begin{eqnarray}
&&
C_{\rm st}(t-t_w) = \lim_{t_w\to\infty} C(t,t_w)  
\\
[5pt]
&& \qquad\qquad {\rm with} \quad C_{\rm st}(t-t_w) = 1-q \; \& \; 
C_{\rm st}(\infty) = 0 
\; , 
\\
[7pt]
&&
C_{\rm ag}(t, t_w) = q \; f\left( \frac{t}{t_w} \right) 
\\
[5pt]
&& \qquad\qquad {\rm with} \quad f(1)=1 \; \& \; f(\infty) = 0
\; , 
\end{eqnarray}
in the long times limits.
We now test whether a scheme like this one also applies for $a\neq 0$.

\begin{figure}[h!]
\centerline{
\includegraphics[scale=0.45]{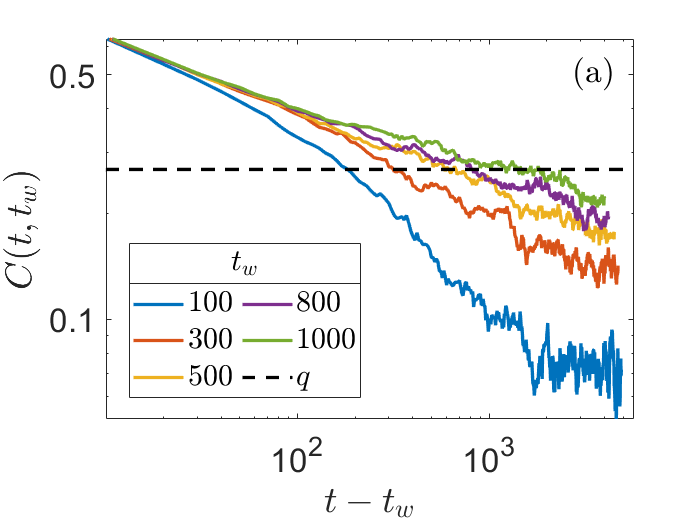}
\includegraphics[scale=0.45]{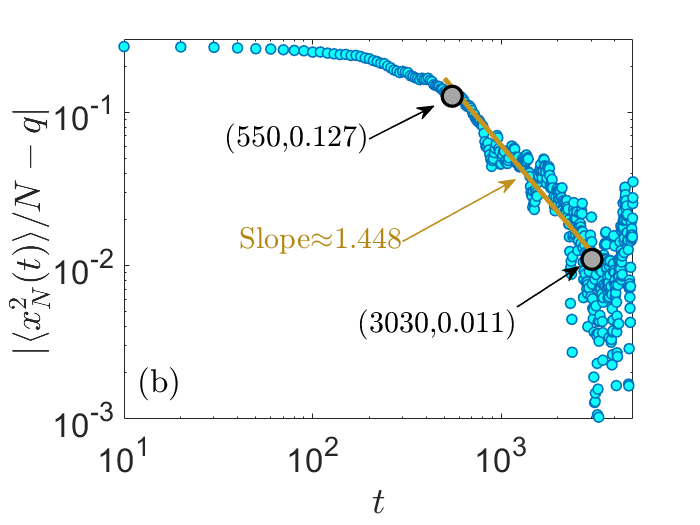}
}
\caption{\small Two-time correlation functions after an instantaneous quench from the 
extended to the condensed phase,  in a model with $N=4000$, $a=0.0625$ at $T=0.5$.
 (a)  Log-log scale. Data averaged over 20 extended initial conditions. 
The $N\to\infty$ Edwards-Anderson parameter $q$ is indicated with a dashed horizontal line.
(b) Approach to condensation of the $N$th mode under the same conditions as in 
panel (a) with $q$ fixed to its theoretical value. The algebraic estimate $t^{-1.448}$
is shown with a straight line. The exponent is very close to the $3/2$ of the  SSK. 
}
\label{fig:aging-apositive}
\end{figure}

Figure~\ref{fig:aging-apositive} displays the decay of the self-correlation in a case with $a>0$, 
$N=4000$ with data averaged over 20 independent initial conditions drawn from the extended 
phase. For this system size the equilibration time is relatively short.  
We see the typical two-step relaxation in (a) and a slow 
approach of the average of the square of the largest mode to its asymptotic value $q$, obtained 
analytically for $N\to \infty$, in (b).
We are limited by not being able to use larger system sizes and reach longer time scale to 
determine the precise scaling form of the self-correlation, but the aging behaviour 
survives for $a>0$.  In App.~\ref{app:GGE-to-SSK} we show similar curves for the SSK ($a=0$) model.

\subsubsection{Rapid relaxation in the condensed phase.}

\begin{figure}[h!]
\begin{center}
    \includegraphics[width=0.45\textwidth]{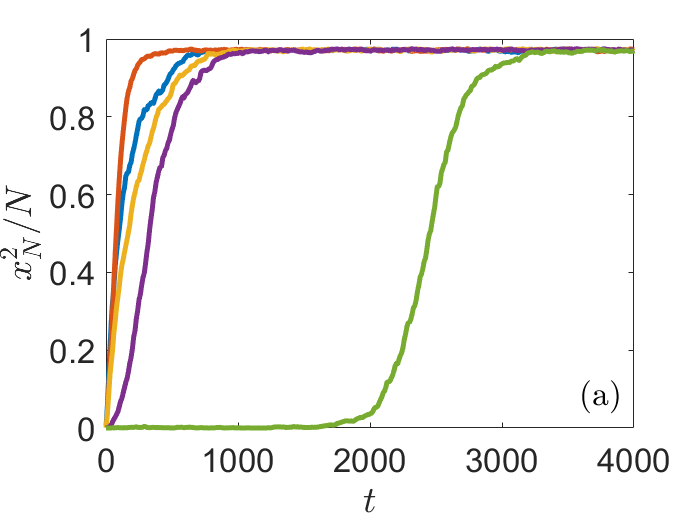}
    \includegraphics[width=0.45\textwidth]{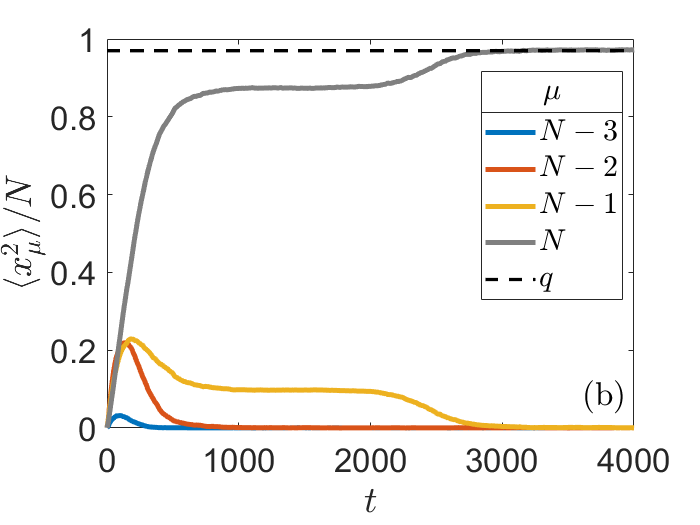}
    \end{center}
    \caption{\small Rapid relaxation in the condensed phase with $a=-0.125$, $T=0.05$ and $N=4000$
    with high temperature initial conditions.
(a) Individual trajectories of the observable $x_N^2 / N$, showing a rapid growth, 
except for the green trajectory, which takes longer to take off.
(b) $\langle x^2_\mu(t)\rangle/N$ averaged over 10 trajectories
for several modes close to the edge. The horizontal dashed line is the analytic value 
for $q=\lim_{N\to\infty} \langle x_N^2\rangle_{\rm GNM}/N$.
}
    \label{fig:rapid-condensed}
\end{figure}

For sufficiently large negative values of $a$ the relaxation in the condensed phase loses its slow character and 
power laws are replaced by exponentials. 
Results for $a=-0.125$ are shown in Fig.~\ref{fig:rapid-condensed}. In panel (a) 
individual trajectories are plotted and all but one very quickly reach the condensation value, which is 
very close to one for this very low noise temperature. One trajectory takes longer to detach 
from zero, and it is the cause of the plateau in panel (b), where the average of $x_N^2$ over 
the ten trajectories  of panel (a) is shown. 

\begin{figure}[h!]
    \begin{center}
    \includegraphics[width=0.45\textwidth]{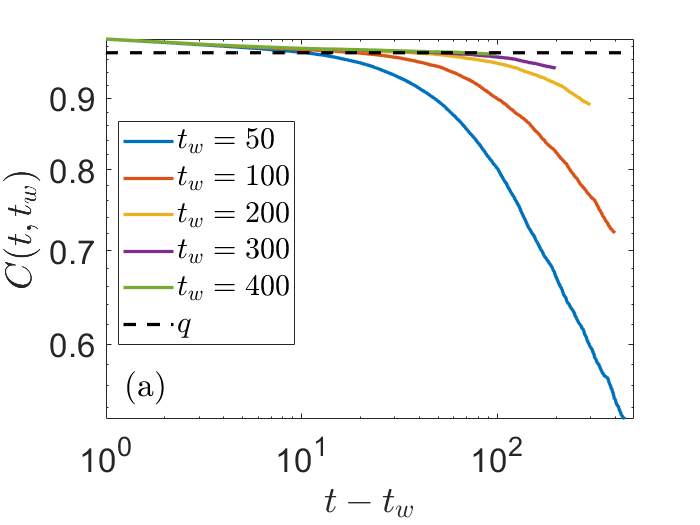}
   \includegraphics[width=0.45\textwidth]{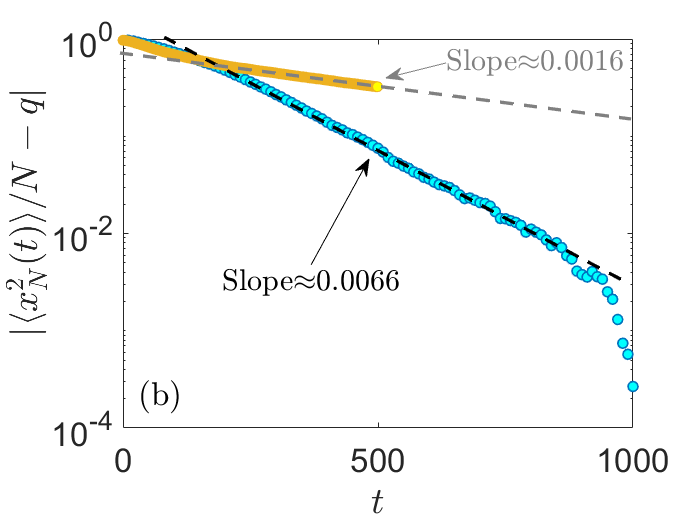}
        \end{center}
        \vspace{-0.25cm}
    \caption{\small 
    Relaxation after a quench from the extended to the condensed phase of a model with $a=-0.125$  at $T=0.05$. 
    (a) Self-correlation function for $N=8000$.
    The Edwards-Anderson parameter $q$ is indicated with a dashed horizontal line.
    (b) Approach to the asymptotic value $qN$, with $q$ the analytic value found for $N\to\infty$, 
    of the normalized square of the largest mode position, 
    $\langle x_N^2\rangle/N$, for $N=4000$ (light blue) and $N=8000$ (yellow) 
    in a linear-log representation.
      }
    \label{fig:aging-anegative}
\end{figure}

In Fig.~\ref{fig:aging-anegative} we display the two-time correlation function 
and averaged approach to its asymptotic value of the largest mode for two system 
sizes. An $N$-dependent exponential relaxation established at times 
$t$ \raisebox{-0.1cm}{$\stackrel{>}{\sim}$} $10^2$, is highlighted in panel (b) 
by using a linear-log scale. It suggests that the 
two-time dependence of the panel (a) is a transient, and that it will 
disappear at longer waiting times, with the correlations rapidly reaching 
$q$ and no longer decaying below it, for these system sizes.

\subsubsection{Moving phase M$_R$}

Taking $a$ to go beyond the phase limits towards the right and left moving phases, we
find rather different dynamic behaviour. Here, we discuss the case in which 
$\bar\gamma_\mu$ has a minimum at $\lambda_{\mu^*} < 2J$, that is to say, the 
dynamics in the phase M$_R$. As anticipated in Sec.~\ref{sec:GGE-Model}, for these 
parameters the equilibrium calculation predicts diverging $\langle p^2_\mu\rangle_{\rm GNM}$, 
for at least some $\mu> \mu^*$. 

\begin{figure}[h!]
\begin{center}
    \includegraphics[width=0.34\textwidth]{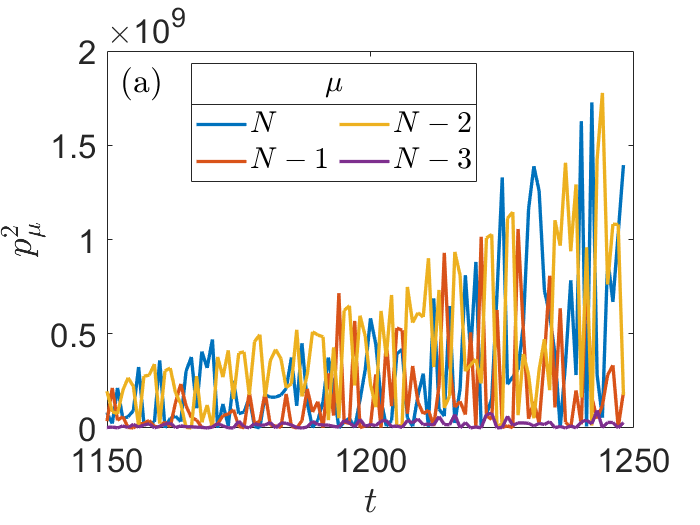}
    \hspace{-0.4cm}
    \includegraphics[width=0.34\textwidth]{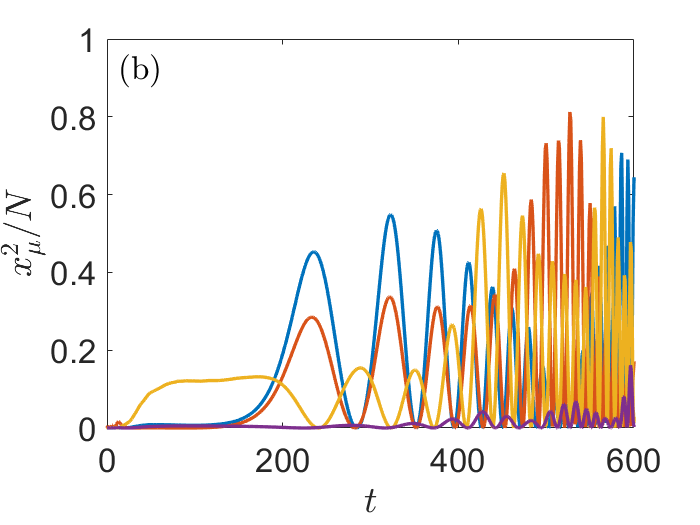}
       \hspace{-0.4cm}
        \includegraphics[width=0.34\textwidth]{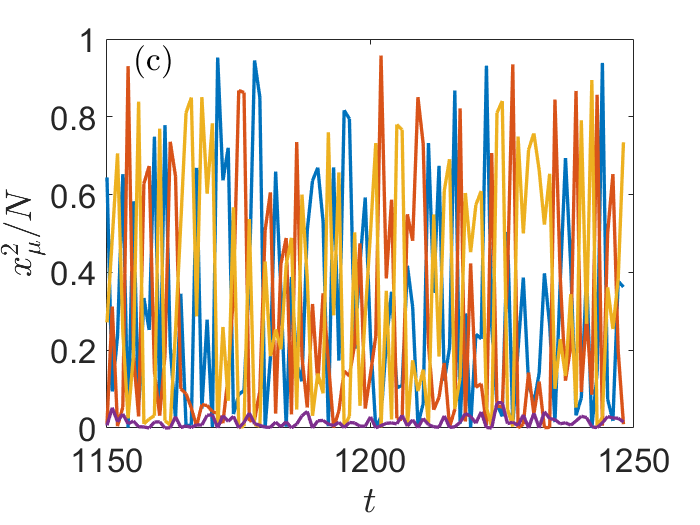}
        \end{center}
            \vspace{-0.25cm}
\caption{\small Zero temperature dynamics in the moving 
    phase M$_R$  for $a = 0.14$, $\gamma=0.1$ and $N = 500$.
(a) Momentum trajectories, $p_\mu^2$, and (b) position trajectories, $x_\mu^2$, 
of a few modes in the right end of the spectrum.
(c) Zoom over long time trajectories obtained using another initial condition, 
where three modes dominate the dynamics. 
}
    \label{fig:fig12}
\end{figure}

The dynamics in the right moving phase are depicted in Fig.~\ref{fig:fig12} for $T=0$.
The modes on the far right of the spectrum show an unbounded growth of the momenta $p_\mu^2$. 
For the initial condition used in (a), the more prominent are  
the last two modes, and to a lesser extent, the third mode from the right.
The positions $x_\mu^2 / N$ display increasingly rapid oscillations, with the two rightmost modes being the most relevant; the third mode decreases in amplitude (b). Depending on the initial condition, the dynamics may vary, with certain modes becoming important, as in (c),  but in all cases the relevant modes are those on the far right, located above the mode that generates the minimum of $\gamma_\mu$. The choice of $\gamma = 0.1$ for this case is justified because in this way the oscillations can be observed more clearly and persist longer before the simulation becomes unstable due to the excessive growth of the momentum. The motion takes place in a subspace of the $N$-dimensional sphere spanned by the modes with $\mu>\mu^*$. 
The mechanism leading to the ever faster motion in the subspace is exemplified in the full solution of the Langevin dynamics of the $N=2$ case in App.~\ref{app:C2}.
Basically, it turns out that the effective damping parameters are proportional to
$-(\bar\gamma_\mu-\bar \gamma_\nu)/(\lambda_\mu-\lambda_\nu)$. As the $\lambda$'s are
chosen in increasing order, $\lambda_1<\lambda_2<\ldots <\lambda_N$, if $\mu < \nu$ then the effective damping  becomes negative whenever $\bar\gamma_\mu < \bar \gamma_\nu$ and the dynamics in the $\mu,\nu$ plane
accelerates without limit. Consequently, although the total energy 
decreases in time,  
$1/(2m) \sum_\mu p_\mu^2$ diverges.

\subsubsection{Moving phase M$_L$}

\begin{figure}[h!]
\begin{center}
    \includegraphics[width=0.34\textwidth]{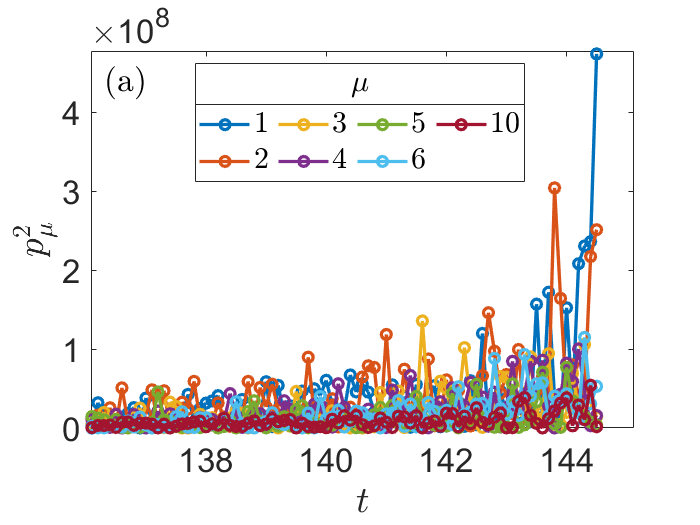}
       \hspace{-0.4cm}
    \includegraphics[width=0.34\textwidth]{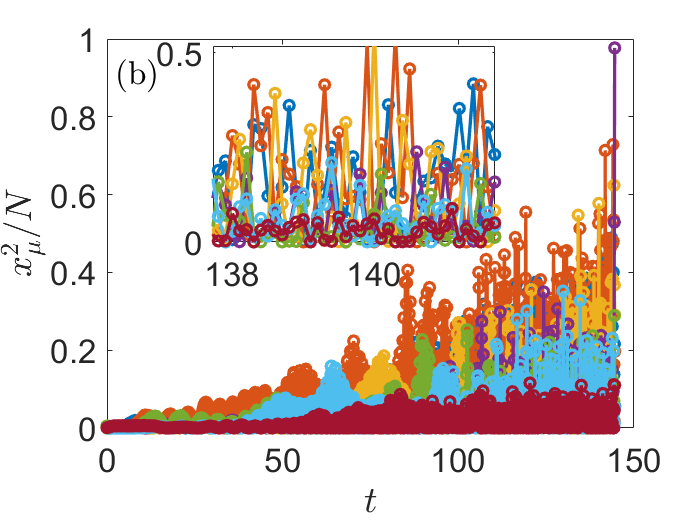}
       \hspace{-0.4cm}
      \includegraphics[width=0.34\textwidth]{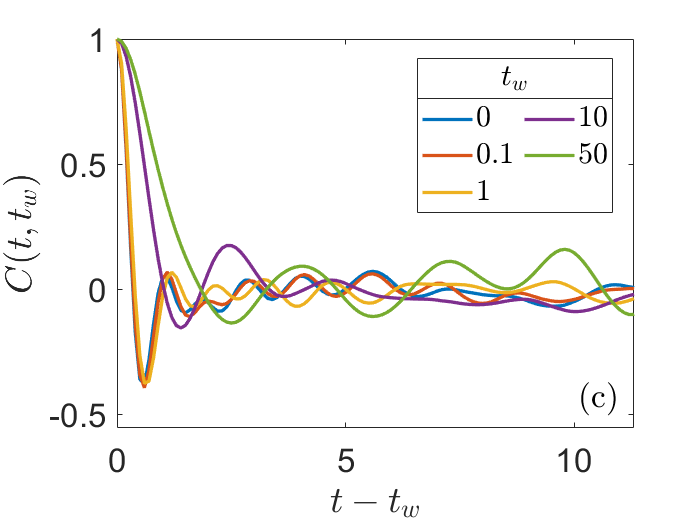}
      \end{center}
          \vspace{-0.25cm}
    \caption{\small 
    Moving phase M$_L$.  Dynamics at  $T=1.8$, in a model with $a=-0.25$ and $N=500$, using 
    a friction coefficient $\gamma=0.1$. 
    (a) Momentum trajectories, $p_\mu^2$, and (b) position trajectories, $x_\mu^2$, 
of a few modes in the left end of the spectrum. Inset: zoom over a late time window.
     (c) Two-time global correlation function, waiting times in
    the key.
       }
    \label{fig:fig13}
\end{figure}

In the left moving phase, M$_L$, the $\bar\gamma_\mu$ as a function of $\lambda_\mu$ 
reach a maximum at a  $-2J<\lambda_{\mu^*}<2J$. 
The most relevant modes are, in this case,  the ones at the left of the spectrum, 
as exemplified in Fig.~\ref{fig:fig13}(a) and (b) where the time-dependent momentum and 
position components $\mu=1, \dots, 6$ are plotted.
These are the  modes located to the left of 
the one that generates the maximum in the profile of $\bar\gamma_\mu$ vs $\lambda_\mu/(2J)$.
They increase the amplitude of their oscillatory behaviour as time elapses. On the contrary, 
the mode $\mu=10$ which is also shown in the figure lies to the right of the maximum, 
$\lambda_{10} > \lambda_{\mu^*}$, and 
its amplitude remains bounded in the course of time.
The global correlation function is displayed in~(c).  The oscillations persist until the simulation becomes unstable because of the divergence of the modes $\mu< \mu^*$. 

A transition from a condensed phase with finite $\langle p_\mu^2\rangle$ for all $\mu$ at low temperature, 
and a moving phase with some diverging $\langle p_\mu^2\rangle$ at high temperature may seem weird.
This is due to the way in which the momenta are coupled to the coordinates in our model. In a model with 
conventional kinetic energy, $\sum_\mu p_\mu^2/(2m)$, the momenta satisfy equipartition and 
$\langle p_\mu^2\rangle = mT$ in Boltzmann equilibrium. Here, instead, the components of the 
momentum vector appear multiplied by a function of the coordinates; in a sense, they have 
position dependent mass. Moreover, they are not bounded. A transition towards a phase with 
infinite $e_{\rm kin}$, can therefore be triggered by a simultaneous change of behaviour of the 
position vector.

\section{Dynamics of the GNM with the $\overline \gamma_\mu$ of the GGE}
\label{sec:sampling}

We now turn to the analysis of the stochastic dynamics of ${\mathcal H}_{\rm GNM}$, 
with parameters $\bar\gamma_\mu$ {obtained from the corrected derivation in 
App.~\ref{app:new-derivation} for the quenches 
of the} isolated Neumann model studied in~\cite{BaCuLoNe22}. We 
call these models ${\mathcal H}_{\rm GGE}$ since 
the parameters $\{\bar\gamma_\mu\}$ are fixed by requiring that the long-term Newtonian evolution of (local)
observables in the original model ${\mathcal H}$ 
is described by the GGE measure, see Table~\ref{table:uno}. 

We investigate whether the 
stochastic dynamics approach the expected stationary state $P_{\rm GGE}$.
We start in Sec.~\ref{subsec:Quench-Inst} 
by recalling which was the quench procedure applied to the isolated  
Neumann model and how the 
parameters $\{\bar \gamma_\mu\}$
of the stationary distribution have to be selected in order to describe 
the asymptotic behaviour with a GGE measure. 
In Sec.~\ref{subsec:role-temperature} we discuss the role played by the temperature of the bath 
in the stochastic dynamics. Then, we perform the stochastic dynamics analysis in Sec.~\ref{subsec:asymptotic} exploring the evolution in the various
phases.

\subsection{Instantaneous quenches of the isolated Neumann model}
\label{subsec:Quench-Inst}

The post-quench dynamics of the isolated classical Neumann model
differ, depending on the initial conditions and the kind of quench performed. 
In Ref.~\cite{BaCuLoNe22} the Neumann model with harmonic constants drawn from 
a Wigner distribution with scale $J_0$ was initialised in conventional, Boltzmann, thermal 
equilibrium at temperature $T_0$. At time $t=0$ the spring constants 
were all instantaneously rescaled  by the factor $J/J_0$. 
The Newtonian dynamics were then studied in the asymptotic long-time limit taken, 
ideally, after the thermodynamic
limit. 

The region of the phase space explored by the Newtonian dynamics is restricted by 
the conservation  of the $N$ independent quantities 
\begin{equation}
I_\mu = x_\mu^2 +\frac{1}{mN} \sum_{\nu (\neq \mu)} \frac{(x_\mu p_\nu - x_\nu p_\mu)^2}{\lambda_\nu-\lambda_\mu} 
\; . 
\end{equation}
These are fixed initially, right after the quench, and remain so ever after. On average over the initial 
distribution and in the infinite size $N\to \infty$ limit,  these constants are given by~\cite{BaCuLoNe22}
\begin{eqnarray}
\langle I_\mu(0^+) \rangle_{i.c.} &=&  
k_1 \; 
 \displaystyle{\frac{b - \lambda_\mu}{a-\lambda_\mu}}
 \label{eq:Imu-GGE-main}
\end{eqnarray}
with $k_1= T_0^2/(J_0 J)$, $b=J(J+J_0)/T_0$, and
\begin{eqnarray}
a= 
\left\{
\begin{array}{ll}
\displaystyle{\frac{J}{J_0}} \left( T_0 + \displaystyle{\frac{J_0^2}{T_0}} \right) & \qquad T_0>J_0
\\
[5pt]
2J & \qquad T_0<J_0
\end{array}
\right.
\end{eqnarray}
 together with
\begin{equation}
\langle I_N(0^+)\rangle_{i.c.} = \left( 1 - \frac{T_0}{J_0} \right)\left( 1 - \frac{T_0}{J}  \right) N + o(N)
\end{equation}
at $T_0<J_0$. 
These expressions depend on the two control parameters $T_0/J_0$ and $J/J_0$.

The long-term Newtonian dynamics reach a steady state 
described by the GGE measure, 
$\propto \exp[-\beta \sum_\nu \bar \gamma_\nu I_\nu- \beta \bar z_{\rm GGE} (\sum_\nu x_\nu^2 -N)]$. 
The time-averages of $\langle x_\mu^2(t)\rangle_{i.c.}$ and $\langle p_\mu^2(t)\rangle_{i.c.}$
reach stationary values which coincide with static averages over this measure~\cite{BaCuLoNe22}.
(Compared to the convention used in~\cite{BaCuLoNe22} for the Lagrange multiplier, 
$\beta \bar z_{\rm GGE} = z_{\rm GGE}$.) 

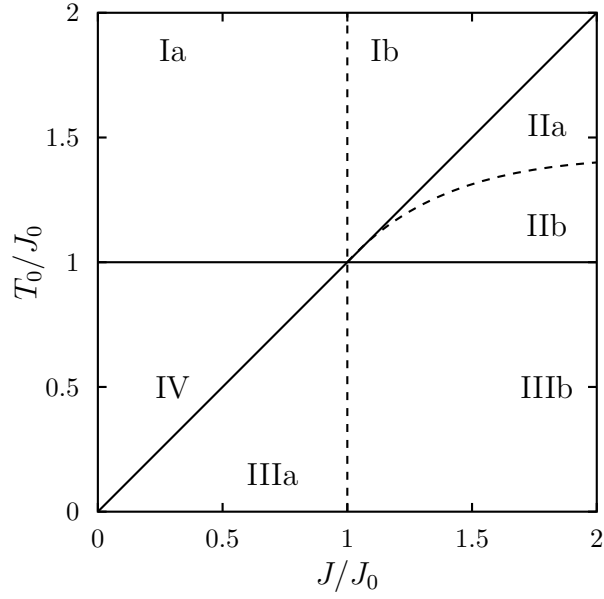
\begin{figure}
\centering
\begin{tikzpicture}[scale=3.3]

\draw[thick] (0,0) rectangle (2,2);

\foreach \x/\xlab in {0/0, 0.5/0.5, 1/1, 1.5/1.5, 2/2}{
    \draw[thick] (\x,0) -- (\x,0.035);
    \draw[thick] (\x,2) -- (\x,1.965);
    \node[below] at (\x,-0.03) {\footnotesize \xlab};
}

\foreach \y/\ylab in {0/0, 0.5/0.5, 1/1, 1.5/1.5, 2/2}{
    \draw[thick] (0,\y) -- (0.035,\y);
    \draw[thick] (2,\y) -- (1.965,\y);
    \node[left] at (-0.03,\y) {\footnotesize \ylab};
}

\node[below=14pt] at (1,0) {$J/J_0$};
\node[rotate=90, above=16pt] at (0,1) {$T_0/J_0$};

\draw[thick] (0,0) -- (2,2);      \draw[thick] (0,1) -- (2,1);      
\draw[thick, dashed] (1,0) -- (1,2);

\draw[thick, dashed] (1,1) .. controls (1.25,1.28) and (1.6,1.37) .. (2,1.4);

\node at (0.3,1.85)  {Ia};
\node at (1.15,1.85) {Ib};
\node at (1.8,1.55)  {IIa};
\node at (1.8,1.15)  {IIb};
\node at (0.3,0.5)   {IV};
\node at (1.8,0.5)   {IIIb};
\node at (0.7,0.15) {IIIa};

\end{tikzpicture}
\caption{\small Dynamic phase diagram of the isolated SSK/Neumann model
   with harmonic constants drawn from a Wigner distribution with scale $J_0$,   
     initialised in thermal Boltzmann 
    equilibrium at temperature $T_0$, and subject to the quench controlled by the parameter 
    $J/J_0$, as explained in the text~\cite{BaCuLoNe22}. 
  The solid line diagonal for $T_0>J_0$ is the phase transition between the phase I where $q=0$ and $z_{\rm GGE}=T_0+J^2/T_0$ and the phase  
  II where $q=0$ and $z_{\rm GGE}=2J$.
  The solid line diagonal for $T_0<J_0$ is the phase transition between the phase
  III with $q\neq 0$ and $z_{\rm GGE}=2J$ and a moving phase
   IV. The distinction Ia-Ib and IIIa-IIIb is made by the vertical dashed line which corresponds to the pure Boltzmann 
   equilibrium at $T_0$ for ${\mathcal H}_{\rm SSK}$ with no quench. The distinction IIa and IIb is marked by the curved dashed line, $T_0/J_0 = (J/J_0)^{1/2}$ where 
   the constants of motion $\langle I_\mu(0^+)\rangle_{i.c.}$ are all equal and the quenched 
   dynamics of the isolated Neumann model has aspects of Boltzmann equilibrium for  ${\mathcal H}_{\rm SSK}$. For example, $\langle p_\mu\rangle_{\rm GGE}$ are all equal. Other 
details on the phase transitions are described in the text.  
          }
    \label{fig:fig14}
\end{figure}

With methods explained in Refs.~\cite{CuLoNePiTa18,BaCuLoNe22}, the analysis of the 
long term dynamics of the infinite size system determines 
a two dimensional dynamic phase diagram, with four phases.
On one axis of this phase diagram, the parameter $T_0/J_0$ controls the kind of initial conditions, 
either randomly located on the sphere for $T_0/J_0>1$, with $\langle x_\mu^2\rangle_{i.c.}={\mathcal O}(1)$
for all $\mu$,  or condensed 
on the North/South pole for $T_0/J_0<1$, with $\langle x_N^2\rangle_{i.c.}=qN$
and all other modes $o(N)$. On the other axis, the parameter $J/J_0$ 
controls whether energy is injected ($<1$) or extracted ($>1$) by the quench. This phase diagram 
is reported in  Fig.~\ref{fig:fig14}.

The Lagrange multiplier enforcing the 
spherical constraint is given by
\begin{eqnarray}
z_{\rm GGE} = \left\{
\begin{array}{ll}
T_0 + \displaystyle{\frac{J^2}{T_0} } & \qquad  J < T_0

\; , 
\\
2J & \qquad J > T_0
\; . 
\end{array}
\right.
\label{eq:zGGE-main}
\end{eqnarray}
In the $N\to\infty$ limit,
the averaged square components of the particle's momentum, which are also the 
mode temperatures, $\langle p^2(\lambda) \rangle_{\rm GGE} = T(\lambda)$, are 
{determined by 
\begin{equation}
\frac{1}{2} \rho(\lambda) (z-\lambda) \langle I(\lambda)\rangle_{i.c.} 
=
\rho(\lambda) T(\lambda) 
\left[
1- \dashint d\lambda' \; \rho(\lambda') \frac{T(\lambda')}{\lambda-\lambda'} 
\right]
\label{eq:T-spectrum}
\end{equation}
which is solved by}
\begin{equation}
2 \pi^2 \, [\rho(\lambda)]^2 \, \langle p^2(\lambda) \rangle_{\rm GGE}^2
= -\left[1+G(\lambda)\right]
+ \sqrt{\left[1+G(\lambda)\right]^2 + 4 [g(\lambda)]^2}
\label{eq:main}
\end{equation}
with 
\begin{eqnarray}
g(\lambda) &=&  \frac{1}{4 J^2}\sqrt{(2J)^2 - \lambda^2}(z-\lambda)k_1 \frac{b-\lambda}{a-\lambda}
\; ,
\\
G(\lambda) &=& \frac{k_1}{2J^2} 
\Big[\lambda^2 - \lambda\left(b+z-a\right)+\left(b-a\right)\left(z-a\right) 
\nonumber\\
&& 
\qquad 
+\frac{\sqrt{a^2-(2J)^2}\left(a-b\right)\left(a-z\right)}{\lambda-a}-2J^2 \Big]
\; , 
\end{eqnarray}
and $\rho(\lambda)$ the Wigner semi-circle law on the interval $[-2J, 2J]$. The components of the 
position vector are then determined by 
\begin{equation}
\langle x^2(\lambda)\rangle_{\rm GGE} = \displaystyle{\frac{\langle p^2(\lambda)\rangle_{\rm GGE}}{z_{\rm GGE}-\lambda} }
\; . 
\end{equation}
The parameter measuring the condensation along the North/South direction is
\begin{eqnarray}
q \equiv \frac{1}{N} \,  \langle x_N^2\rangle =  
\left\{
\begin{array}{ll}
0 & \qquad T_0 > J_0
\\
\left( 1-  \displaystyle{ \frac{T_0}{J_0}} \right)^{1/2} \left( 1-  \displaystyle{ \frac{T_0}{J} } \right)^{1/2} 
& \qquad T_0<\min(J,J_0)
\end{array}
\right.
\end{eqnarray}

Some details about the phase transitions in Fig.~\ref{fig:fig14} are the following.
At the transition between I and II, $J=T_0>J_0$ and $\langle p_N^2\rangle =0$.  
The parameter $q$ vanishes at the transition between II and III, $T_0=J_0<J$.
At the transition between I and IV, $T_0=J_0$ and $\langle p_N^2\rangle \to \infty$.
At the transition between III and IV, $J=T_0<J_0$ and $\langle p_N^2\rangle \to \infty$ as well.
On the dashed curve in phase II all the $\langle I_\mu(0^+)\rangle_{i.c.}$ turn out of be equal and 
the averaged square momenta spectrum flat, $\langle p^2_\mu\rangle_{\rm GGE} = {\rm cst}$.

The $N$ requirements that fix the parameters $\beta\bar\gamma_\mu$ of the GGE measure that describes the 
asymptotic dynamics of the quenched isolated model are
\begin{equation}
\langle I_\mu(0^+) \rangle_{i.c.} = \langle I_\mu \rangle_{\rm GGE} \qquad  \mbox{for all} \; \mu
\; . 
\end{equation}
{When performing this study we realized that there was a mistake in the 
derivation of the multipliers $\gamma(\lambda)$ in Sec. 5.5 
of Ref.~\cite{BaCuLoNe22}. We explain the 
correct derivation in App.~\ref{app:new-derivation}. The expressions given in that paper were correct only in Boltzmann equilibrium, $J=J_0$,  
and on the special line, $T_0^2 = JJ_0$. They did not invalidate the evaluation of the 
mode temperatures from Eq.~(\ref{eq:T-spectrum}) since this equation is independent of 
$\gamma(\lambda)$. In the present study, though, it is necessary to use the 
correct $\bar\gamma_\mu(\lambda_\mu)$ to get the target $T(\lambda_\mu)$ with the Langevin process.
}

It is clear from the expressions derived in App.~\ref{app:new-derivation} 
that the $\bar\gamma_\mu$ are  
not simple quadratic functions of the $\lambda_\mu$, as the ones that we used in the  analytic and 
numeric studies of Sec.~\ref{sec:numerics}. Moreover, they are proportional to the parameter $T$, 
which we have (arbitrarily at this point) factorized from the parameters $\gamma_\mu$ used in~\cite{BaCuLoNe22}. 

Let us give some details on the dependence of $\{ \bar\gamma_\mu\}$ on 
the spring constants.
The threshold $J=J_0$, shown as a vertical dashed line in Fig.~\ref{fig:fig14}, corresponds to no quench performed
and, since the initial conditions were taken from Boltzmann equilibrium of the model ${\mathcal H}_{\rm SSK}$ at temperature $T_0$, 
the Newtonian dynamics proceed in the same Boltzmann equilibrium. Thus, 
$\beta \bar\gamma_\mu = -\lambda_\mu/(2T_0)$ for $J=J_0$ and any $T_0$. 

The $\bar\gamma_\mu(\lambda_\mu)$  for four representative sets of parameters 
$(J/J_0, T_0/J_0)$ 
are plotted in Fig.~\ref{fig:phaseI} (a)-(c) in phase I, Fig.~\ref{fig:phaseIII}
(a)-(c) in phase III, Fig.~\ref{fig:phaseIIa} (a), (d) in phase IIa.
Here, we added a distinction within I, with Ia being 
to the left of the equilibrium vertical line $J=J_0$ and Ib to the right of it. 
Similarly, we distinguished IIIa and IIIb with the same criterion. 
We can already reckon an analogy between this phase diagram
and the phases of the quadratic model displayed in Fig.~\ref{fig:phase-diagram}. 
it will be discussed in  detail below. 
The behaviour of $\bar\gamma_\mu(\lambda_\mu)$ 
in sector IV resembles the one of the right moving phase M$_R$ of the quadratic model.

In the numerical studies that we will discuss below, we 
{use as $\gamma_\mu$ the solutions to the analytic large $N$ equation for 
$\gamma(\lambda)$~\cite{BaCuLoNe22} derived in App.~\ref{app:new-derivation}
evaluated} at the discrete values $\lambda_\mu$ sampled from the Wigner-Dyson
semi-circle. Like this, we get the pairs $(\lambda_\mu, \gamma_\mu)$. This construction avoids the 
large fluctuations in $\gamma_\mu$  found when solving the finite $N$ equations.

 \subsection{The role of the bath temperature}
 \label{subsec:role-temperature}

The Langevin 
equation for a generic ${\mathcal H}_{\rm GGE} = \sum_\nu \bar \gamma_\nu I_\nu$ reads
\begin{eqnarray}
\dot x_\mu &=&   \displaystyle{ \frac{\delta}{\delta p_\mu}} \sum_\nu  \bar\gamma_\nu I_\nu 
= k_BT \displaystyle{ \frac{\delta}{\delta p_\mu}} \sum_\nu  \gamma_\nu I_\nu
\; , 
\label{eq:Langevin-1}\\
\dot p_\mu &=& -  \displaystyle{ \frac{\delta}{\delta x_\mu}} \sum_\nu  \bar\gamma_\nu I_\nu -  z_{\rm GGE} x_\mu + P_\mu^\nu (\xi_\nu - \gamma \dot x_\nu)
\nonumber\\
&=& - k_BT \displaystyle{ \frac{\delta}{\delta x_\mu}} \sum_\nu  \gamma_\nu I_\nu -  z_{\rm GGE} x_\mu + P_\mu^\nu (\xi_\nu - \gamma \dot x_\nu)
\; , 
\label{eq:Langevin-2}
\end{eqnarray}
with $\xi$ a white noise, with delta-correlation, $\langle \xi_\mu(t) \xi_\nu(t') \rangle = 2\gamma k_BT \delta_{\mu\nu} \delta(t-t')$.
Apart from the spherical constraint, this equation samples the measure 
\begin{equation}
P 
\propto 
e^{-\beta \sum_\mu \bar\gamma_\mu I_\mu }
=
e^{- \sum_\mu \gamma_\mu I_\mu }
\label{eq:GGE-measure-last}
\end{equation}
with $\gamma_\mu(J/J_0, T_0/J_0)$ the parameters determined 
{by the corrected derivation in App.~\ref{app:new-derivation}, independently of 
$\beta$, for the quench protocol in}~\cite{BaCuLoNe22}. 
Therefore, we can sample the measure~(\ref{eq:GGE-measure-last}) using the stochastic dynamics (\ref{eq:Langevin-1})-(\ref{eq:Langevin-2})
at any temperature $T$.
We confirm this assertion in App.~\ref{app:temperature} where we show long-term results for a particular 
choice of parameters $\bar\gamma_\mu$ and three temperatures of the bath.

\subsection{Asymptotic dynamics}
\label{subsec:asymptotic}

In this section we study the stochastic dynamics and we check whether, in the large-$N$ limit, the long-time noise 
and initial-condition averages of trajectories generated by the Langevin dynamics of the model ${\mathcal H}^z_{\rm GGE}$ converge to their static counterparts, 
\begin{eqnarray}
 \lim\limits_{t\to\infty} \langle x_\mu^2(t)\rangle_{\xi, i.c.} =  \langle x_\mu^2\rangle_{\rm GGE}
 \; , 
\\
 \lim\limits_{t\to\infty}\langle p_\mu^2(t)\rangle_{\xi, i.c.}  = \langle p_\mu^2\rangle_{\rm GGE} 
 \; , 
 \\
 \lim\limits_{t\to\infty} \langle x_\mu(t) p_\mu(t)\rangle_{\xi, i.c.} = \langle x_\mu p_\mu \rangle_{\rm GGE} 
 \; , 
 \end{eqnarray}
 (neglecting possible finite size fluctuations).

First, we will discuss the behaviour of the model for parameters
in the phases I, III and IV of the phase diagram in Fig.~\ref{fig:fig14} which, we 
recall, represents the Newtonian dynamics of the SSK/Neumann model and its 
representation in terms of the GGE. Next, we will explain what happens in 
phase II and why the behaviour is different there.

\subsubsection{Phase I.}
\begin{figure}[h!]
\begin{center}
\includegraphics[width=0.32\textwidth]{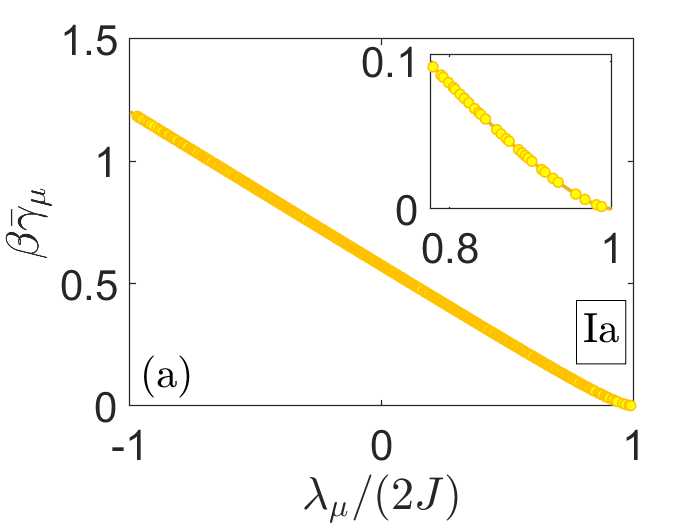}
\includegraphics[width=0.32\textwidth]{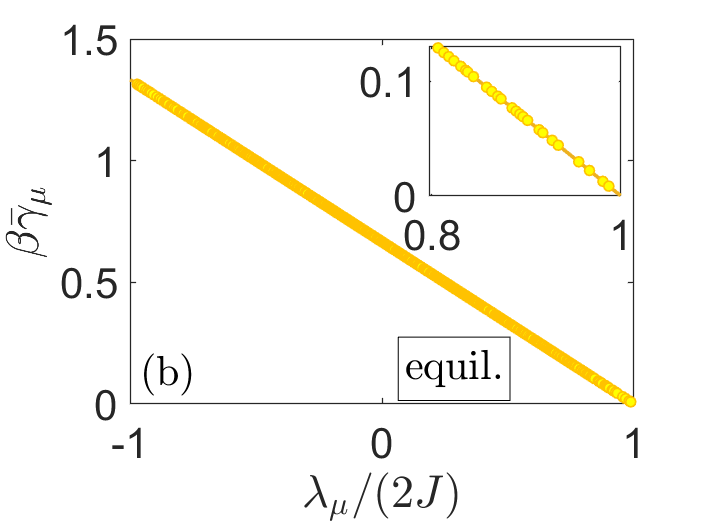}
\includegraphics[width=0.32\textwidth]{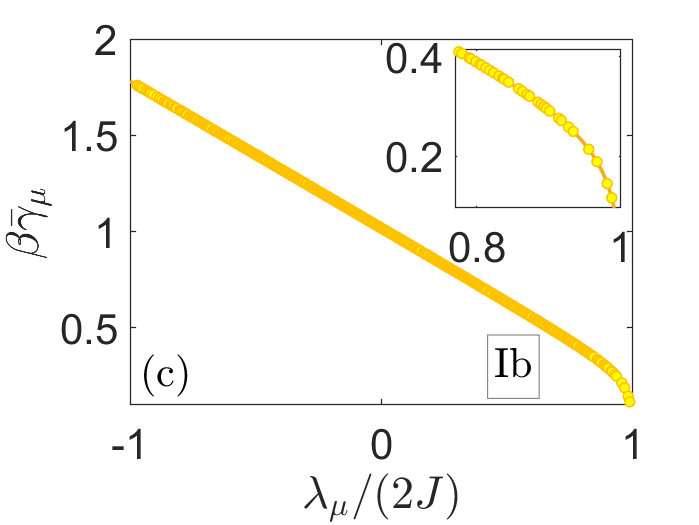}
\end{center}
\begin{center}
\includegraphics[width=0.32\textwidth]{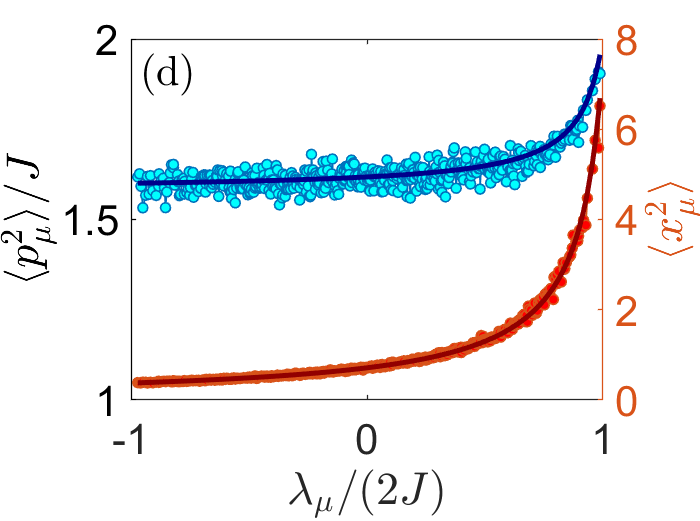}
\includegraphics[width=0.32\textwidth]{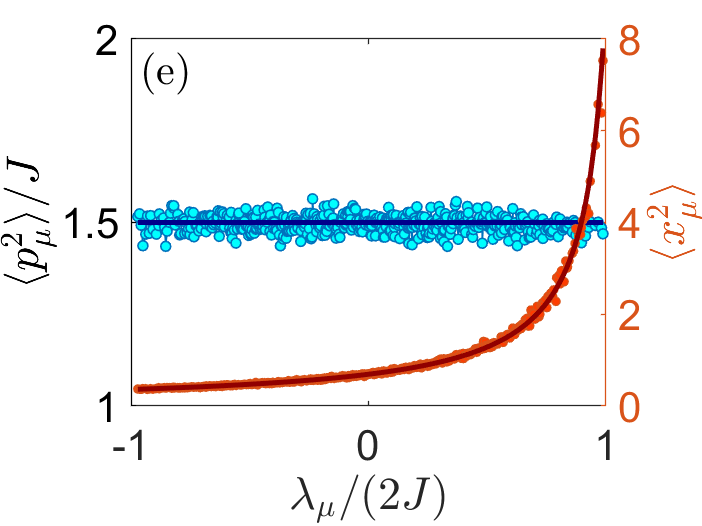}
\includegraphics[width=0.32\textwidth]{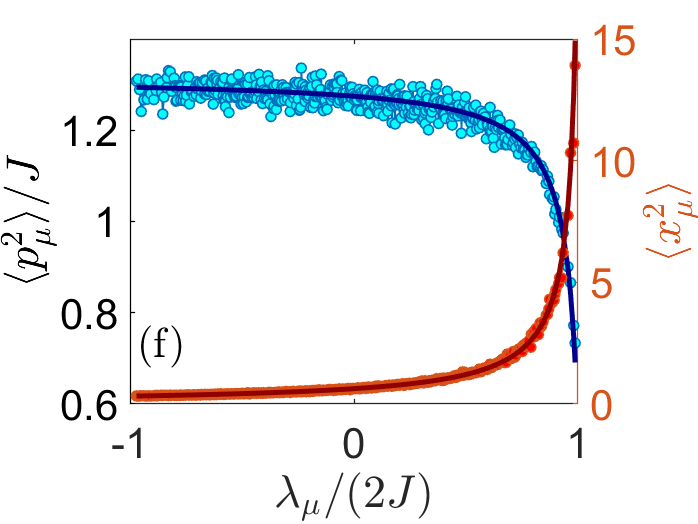}
        \end{center}
           \caption{\small Phase I. 
    First row: the parameters $\beta\bar\gamma_\mu$ as functions of $\lambda_\mu/(2J)$ in three points in phase I 
    of the phase diagram of  Fig.~\ref{fig:fig14}: 
    $J/J_0=0.9$ (a), $1$ (b), and $1.3$ (c). $T_0/J_0=1.5$ and $T=T_0$ 
    in all cases. $N=500$. 
    {In equilibrium (b) $J=J_0=1$, $\beta\bar\gamma_\mu=(2J-\lambda_\mu)/(2T_0)$, 
    so that $\beta {\mathcal H}_{\rm GGE} = \beta_0 {\mathcal H}_{\rm NM/SSK} +\beta_0 J N$.
    For $T_0=1.5$, $\beta\bar\gamma_\mu = (2J/3)[1-\lambda_\mu/(2J)]$.
    } The insets show magnifications close to the right edge of the spectrum.
    Second row: the observables $\langle p_\mu^2 \rangle$ and $\langle x_\mu^2 \rangle$ for the 
    same parameters used in the first row. The data points are the long time limit of the Langevin evolution 
    at $T=T_0 = 1.5 \, J_0$, averaged over time between $t=100$ and $t=1000$. 
    In (e) the system reaches conventional Boltzmann equilibrium 
    with ${\mathcal H}_{\rm NM/SSK}$, and $\langle p_\mu^2 \rangle/J = mT/J =1.5$. The solid lines are the 
    static averages $\langle x_\mu^2 \rangle_{\rm GGE}$ and  $\langle p_\mu^2 \rangle_{\rm GGE}$.
   }
    \label{fig:phaseI}
    \end{figure}

Phase I is studied in Fig.~\ref{fig:phaseI}.
In this phase, the parameters $\bar\gamma_\mu(\lambda_\mu)$ take different forms depending on 
$J/J_0<1$ (Ia), $J/J_0=1$ (equilibrium) or $J/J_0>1$ (Ib), see the phase diagram in Fig.~\ref{fig:fig14}. The $\lambda_\mu$ dependence of  $\overline \gamma_\mu$ is displayed in panels (a), (b), (c), respectively, including a zoom over the right-edge of the spectrum in the insets. Since $N$ is finite, the maximum $\lambda_N$ does not reach $2J$ but is slightly smaller than it. In phase Ia (a), the curve $\overline\gamma_\mu(\lambda_\mu)$ has an incipient minimum close to the 
right edge which will develop for larger $N$.
In the context of the Newtonian dynamics of the isolated Neumann model, the threshold $J=J_0$ corresponds to no quench performed
and, since the initial conditions were taken from Boltzmann equilibrium at temperature $T_0$, 
the Newtonian dynamics proceed in the same Boltzmann equilibrium. Thus, 
$\beta \bar\gamma_\mu = -\lambda_\mu/(2T_0)$ for $J=J_0$ and any $T_0$ (b). This linear 
functional form separates the two trends shown in the panels (a) and (c). In the latter case, 
phase Ib, the $\overline\gamma_\mu$ slightly bends downwards close to the right edge.

In the Langevin simulations we used initial conditions built with the procedure 
explained in Sec.~\ref{subsec:numerical-procedure}, with no mode carrying macroscopic weight and rather high initial kinetic energy. We then evolved these initial conditions 
with the Langevin equations for ${\mathcal H}_{\rm GGE}$ with
$\bar\gamma_\mu(T_0/J_0, J/J_0)$, at inverse temperature $\beta=\beta_0$. 
The time-averaged long-time results are shown in the figure with blue data points
for the momenta and red data points for the coordinates. The black curves are 
the analytic predictions for the asymptotic expressions,
obtained from the static calculation with the GGE measure for $N\to\infty$. 

In (e) the system reaches conventional Boltzmann equilibrium 
    with ${\mathcal H}^z_{\rm SSK/NM}$ at $T=T_0$. 
In the two non-trivial cases shown falling in Ia (d) and Ib (f)  neither the position 
nor the momentum condense and all modes take finite 
values. This is similar to the behaviour found in phase 
E with $a > 0$ of the quadratic model.

The momenta, $\langle p^2_\mu\rangle$, increase monotonically with $\lambda_\mu$ 
with positive curvature in Ia (d) while they decrease monotonically in Ib (f). 
The crossover occurs at $J=J_0$, where the spectrum of averaged square
momenta is simply flat.

The coordinates,
$\langle x^2_\mu\rangle$,  increase monotonically in Ia (d) and Ib (f). 
They are well captured by the 
stochastic data, apart from some deviations at the right end of the spectrum which we ascribe to 
finite size effects and poor numerical precision. 

Note that the fact that in Ia (a) the $N\to\infty$ 
function $\overline\gamma(\lambda)$ 
has a minimum close to the edge does not pose any problem: 
the stochastic dynamics converge to the expected
finite asymptotic expressions.
Indeed, in Ia, panels (a) and (d), there is no divergence of the averaged momentum square 
although there is an incipient minimum in $\bar \gamma_\mu(\lambda_\mu)$. For these parameters, we would 
need to use much larger system sizes to see the minimum appear.
Using a lower $J/J_0 = 0.8$ and still $T_0/J_0=1.5$
       the minimum is already visible for $N=500$. Nevertheless, the modes at the right of the minimum are still 
       stable with no momentum divergence,  and the Langevin dynamics converge 
       to the expected finite asymptotic limit.
This is similar to what we obtained in the reentrant extended phase of the quadratic model, 
discussed in Sec.~\ref{subsec:extended-quadratic}.

\subsubsection{Phase II.}

\begin{figure}[h!]
\begin{center}
\includegraphics[width=0.32\textwidth]{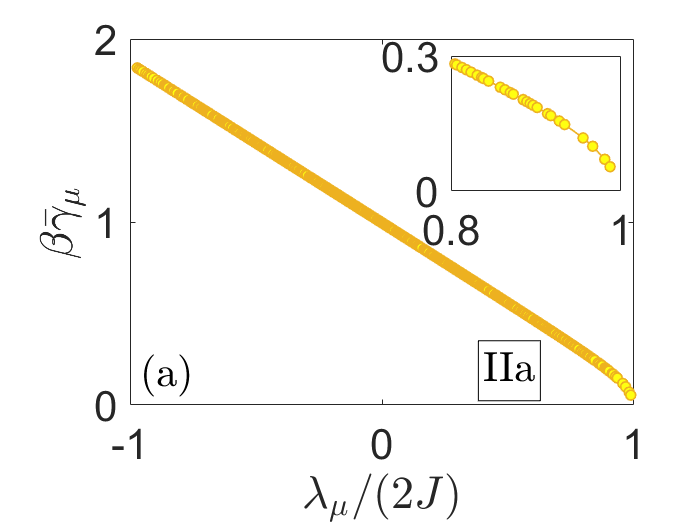}
\includegraphics[width=0.32\textwidth]{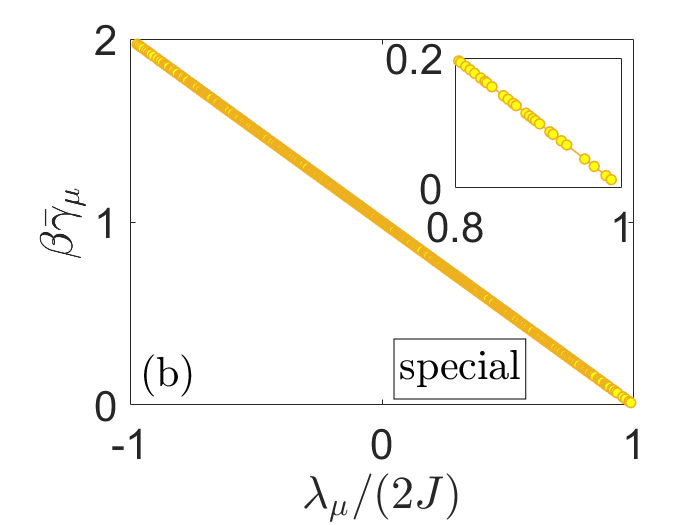}
\includegraphics[width=0.32\textwidth]{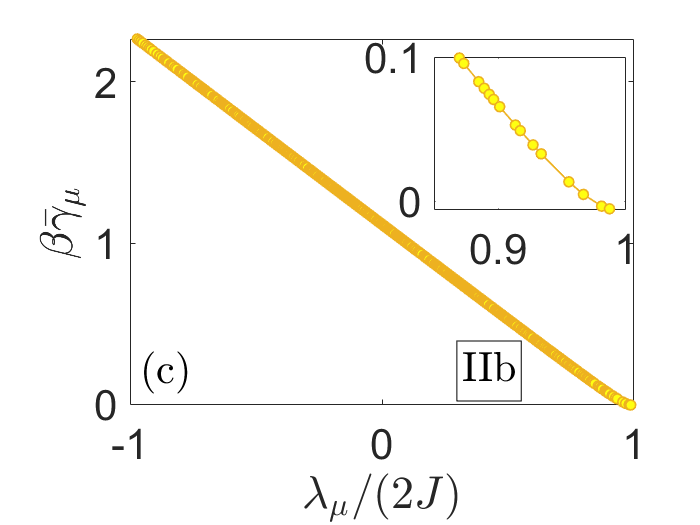}
\end{center}
\begin{center}
\includegraphics[width=0.32\textwidth]{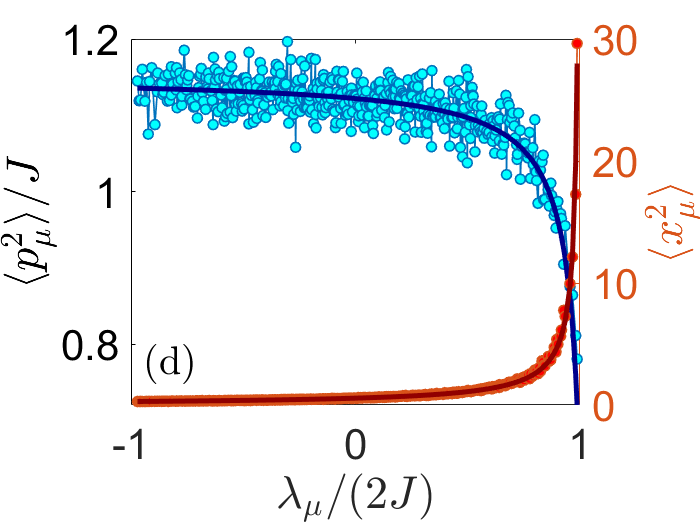}
\includegraphics[width=0.32\textwidth]{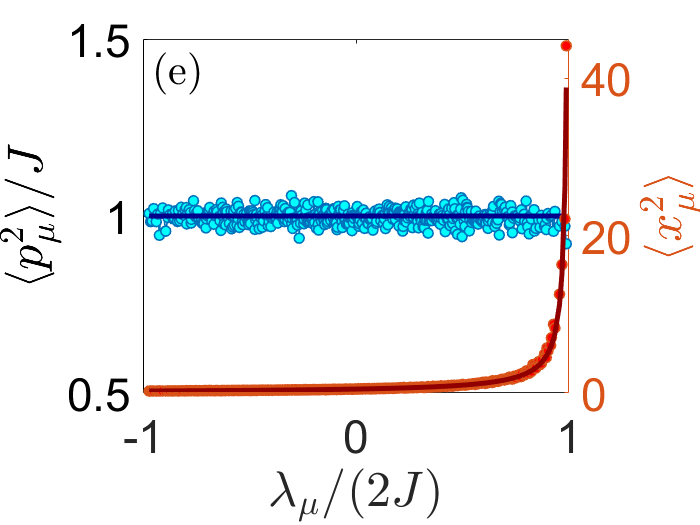}
\includegraphics[width=0.32\textwidth]{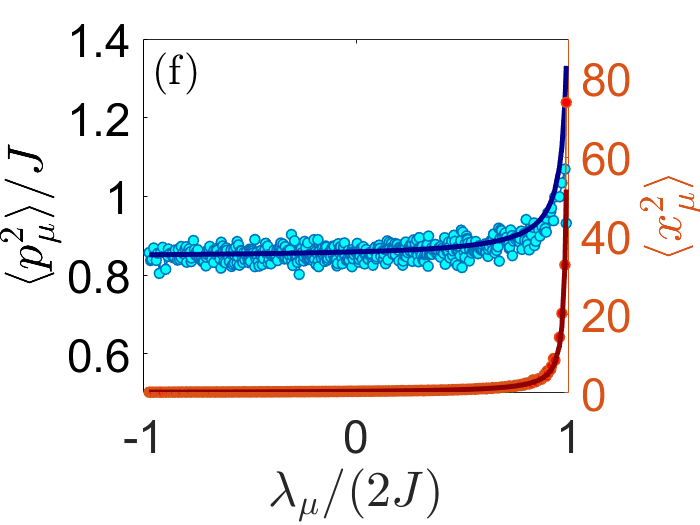}
\end{center}
    \caption{\small {Phase II. $J_0=1$ and $J=1.7$ in all cases.
    $T_0=1.5$ in IIa (a) and (d), $T_0=\sqrt{1.7}$ on the special line (b) and (e), and 
    $T_0=1.1$ in IIb (c) and (f). The system size is $N=500$. In the 
    first row: the parameters $\beta\bar\gamma_\mu$ as functions of $\lambda_\mu/(2J)$.
    In IIa, $\gamma(\lambda)$ has negative 
    curvature close to the edge $\lambda/(2J)=1$, on the special line, $\gamma(\lambda)$ is linear
    in $\lambda$, while in IIb, $\gamma(\lambda)$  has positive curvature at this end.  In the second row: 
    the data points are the long time limit of the Langevin evolution while the solid lines are the 
    GGE predictions.}
    }
    \label{fig:phaseIIa}
    \end{figure}

{
Three cases in phase II are studied in Fig.~\ref{fig:phaseIIa}, where we display 
$\beta \bar\gamma_\mu$ (a)-(c), and $\langle p^2_\mu\rangle$ together with $\langle x_\mu^2 \rangle$ (d)-(f), 
against $\lambda_\mu/(2J)$. The main difference in $\gamma_\mu(\lambda_\mu)$
is the curvature close to the limit $\lambda_\mu = 2J$: negative in phase IIa, zero on the special line, and positive in 
phase IIb. This is reproduced by the sign of the derivative of $\langle p_\mu^2\rangle$ with respect to 
$\lambda_\mu$ close to the same edge. The mechanism for this fact is similar, though not equal, to the one 
for the quadratic model. Although there is here an integral relation between $d^2\gamma(\lambda)/d\lambda^2$ and 
$dT(\lambda)/d\lambda$, the contribution to the integral of modes close to the edge is magnified by a diverging 
factor and,  
in practice, the relation becomes local there.
The agreement between stochastic (light blue and red data points) and 
asymptotic static results (dark solid lines) in panels (d)-(f)
is very good.}

\subsubsection{Phase III.}

\begin{figure}[h!]
\begin{center}
 \includegraphics[width=0.32\textwidth]{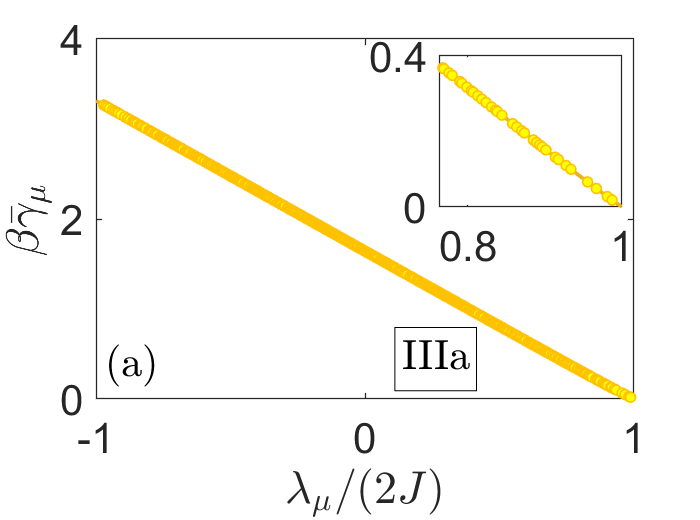}
\includegraphics[width=0.32\textwidth]{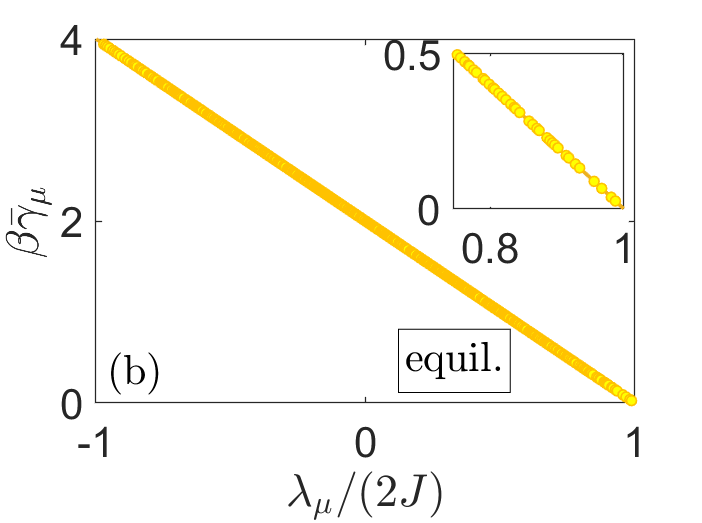}
\includegraphics[width=0.32\textwidth]{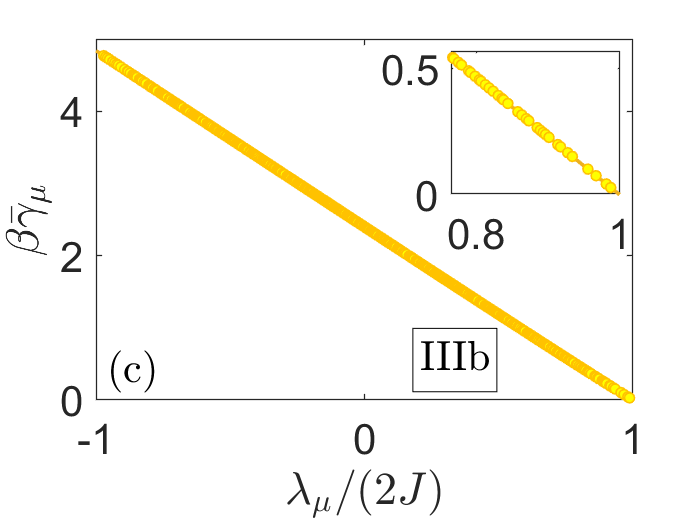}
\end{center}
\begin{center}
\includegraphics[width=0.32\textwidth]{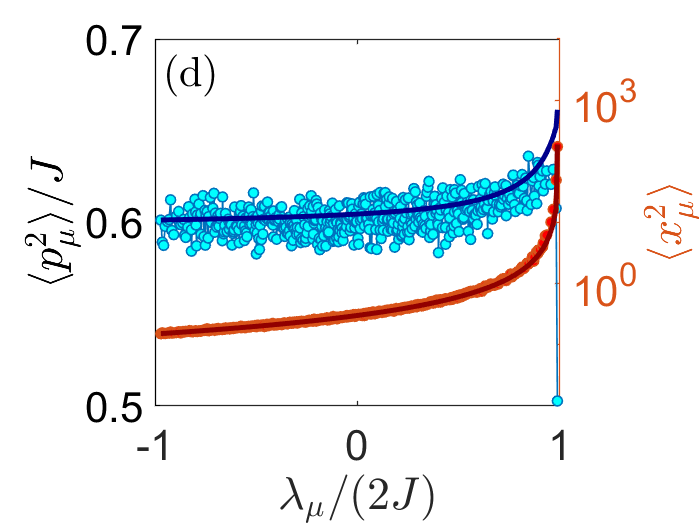}
\includegraphics[width=0.32\textwidth]{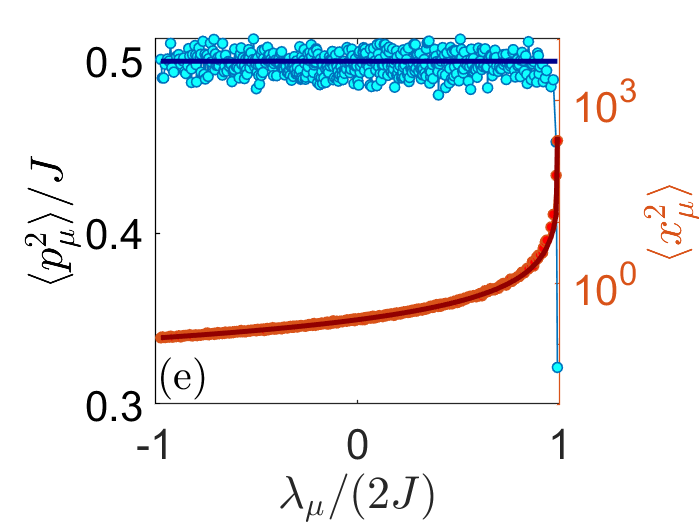}
\includegraphics[width=0.32\textwidth]{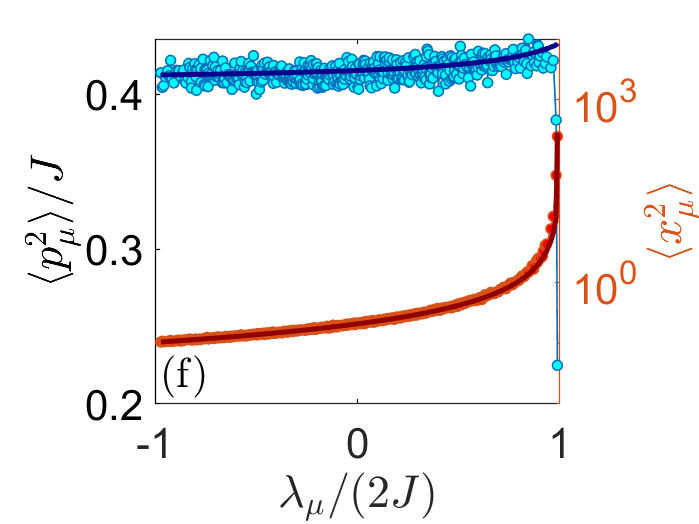}
\end{center}
    \caption{\small Phase III. The parameters $\beta\bar\gamma_\mu$ in three points in phase III of Fig.~\ref{fig:fig14}: 
    $T_0/J_0=0.5$ and $J/J_0=0.7$ (a) and (d), $1$ (b) and (e), and $1.5$ (c) and (f). 
    $N=500$. The stochastic dynamics evolve from initial conditions with macroscopic weight on the 
    $N$th direction of space and rather low kinetic energy. $\langle p_N^2\rangle$ suffers from strong finite size effects, 
    just as in Fig.~\ref{fig:hardpN}(c) for the quadratic model.
     }
    \label{fig:phaseIII}
    \end{figure}

Three cases with condensation are displayed in Fig.~\ref{fig:phaseIII}. 
Phase III is the counterpart of the condensed phase C of the quadratic model for $a>0$.
The $\bar\gamma_\mu(\lambda_\mu)$ are close to linear, as shown in (a)-(c) and are monotonically 
decreasing in the full extent of this phase, to the right and to the left of the equilibrium 
dashed vertical line.

In the stochastic dynamics, the initial conditions were chosen to be of condensed kind, 
as the ones described in Sec.~\ref{subsec:numerical-procedure}, 
with $T_0=0.5$. We make this choice to avoid the slow relaxation that  is characteristic of this phase for extended random initial conditions.

{The bulk momentum spectra vary only slightly with $\lambda_\mu$ in the two non-equilibrium cases (d) and (f), whereas the large-$N$ bulk prediction is flat in equilibrium (e). The isolated last mode requires separate consideration.}
The coincidence between
stochastic and asymptotic static results is very accurate, including the divergence of 
$\langle x^2_N \rangle$. The $N$th component of the momentum, $\langle p_N^2\rangle$, takes a smaller value 
than expected. This mismatch is similar to the one we observed in the quadratic model, see Fig.~\ref{fig:hardpN}, 
and larger system sizes are presumably needed to get closer to the analytic prediction.

\subsubsection{Phase IV.}
    
    The phase  named IV in~\cite{BaCuLoNe22} is similar to the moving phase M$_R$ for the quadratic model. A
spectrum of $\bar\gamma_\mu$ with a minimum at $\lambda_\mu^*$ close to $\lambda_N$ leads to  
divergent $\langle p_{\mu > \mu^*}^2\rangle_{\rm GGE}$, which are the hallmarks of these moving phases.
The stochastic dynamic behaviour of these modes strongly 
resemble the ones displayed in Fig.~\ref{fig:fig13} and we do not 
show them here.

   \section{Conclusion}
\label{sec:conclusions}

In this paper we studied the stochastic Langevin dynamics of a model defined by the 
Hamiltonian ${\mathcal H}_{\rm GNM} = \sum_\mu \bar\gamma_\mu I_\mu$ with 
$I_\mu = x_\mu^2 +1/(mN) \sum_{\nu(\neq \mu)} (x_\mu p_\nu -x_\nu p_\mu)^2/(\lambda_\nu-\lambda_\mu)$.
This model is a generalization of the integrable SSK/Neumann model of classical mechanics, which is
recovered for $\bar\gamma_\mu=-\lambda_\mu/2$.  For any choice of the 
parameters $\bar\gamma_\mu$, the model remains integrable under Newtonian dynamics.

Our interest was to use the Langevin stochastic dynamics to sample the distribution
$\exp(-\beta {\mathcal H}^z_{\rm GNM})/Z_{\rm GNM}$ which has two interpretations:
\begin{itemize}
\item 
It is the {conventional} canonical Boltzmann equilibrium 
measure at inverse temperature $\beta$ of the model ${\mathcal H}^z_{\rm GNM}$
with arbitrary $\bar\gamma_\mu$, {provided that its partition function is finite.}
\item
It is the Generalized Gibbs Ensemble measure which describes local observables
of the SSK/Neumann model evolved with Newtonian dynamics, if the 
$\bar\gamma_\mu$ are tuned to be fixed by the conditions 
$\langle I_\mu(0^+)\rangle_{i.c.} = \langle I_\mu\rangle_{\rm GGE}$ in the
quenching process. 
\end{itemize}

Our results are twofold. 
\begin{itemize}
\item
We presented the study of the canonical equilibrium and stochastic dynamics of one such GNM, 
for the choice $\bar\gamma_\mu =(a/J) \lambda_\mu^2 - b\lambda_\mu$. We 
determined the canonical phase diagram of the model in the $(a, T/J)$ plane fixing $b=1/2$.
We elucidated the  main characteristics
of the phase transitions and each of the phases. 

Two phases are similar 
to the well-known extended (paramagnetic) and condensed (ordered) phases of
the SSK model, which is recovered for $a=0$. The addition of non-trivial momentum 
dependencies in ${\mathcal H}_{\rm GNM}$ does not 
modify their qualitative properties. However, for $a\neq 0$ and sufficiently 
large in absolute value, two unstable regions appear in this 
kinetic problem which we called ``moving phases", $M_L$ and $M_R$. In these phases the averaged squared momenta of unstable 
modes diverge and the particle's motion gets faster and faster on the 
corresponding subspace of the sphere.

We confirmed these results 
by solving numerically the Langevin dynamics with white noise and dissipation.
We thus found that 
in the long term the stochastic evolution reaches the 
expected asymptotic limits. Depending on the phases being extended, condensed or moving
different kinds of relaxation or never ending motion were found. Of special interest
are
\begin{itemize}
    \item 
    the slow relaxation with aging phenomena 
found for quenches of the open model from extended initial conditions to the 
condensed phase. 
\item 
the divergence of some noise-averaged components of the momentum squared 
in the moving phases.
\end{itemize}

\item
We proposed to use Langevin methods to sample the GGE measure realised by an 
integrable classical model evolving in isolation, after a quench. We tested this proposal
by using the Hamiltonian ${\mathcal H}_{\rm GGE}$ with  
 the $\bar\gamma_\mu$ fixed to be the ones determined from the analysis of the 
isolated SSK/Neumann model with Newton's dynamics. 
{We found numerical agreement between the sampled observables and the target GGE predictions in the stable regimes studied. 
A conventional canonical probability interpretation of the measure with Hamiltonian ${\mathcal H}_{\rm GGE}$ 
at inverse temperature~$\beta$
additionally requires a finite partition function.}
However, as for the square $\bar\gamma_\mu$ model,
we observed:
\begin{itemize}
    \item 
    Slow relaxation with aging phenomena for quenches of the open model 
    from extended initial conditions to the condensed phase (not shown). We avoided them 
    by using condensed initial conditions.
        \item 
    The divergence of some noise-averaged components of the momentum squared 
    in the moving phase. This is a consequence of the fact that the effective friction of
the Langevin dynamics changes sign, {triggering a dissipation induced instability} which accelerates the dynamics instead of slowing it down.
\end{itemize}
The Hamiltonian ${\mathcal H}_{\rm GGE}$, although being integrable under Newton's 
dynamics, it still has a relatively complex structure with {a family of} 
$N$ {antipodal pairs of axis-aligned stationary configurations
with $\vec p=0$},
similarly to the better known SSK/Neumann model. This energy landscape induces 
slow stochastic relaxation for several sets of parameters.

\end{itemize}

As discussed  in Section~\ref{sec:531} slow dynamics may be present in whole regions
of parameter space, where diverging times  with $N$, the number of 
degrees of freedom, may be needed to reach this limit. 
We found slow dynamics towards condensation in the model ${\mathcal H}_{\rm GGE}$, 
similarly to what happens in the SSK/Neumann model. This tells us that in some cases sampling GGE measures 
with stochastic methods can be 
tricky, since extremely long time limits might be needed to do it correctly.

From a general perspective, the SSK/Neumann model shares many points in common with the $d$-dimensional $O(N)$ field theory in the large $N$ limit.
This relation has been realised in the context of coarsening phenomena~\cite{Bray94,Zannetti15,Crisanti-etal20} 
and how they compared to the dynamics of simple spin models with disorder
like the SSK one. Recently, the GGE description of the 
quantum $O(N)$ model was clarified~\cite{Defenu25b}. The idea of applying stochastic methods
to sample GGE measures could be applied to this and other models as well.

\vspace{1cm}

\noindent
{\bf Acknowledgements}
Financial support is acknowledged from 
ANR THEMA ANR-19-CE30-0014 (L. F. Cugliandolo and R. Gonz\'alez-Albaladejo), 
the Capes-Cofecub project PE3207237P (L. F. Cugliandolo, R. Gonz\'alez-Albaladejo and D. A. Stariolo), 
CNPq-fellowship 307337/2021-9 and grant Universal 402487/2023-0 (D. A. Stariolo)
and the 
ECOS SUD – Francia/Argentina A26E02: Thermal and Active Avalanches in Disordered Materials, 
the CNRS IRPs COMplexité en grande dimension : l'Apport de la Physique Statistique 
and Activation and Activity in Disordered Materials (L. F. Cugliandolo and G. S. Lozano).
{
During the preparation of this manuscript, the authors used Claude Sonnet 5.5 to check the derivation in 
App.~\ref{app:new-derivation} and ChatGPT was occasionally consulted on specific technical questions.  LLM were used to check English grammar in some parts of the text.}
    
\appendix
\setcounter{section}{0} \renewcommand{\thesection}{\Alph{section}}
\makeatletter
\renewcommand\@seccntformat[1]{\csname the#1\endcsname.\hspace{1em}}
\makeatother

\section{The Kramers equation}
\renewcommand{\thesection}{\Alph{section}}
\newcommand{\appref}[1]{App.~\ref{#1}}
\label{app:Kramers}
\renewcommand{\theequation}{A.\arabic{equation}}
\setcounter{equation}{0}

The presence of the  constraint which forces the particle  to 
remain on the sphere, 
has transformed the Langevin equation in one 
with multiplicative white noise. 
We should then reconsider the derivation of the Kramers equation
taking into account this uncommon structure~\cite{GoBaTs14,ArBaCuGoLo14,ArBaCuGoLo16,Schwartz}. 

Exploiting the Markov character of the process, the infinitesimal evolution 
of the probability density of finding the particle within the infinitesimal volume $[\vec x, \vec x+\Delta \vec x]$ 
with momentum $[\vec p, \vec p+\Delta \vec p]$ can be written as
\begin{eqnarray}
P(\vec x, \vec p; t+\Delta t) 
= 
\int D\vec x_t \int D\vec p_t \int D\vec \xi_t \; P(\vec \xi_t) \; 
\nonumber\\
\qquad\qquad\qquad\qquad\qquad\qquad
\delta(\vec p_{t+\Delta t} - \vec p)\delta(\vec x_{t+\Delta t} - \vec x) \; 
P(\vec x_t, \vec p_t; t)
\; , 
\label{eq:update}
\end{eqnarray}
where we indicate the discrete time dependencies as lower sub-scripts and we defined $\Delta t$ as the 
infinitesimal time-step.
We also use the compact notations $D\vec x_t = \prod_{\nu=1}^N dx^\nu_t$, and similarly for 
$D\vec p_t$ and $D\vec \xi_t$. The vectorial Dirac delta represents the product of 
deltas on all coordinates, $\delta(\vec p_{t+\Delta t} - \vec p) = \prod_{\nu=1}^N \delta(p^\nu_{t+\Delta t} - p^\nu)$, 
with an analogous expression for the delta on the position variable. 
The Gaussian weight of the white noise reads
\begin{equation}
P(\vec \xi_t) = \prod_{\nu=1}^N \frac{\Delta t^{1/2}}{\sqrt{4\pi\gamma k_BT}} \; e^{- \frac{1}{2} (\xi^\nu_t)^2 \frac{\Delta t}{2\gamma k_BT}}
\; .
\end{equation}
A derivation of the Kramers equation for a generic ${\mathcal H}$ and unconstrained dynamics, 
\begin{eqnarray}
&& 
\frac{\partial P(\vec x, \vec p; t)}{\partial t}
=
-\frac{\partial }{\partial p^\nu}
\Big\{ 
 \Big[ 
 - \frac{\partial {\mathcal H}}{\partial x_\nu}  - \gamma \frac{\partial {\mathcal H}}{\partial p_\nu}
\Big]  P(\vec x, \vec p; t) 
\Big\}
-
\frac{\partial }{\partial x^\nu} 
\Big[ 
 \frac{\partial {\mathcal H}}{\partial p_\nu} \; P(\vec x, \vec p; t)
\Big]
\nonumber\\
&&
\qquad\qquad\qquad
+ \gamma k_BT \; \frac{\partial^2 P(\vec x, \vec p; t) }{\partial p^\nu\partial p_\nu} 
\; , 
\end{eqnarray}
can be found in textbooks~\cite{vanKampen2007}.
This equation admits a stationary solution of the expected form $e^{-\beta {\mathcal H}}/{\mathcal Z}_{\rm GB}$. Indeed, setting the left-hand-side to zero
one can simply check that the exponential form sets to zero the right-hand-side for any  $\gamma$, $k_BT$  
and ${\mathcal H}$. The GNM form ${\mathcal H}_{\rm GNM} = \sum_\nu \bar\gamma_\nu I_\nu$ with 
$P_{\rm GNM} = Z^{-1}_{\rm GNM} \exp(-\beta \sum_\nu \bar\gamma_\nu I_\nu)$  is just a particular case. 
However, subtleties linked to the constraint appear in the treatment of the 
spherical models of interest and we now revisit the derivation of the Kramers equation for these models.

We start from Eq.~(\ref{eq:update}). 
The treatment of the delta over the coordinates does not present any difficulty. 
The analysis of the delta over the momenta needs more care. First, as we have 
already argued, the coordinate evolves regularly, since {$\Delta x^\mu_t = \partial {\mathcal H}/\partial p_\mu \; \Delta t$}. 
Thus, 
the projector ${\rm P}^\mu_\nu(\overline x_t^\mu \overline x_t^\nu)$, which in principle should be evaluated
at $\overline x^\rho_t = x^\rho_t + \alpha \Delta x^\rho_t$,
 can be safely evaluated at the pre-points
$x^\mu_t x^\nu_t$. This allows one to integrate over the thermal noise  but 
modifies the term with the double derivative with respect to the momenta:
\begin{equation}
 \frac{\partial^2}{\partial p^\eta \partial p_\zeta} 
 \Big[ {\rm P}^\eta_{\, \rho} {\rm P}^\rho_{\, \zeta} P(\vec x, \vec p, t)\Big]
=
 {\rm P}^\eta_{\, \zeta} \, 
\frac{\partial^2}{\partial p^\eta \partial p_\zeta} P(\vec x, \vec p, t)
\; , 
\end{equation}
where we used that ${\rm P}^\eta_{\, \rho}$ only depends on $\vec x$ and 
${\rm P}^\eta_{\, \rho} {\rm P}^\rho_{\, \zeta} = {\rm P}^\eta_{\, \zeta}$.
The Kramers equation then reads
\begin{eqnarray}
&& 
\frac{\partial P(\vec x, \vec p; t)}{\partial t}
=
-\frac{\partial }{\partial p^\mu}
\left\{ 
 \left[ 
- \frac{\partial{\mathcal H}}{\partial x_\mu}
 - \gamma {\rm P}^\mu_{\, \nu} \frac{\partial {\mathcal H}}{\partial p_\nu}
\right] P(\vec x, \vec p; t) 
\right\}
\nonumber\\
&&
\qquad\qquad\qquad
-
\frac{\partial }{\partial x^\mu} 
\left[ 
\frac{\partial {\mathcal H}}{\partial p_\mu} P(\vec x, \vec p; t)
\right]
+ \, \gamma k_BT \; {\rm P}^\mu_{\, \nu} \;  \frac{\partial^2 P(\vec x, \vec p; t) }{\partial p^\mu\partial p_\nu} 
\; .
\qquad
\end{eqnarray}
For a Hamiltonian with a conventional kinetic term with a quadratic dependence on the momenta,  
$\partial {\mathcal H} / \partial p^\rho = p_\rho/m$ and ${\rm P}^\rho_\nu \, \partial {\mathcal H} / \partial p^\rho =
{\rm P}^\rho_\nu \, p_\rho/m = p_\nu/m  = \partial {\mathcal H}/ \partial p^\nu$, simplifying a bit the 
terms within the square brackets. We have included the projector in front of the friction term here.

When the spherically constrained Boltzmann measure
$Z^{-1}_{\rm GB} \, e^{-\beta {\mathcal H}}$
is normalizable, it is the equilibrium measure of the finite $N$ dynamics. Convergence to it
requires a separate dynamical argument.
(In some cases, like the SSK model, 
the Langevin dynamics 
fails to equilibrate an infinite size system at low temperatures 
unless diverging times with $N$ are considered~\cite{CuDe95a,CuDe95b}.)

We  now check whether the Kramers equation (\ref{eq:kramers}) admits a stationary solution of Gibbs-Boltzmann form, 
with ${\mathcal H}$ generic. 
Let us first focus on the terms proportional to the friction $\gamma$:
\begin{equation}
\gamma {\rm P}^\mu_{\, \nu} \frac{\partial }{\partial p^\mu} \left[  \frac{\partial {\mathcal H}}{\partial p_\nu } P(\vec x, \vec p; t) 
+ k_BT \;  \;  \frac{\partial P(\vec x, \vec p; t) }{\partial p_\nu} 
\right]
\; . 
\end{equation}
The factors within the square brackets yield
\begin{eqnarray}
&& 
\left[  \frac{\partial {\mathcal H}}{\partial p_\nu } + k_BT (-\beta)    \frac{\partial {\mathcal H}}{\partial p_\nu } \right] P(\vec x, \vec p) = 0
\; . 
\end{eqnarray}
The remaining terms are
\begin{eqnarray}
&& 
\frac{\partial}{\partial p_\mu} 
\left[ \frac{\partial {\mathcal H}}{\partial x^\nu } 
 P(\vec x, \vec p) \right] 
- \frac{\partial}{\partial x_\mu} \left[\frac{\partial {\mathcal H}}{\partial p^\nu }  P(\vec x, \vec p) \right]
 \label{eq:intermediate0}
\end{eqnarray}
which also vanish identically using $  P(\vec x, \vec p) \propto \exp(-\beta {\mathcal H})$.
    
\section{The SSK/Neumann model}
\renewcommand{\thesection}{\Alph{section}}
\setcounter{equation}{0}
\label{app:GGE-to-SSK}
\renewcommand{\theequation}{B.\arabic{equation}}
\setcounter{figure}{0}
\renewcommand{\thefigure}{B.\arabic{figure}}

In this Appendix we show how to obtain the SSK/Neumann model as a particular case of the 
GNM one (App.~\ref{app:B1}) and we show some numerical results for the stochastic dynamics 
of the SSK/Neumann model (App.~\ref{app:B2}).

\subsection{From the GNM to the SSK/Neumann model}
\label{app:B1}

We here show that the generic model ${\mathcal H}_{\rm GNM}$ reduces to the
SSK/Neumann one for $\bar\gamma_\mu=-\lambda_\mu/2$ and using the spherical constraint. Indeed, 
\begin{eqnarray}
{\mathcal H}_{\rm GNM}&=& 
-\frac{1}{2} \sum_\mu \lambda_\mu I_\mu 
\nonumber\\
&=&
-\frac{1}{2} \sum_\mu \lambda_\mu x_\mu^2
-
 \frac{1}{2mN} \sum_{\mu \neq \nu} \lambda_\mu 
 \frac{x_\mu^2 p_\nu^2 + x_\nu^2 p_\mu^2 - 2 x_\mu p_\mu x_\nu p_\nu}{\lambda_\nu - \lambda_\mu}
\end{eqnarray}
Let us focus on the double sum in the second term:
\begin{eqnarray}
{\rm 2nd \ Term} &=& 
- \frac{1}{2mN} \sum_{\mu \neq \nu} (\lambda_\mu -\lambda_\nu+\lambda_\nu) 
\frac{x_\mu^2 p_\nu^2 + x_\nu^2 p_\mu^2 - 2 x_\mu p_\mu x_\nu p_\nu}{\lambda_\nu - \lambda_\mu}
\nonumber\\
&=& 
 \frac{1}{2mN} \sum_{\mu \neq \nu} 
\left( x_\mu^2 p_\nu^2 + x_\nu^2 p_\mu^2 - 2 x_\mu p_\mu x_\nu p_\nu \right)
\nonumber\\
&& 
- \frac{1}{2mN} \sum_{\mu \neq \nu} \lambda_\nu
\frac{x_\mu^2 p_\nu^2 + x_\nu^2 p_\mu^2 - 2 x_\mu p_\mu x_\nu p_\nu}{\lambda_\nu - \lambda_\mu}
\nonumber\\
&=& 
2e_{\rm kin} N  - {\rm 2nd \ Term}
\end{eqnarray}
and the 2nd Term equals $e_{\rm kin} N$. Thus, 
\begin{equation}
{\mathcal H}_{\rm GNM} = - \frac{1}{2} 
\sum_\mu \lambda_\mu I_\mu 
= 
-\frac{1}{2} \sum_\mu \lambda_\mu x_\mu^2
+ \frac{1}{2m} \sum_\mu p_\mu^2
= {\mathcal H}_{\rm SSK/NM}
\; . 
\end{equation}
    
\subsection{Numerical results}
\label{app:B2}

In order to estimate the precision of the numerical integration of the dynamics of the GNM model, we 
present here some numerical results for the relaxation of the SSK/Neumann model with ($m \neq 0$) and without 
($m =  0$) inertia, after a quench from the high temperature phase to the low one. 

\begin{figure}[h!]
\hspace{0.5cm} (a) \hspace{4.25cm} (b) \hspace{4.25cm} (c)
\\
\centerline{
\includegraphics[scale=0.28]{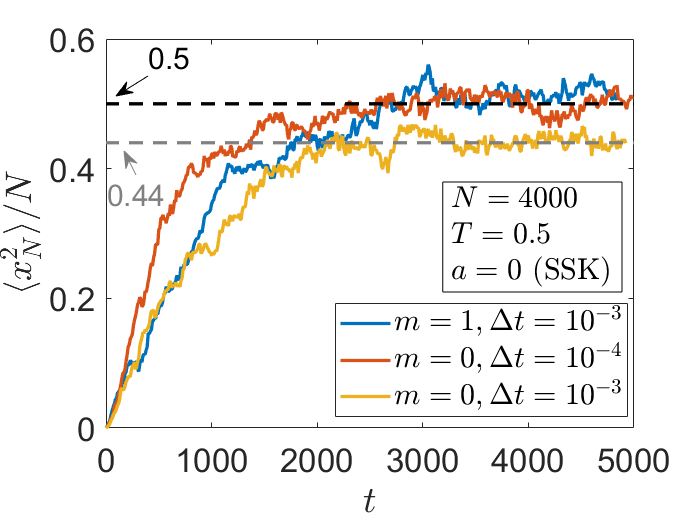}
\includegraphics[scale=0.28]{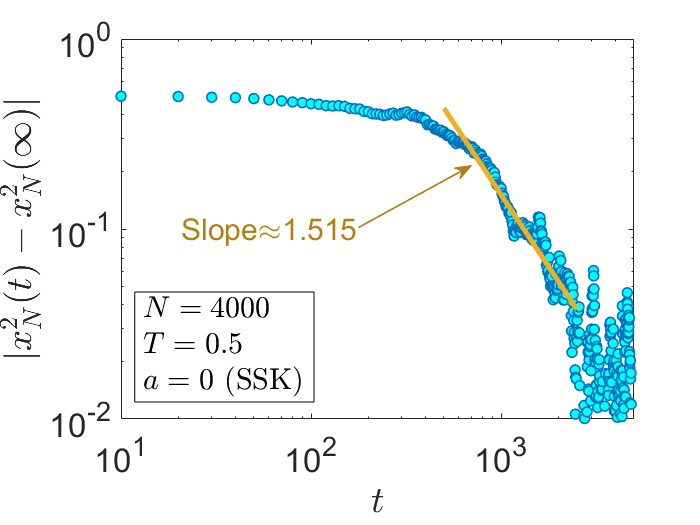}
\includegraphics[scale=0.28]{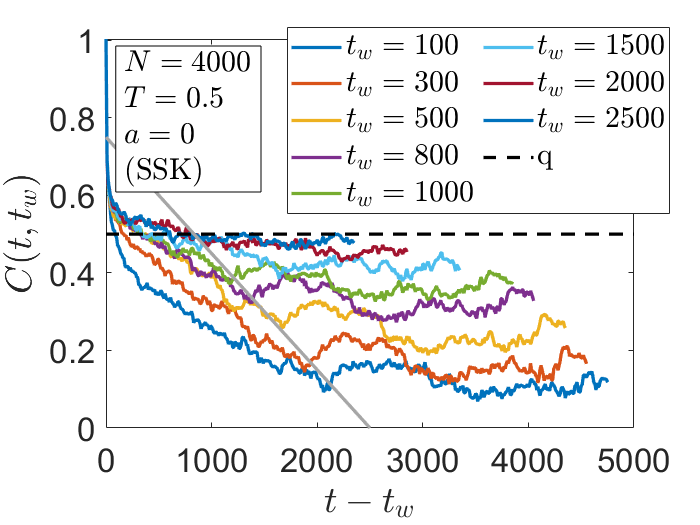}
}
\\
\centerline{(d) \hspace{4.25cm} (e) \hspace{4cm}}
\\
\centerline{
\includegraphics[scale=0.28]{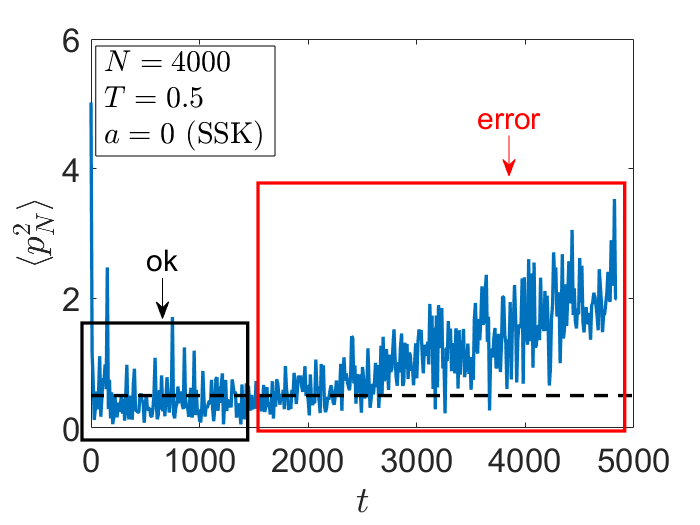}
\includegraphics[scale=0.28]{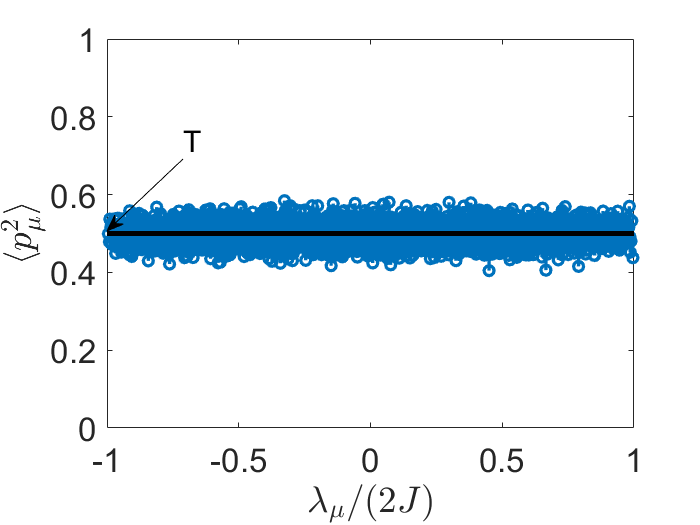}
}
\caption{\small The relaxation dynamics of the SSK/Neumann model from a 
disordered initial condition evolved at $T=0.5$. (a) Approach of 
$\langle x_N^2\rangle $ to its asymptotic value $qN$ for under-damped ($m=1$ with time-step 
$\Delta t =0.001$) 
and  over-damped ($m =  0$ with time-step $\Delta t = 0.0001$ in red and $\Delta t = 0.001$ in orange) dynamics. 
 (b) Time dependent distance
from the asymptotic value $qN$ in the under-damped case, 
compared to the analytic prediction of the over-damped case, $t^{-3/2}$. In (c)-(e) under-damped dynamics. 
(c)  Self-correlation 
function. The inclined grey line separates the out of equilibrium relaxation (to the left) from 
the equilibrium dynamics (to the right). (d) Time dependence of the average of the square of the $N$th momentum 
component.
(e) Spectrum of the square of the asymptotic values of the 
components of the momentum compared to the working temperature.
$\gamma=5, \, J=1$.
}
\label{fig:app-G}
\end{figure}

In Fig.~\ref{fig:app-G}(a) we see that, for the system size $N=4000$, the $N$th component 
of the position reaches the expected statistics, with  $\langle x_N^2\rangle = qN = (1-T/J) N$,
both for the under-damped and over-damped dynamics. However, there is an important 
difference. In the under-damped case the expected result is obtained for a time 
step $\Delta t = 0.001$  while in the over-damped case a smaller time-step is needed, 
$\Delta t = 0.0001$.
In both cases, the asymptote is reached at a time of the order of $t_{\rm eq}(N,T) \sim 2500$.
In the under-damped case, more or less simultaneously,  
the square of the $N$th component of the momentum detaches from $T=0.5$ and starts increasing,
showing that the errors incurred by the numerical algorithm start adding up [panel (c)]. 

The approach of the square of the $N$th component of the position to its asymptotic value 
follows the law $t^{-3/2}$ (derived in the over-damped limit~\cite{CuDe95a}), both with and without inertia 
[panel (b)].

The self-correlations decay in a way that depends on the waiting-time $t_w$, as long
as $t_w < t_{\rm eq}(N,T) \sim 2500$. Beyond this waiting-time the correlation 
saturates to $q$.  Moreover, for $t_w < t_{\rm eq}(N,T) \sim 2500$, when the 
total time $t$, which is used in the horizontal axis to build the time-delay $t-t_w$, 
gets close to $t_{\rm eq}(N,T) \sim 2500$ the curves can no longer be trusted to 
study the out of equilibrium relaxation. Indeed, although rather noisy, we see that 
all curves tend to saturate to a plateau when this limit, which is shown with a 
diagonal grey line,  is reached. The height of this plateau progressively gets closer to 
$q$ for increasing $t_w$. 
     
\section{The $N=2$ GNM}
\renewcommand{\thesection}{\Alph{section}}
\label{app:N2}
\setcounter{equation}{0}
\renewcommand{\theequation}{C.\arabic{equation}}
\setcounter{figure}{0}
\renewcommand{\thefigure}{C.\arabic{figure}}

As a fully solvable model we discuss the case in which the particle is constrained to move 
on a circle in two dimensions. Although one cannot claim any statistical physics 
behaviour for $N$ finite, a number of special properties are captured by
this simple example. The Hamiltonian of the GNM in the $N=2$ case reads
\begin{equation}
{\mathcal H}_{\rm GNM}=\sum_\mu \bar\gamma_\mu I_\mu
= \bar\gamma_1 x_1^2 +\bar \gamma_2 x_2^2 
- \frac{1}{2m} \frac{\bar\gamma_1-\bar\gamma_2}{\lambda_1-\lambda_2}
(x_1 p_2 - x_2 p_1)^2
\; , 
\label{eq:HGGE-N2}
\end{equation}
and we recall our convention, $\lambda_2>\lambda_1$.

\subsection{Stationary points}
\label{app:C1}

We analyze the extrema of the energy of the $N=2$ GNM
defined  in (\ref{eq:HGGE-N2}), to which we add the Lagrange multiplier
to enforce the spherical constraint
\begin{equation}
 {\mathcal H}^z_{\rm GNM}  =  {\mathcal H}_{\rm GNM} + \frac{z_{\rm GNM}}{2} (x^2 -2)
 \; . 
\end{equation}

Some particular cases and properties are the following.
\begin{itemize}
\item
For $\bar\gamma_\mu=-\lambda_\mu/2$, as in the SSK/Neumann case, 
 \begin{eqnarray}
 {\mathcal H}_{\rm GNM}
 &=& -\frac{1}{2}
\lambda_1 x_1^2 -\frac{1}{2}
\lambda_2 x_2^2 
+\frac{1}{2m}
 \; (p^2_1+ p^2_2 ) 
 \; , 
\end{eqnarray}
and we recover  the usual kinetic energy. 
\item
 For $\bar\gamma_1=-\lambda_1^2$,  $\bar\gamma_2=-\lambda_2^2$, the last term in ${\mathcal H}_{\rm GNM}$ reads
 \begin{equation}
-\frac{1}{2m}
 \frac{\lambda^2_1-\lambda^2_2}{{\lambda_2-\lambda_1}} \; (p_1 x_2 - p_2 x_1)^2
=
\frac{1}{m}
(\lambda_1+\lambda_2) \; p^2
 \end{equation}
and its sign is decided by $\lambda_1+\lambda_2$. Thus, if, for example, $\lambda_1=-1.5$ and 
$\lambda_2=1$, this contribution looks like a usual kinetic  energy but with negative 
sign. It can therefore be made as negative as wanted, and thus reduce the total energy as desired, 
by taking large values of $p_1$ and/or $p_2$. The minimal energy is thus achieved by 
a dynamic state.
\item
For $\bar\gamma_1>0$,  $\bar\gamma_2>0$ and $\lambda_2>\lambda_1$, we need to 
take $\bar\gamma_1>\bar\gamma_2$ to ensure that the last term in ${\mathcal H}_{\rm GNM}$ be positive definite. 
\item
If $\bar\gamma_1<\bar\gamma_2$ and $\lambda_1 <\lambda_2$,  the term involving the momenta can be taken to minus infinity by $p_1 \to \infty$
and/or $p_2\to \infty$. The normalization of $\vec x$ 
\begin{equation}
x^2_1 + x_2^2 = 2  \qquad \Rightarrow \qquad x_2 = \pm \sqrt{2-x_1^2}
\end{equation}
which is finite  (as $x_2$) and the  orthogonality with $\vec p$ implies 
\begin{equation}
p_1 = - \frac{ \sqrt{2-x_1^2}}{x_1} \, p_2
\end{equation}
so both components of the momentum diverge respecting this proportionality. 
Dynamically, in the minimal energy state the particle turns with infinite momentum around the circle.
\end{itemize}

\begin{figure}[h!]
\begin{center}
\begin{tikzpicture}
\draw[black, thick, <->] (-3.5,0) -- (3.5,0);
\draw[black, thick, <->] (0,-3.5) -- (0,3.5);
\draw[black, thick, ->,  in=180, out=0, looseness=5] (0,0) -- (1.43,1.43);
\draw[black, thick, ->, in=180, out=0, looseness=5] (1.76,1.08) -- (1.08,1.76);
\draw[gray, dashed, -] (1.43,0) -- (1.43,1.43);
\draw[gray, dashed, -] (0,1.43) -- (1.43,1.43);
\draw (2,0) circle (5pt);
\draw (-2,0) circle (5pt);
\fill (1.43,1.43) circle (2pt);
\node[star, star points=5, star point ratio=1.7, minimum size=0.1cm, fill=black] at (0,2) {};
\node[star, star points=5, star point ratio=1.7, minimum size=0.1cm, fill=black] at (0,-2) {};
\draw (0,0) circle (2cm);
\node at (3.5,-0.5) {$\vec v_1$};
\node at (-0.5,3.5) {$\vec v_2$};
\node at (1.43,-0.5) {$x_1$};
\node at (-0.5,1.43) {$x_2$};
\node at (0.8,0.3) {$\theta$};
\end{tikzpicture}
\end{center}
\caption{\small The $N=2$ configuration space of the GNM. The vertical positions 
signalled with the filled stars are stable, 
while the horizontal ones with large empty circles are unstable, in cases with $\bar\gamma_1>\bar\gamma_2$ and
$\lambda_2>\lambda_1$. The small dot on the circle represents a generic position of the 
particle $\vec x = (x_1, x_2) = \sqrt{2} (\cos \theta \, \hat e_1 + \sin\theta \, \hat e_2)$. 
The arrow attached to it sketches its velocity. The circle has radius $\sqrt{2}$. }
\label{fig:sketchN2}
\end{figure}
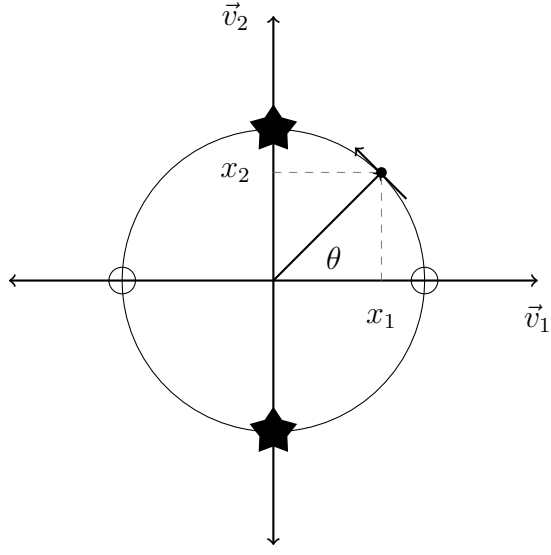

Let us examine the stationary states and their stability for generic $\bar\gamma_1, \bar\gamma_2$. 
The extremisation conditions read 
\begin{eqnarray}
0= 2\bar\gamma_1 x_1 -  \frac{1}{m} \frac{\bar\gamma_1-\bar\gamma_2}{\lambda_1-\lambda_2} 
\; (x_1 p_2-x_2 p_1) p_2 + z_{\rm GNM}(\vec x, \vec p) x_1 
\; , 
\\
[4pt]
0=2\bar\gamma_2 x_2 +   \frac{1}{m}  \frac{\bar\gamma_1-\bar\gamma_2}{\lambda_1-\lambda_2} 
\; (x_1 p_2-x_2 p_1) p_1 + z_{\rm GNM}(\vec x, \vec p) x_2
\; , 
\\
[4pt]
0=   \frac{1}{m}  \frac{\bar\gamma_1-\bar\gamma_2}{\lambda_1-\lambda_2} ( x_1 p_2-x_2 p_1) x_2 
\; , 
 \\
 [4pt]
0=  - \frac{1}{m}  \frac{\bar\gamma_1-\bar\gamma_2}{\lambda_1-\lambda_2} (x_1 p_2-x_2 p_1) x_1 
\; .
\end{eqnarray} 
The variation of $z_{\rm GNM}$ is unnecessary since after enforcing the spherical constraint those terms
vanish. 
$x_1$ and $x_2$ cannot be simultaneously zero. 
We identify two kinds of solutions, both with zero momentum:
\begin{itemize}
\item
{\it Non-degenerate $\bar\gamma_\mu$, $\bar\gamma_1\neq \bar\gamma_2$.}
As a first guess we set $x_2=\sqrt{2}$ and $x_1=0$. The third equation implies
$p_1=0$.   The first and forth equations are then identically satisfied. The second equation simplifies to 
$
[2\bar\gamma_2 + z_{\rm GNM}(\vec x, \vec p)] \sqrt{2} 
= 0 
$, which implies $z_{\rm GNM} = -2\bar\gamma_2 
$
For the moment we have  $\vec x = (0,\sqrt{2})$ and $\vec p = (0,p_2)$. The further condition 
$\vec x \cdot \vec p=0$ implies $p_2=0$ as well. Therefore, $\vec x = (0,\sqrt{2})$, $\vec p = (0,0)$, $z_{\rm GNM}=-2\bar\gamma_2$
and  $\vec x = (\sqrt{2},0)$, $\vec p = (0,0)$,  $z_{\rm GNM}=-2\bar\gamma_1$ are stationary states, 
together with their reversed $\vec x \mapsto -\vec x$. These four stationary states 
are highlighted with two filled stars and open circles 
in Fig.~\ref{fig:sketchN2} which shows the two dimensional configurational space. 

\item
{\it Degenerate $\bar\gamma_\mu$, $\bar\gamma_1 =\bar \gamma_2$.}
Can one have stationary states with both $x_1\neq 0$ and $x_2\neq 0$? The third and fourth equations  imply 
$p_1x_2 -p_2x_1=0$. The orthogonality between position and momentum reads $x_1 p_1 + x_2 p_2 =0$.
Putting these two together, one has $p_2 (x^2_1+x_2^2) =0$, and this implies $p_2=0$ which in turn implies $p_1=0$.
Going now to the first and second equations, 
assuming $x_1 \neq 0$ and $x_2\neq 0$, 
one finds that $2\bar\gamma_1+z_{\rm GNM} = 2\bar\gamma_2+z_{\rm GNM}=0$, 
which is impossible, unless $\bar\gamma_1=\bar\gamma_2$. Focusing now on this special 
case, for $N=2$ the momentum, $\vec p$, disappears from the 
Hamiltonian and one is left with only an isotropic quadratic potential energy plus the constraint. There is no 
dynamics, unless a dissipative term is added to the evolution. Finally, for $\bar\gamma_1=\bar\gamma_2$, 
all positions on the circle solve the extremisation equations, with $z_{\rm GGE} = -2 \bar\gamma_1 =-2\bar\gamma_2$. 
The energy is constant, ${\mathcal H}_{\rm GNM}= 2\bar\gamma_1$.
We note that this case {\it is not} realised in the SSK/Neumann model, since  
$\bar\gamma_\mu = -\lambda_\mu/2$ and the $\lambda_\mu$s are taken to be all different, 
either because they are the eigenvalues of a matrix belonging to the GOE ensemble, or because 
they are taken to be different to ensure integrability.

\end{itemize}

Concerning the stability of these stationary states, derived from the 
signs of the eigenvalues of the Hessian of ${\mathcal H}_{\rm GNM}$, the 
 four by four symmetric matrix of its second derivatives.
 We find the following.
\begin{itemize}
\item
{\it Non-degenerate $\bar\gamma_\mu$, $\bar\gamma_1\neq \bar\gamma_2$.}
A simple calculation shows that, once evaluated at any of the four stationary states, 
the Hessian is diagonal. For $\vec x = (0,\pm \sqrt{2})$, $\vec p=\vec 0$, $z_{\rm GNM}=-2\bar\gamma_2$,  
the elements are
\begin{eqnarray}
\frac{\partial^2 {\mathcal H}_{\rm GNM}}{\partial x_1^2} = 2 \bar\gamma_1 - 2 \bar\gamma_2
\; , 
\qquad\qquad\qquad
\frac{\partial^2 {\mathcal H}_{\rm GNM}}{\partial x_2^2} =  0
\; , 
\\
\frac{\partial^2 {\mathcal H}_{\rm GNM}}{\partial p_1^2} =  -  \frac{2}{m} \frac{\bar\gamma_1- \bar\gamma_2}{\lambda_1-\lambda_2}
\; , 
\qquad\qquad\quad
\frac{\partial^2 {\mathcal H}_{\rm GNM}}{\partial p_2^2} =  0
\; . 
\end{eqnarray}
These stationary states (the filled stars in Fig.~\ref{fig:sketchN2})
are therefore (marginally) stable if and only if $(\bar\gamma_1- \bar\gamma_2)/(\lambda_1-\lambda_2)<0$ 
and $\bar\gamma_1- \bar\gamma_2>0$
and unstable otherwise. Concerning  $\vec x = (\pm \sqrt{2},0)$, $\vec p=\vec 0$, $z_{\rm GNM}=-2\bar\gamma_1$,
they are (marginally) stable in the opposite case and unstable otherwise.
 
\item
{\it Degenerate $\bar\gamma_\mu$, $\bar\gamma_1 = \bar\gamma_2$.}
In this case, the moments disappear from ${\mathcal H}_{\rm GNM}$. The Hessian on any 
$\vec x$ is still diagonal since there are no terms coupling $x_1$ and $x_2$ in ${\mathcal H}_{\rm GNM}$. Then,
\begin{eqnarray}
\frac{\partial^2 {\mathcal H}_{\rm GNM}}{\partial x_1^2} =
\frac{\partial^2 {\mathcal H}_{\rm GNM}}{\partial x_2^2} = 0
\; , 
\end{eqnarray}
and the potential is completely flat around any $\vec x$ on the circle, just because the energy is constant.

\end{itemize}

\subsection{Stochastic dynamics}
\label{app:C2}

The  Langevin equations for the model  
${\mathcal H}_{\rm GNM}=\sum_\mu \bar\gamma_\mu I_\mu$ with $N=2$ read
\begin{eqnarray} 
&& \dot x_1  = \frac{1}{m}(x_1p_2-x_2p_1) \, x_2 \! \left(\frac{\bar\gamma_1-\bar\gamma_2}{\lambda_1-\lambda_2}\right) 
\; , 
\ \ \ 
\label{eq:dotx1} \\
&& \dot x_2 = \frac{1}{m}(x_2p_1-x_1p_2)\, x_1 \! \left(\frac{\bar\gamma_2-\bar\gamma_1}{\lambda_2-\lambda_1}\right) 
\; , 
\ \ \ 
\label{eq:dotx2} \\
&& \dot p_1 =  -2\bar\gamma_1 x_1 - (x_2p_1-x_1p_2) \, \frac{p_2}{m} \! \left(\frac{\bar\gamma_1-\bar\gamma_2}{\lambda_1-\lambda_2}\right)
\! -\! z_{\rm GNM} x_1 \! - \! \gamma \dot x_1 + \xi_1 
\; , 
\ \ \
 \label{eq:dotp1}\\
&& \dot p_2 = -2\bar\gamma_2 x_2 - (x_1p_2-x_2p_1) \frac{p_1}{m} \! \left(\frac{\bar\gamma_2-\bar\gamma_1}{\lambda_2-\lambda_1}\right)
\! - \! z_{\rm GNM} x_2 \! - \! \gamma \dot x_2 + \xi_2 
\; , 
\ \ \
\label{eq:dotp2}
\end{eqnarray}
where $z_{\rm GNM}$ depends on $\vec x$ and $\vec p$. We keep the convention $\lambda_1 < \lambda_2$.
We note that  $\dot {\vec x}$ is perpendicular to $\vec x$ for any $\{\bar\gamma_\mu\}$ and $\{\lambda_\mu\}$. 
Another interesting relation is 
\begin{equation}
\vec p \wedge \dot{\vec x} = - 
\frac{\bar\gamma_2 - \bar\gamma_1}{\lambda_2-\lambda_1} \frac{1}{m} (\vec x\cdot \vec p) 
(\vec x \wedge \vec p)
\; . 
\end{equation}
A possible solution to this equation is $\vec p \wedge \dot{\vec x} = (\vec x\cdot \vec p) =0 $. 
In conclusion, for $N=2$, $\dot{\vec x}$ is parallel to $\vec p$ and both are perpendicular to 
$\vec x$. 

In the main text we showed that 
the Lagrange multiplier is given by Eq.~(\ref{eq:z-in-the-end}),
\begin{eqnarray}
z_{\rm GNM}  &=& \frac{1}{N} \sum_\nu \frac{\partial {\mathcal H}_{\rm GNM}}{\partial p_\nu} p_\nu - \frac{1}{N} \sum_\nu \frac{\partial {\mathcal H}_{\rm GNM}}{\partial x_\nu} x_\nu 
=
- \frac{2}{N} \sum_\nu \bar\gamma_\nu x^2_\nu
\; .
\end{eqnarray}

Configuration space is the circle $x_1^2+x_2^2=2$. Changing to polar coordinates, 
\begin{eqnarray}
&& x_1 = r\cos{\theta}=\sqrt{2}\cos{\theta}  
\; , 
\\
&& x_2 = r\sin{\theta}=\sqrt{2}\sin{\theta} 
\; , 
\end{eqnarray}
and
\begin{eqnarray}
&& \dot x_1 = \dot r\cos{\theta}-r\sin{\theta}\dot \theta = -\sqrt{2}\sin{\theta}\, \dot \theta
\; , 
\\
&& \dot x_2 = \dot r\sin{\theta}+r\cos{\theta}\dot \theta = \sqrt{2}\cos{\theta} \, \dot \theta
 \; .
\end{eqnarray}
(We use standard differentiation rules as with Stratonovich calculus.)
We can easily verify that $x_1\dot x_1+x_2 \dot x_2=0$, which is a consequence of the spherical constraint.
Substituting in Eqs.~(\ref{eq:dotx1}) and (\ref{eq:dotx2}) we obtain
\begin{eqnarray}
&& \dot \theta =  \frac{1}{m}(x_2p_1-x_1p_2)\left(\frac{\bar\gamma_2-\bar\gamma_1}{\lambda_2-\lambda_1}\right)
\nonumber\\
&& 
\;\;  =
\frac{\sqrt{2}}{m}(p_1\sin{\theta}-p_2\cos{\theta})\left(\frac{\bar\gamma_2-\bar\gamma_1}{\lambda_2-\lambda_1}\right)
\; . \qquad
\label{eq:dottheta} 
\end{eqnarray}
Then,
\begin{equation}
\ddot \theta =
\frac{\sqrt{2}}{m}\left(\frac{\bar\gamma_2-\bar\gamma_1}{\lambda_2-\lambda_1}\right)
(p_2\sin{\theta}\ \dot\theta+p_1\cos{\theta}\ \dot\theta + \dot p_1\sin{\theta} - \dot p_2 \cos{\theta})
\; . 
\label{eq:ddottheta}
\end{equation}
Note that from Eq.~(\ref{eq:dottheta}):
\begin{equation}
x_2p_1-x_1p_2 = m\left(\frac{\lambda_2-\lambda_1}{\bar\gamma_2-\bar\gamma_1}\right)\dot \theta
\; . 
\end{equation}
Replacing now in Eqs.~(\ref{eq:dotp1}) and (\ref{eq:dotp2}) and multiplying by $\sin{\theta}$ and
$\cos{\theta}$, respectively, 
\begin{eqnarray}
\sin{\theta}\,\dot p_1 &=&  -2\sqrt{2}\bar\gamma_1 \sin{\theta}\cos{\theta} - \sin{\theta}\,p_2\, \dot\theta
+\sqrt{2}\gamma \sin^2{\theta} \dot\theta 
\nonumber \\
& & - \sqrt{2}z_{\rm GNM} \sin{\theta}\cos{\theta}  + \, \xi_1\sin{\theta} \; ,  \\
\cos{\theta}\, \dot p_2 &=&  -2\sqrt{2}\bar\gamma_2 \sin{\theta}\cos{\theta} + \cos{\theta}\,p_1\, \dot\theta
-\sqrt{2}\gamma \cos^2{\theta} \dot\theta \nonumber \\
& & - \sqrt{2}z_{\rm GNM} \sin{\theta}\cos{\theta}  + \, \xi_2\cos{\theta} \; . 
\end{eqnarray}

\subsection{Angular motion}
\label{app:C21}

Plugging these results in Eq.~(\ref{eq:ddottheta}) we derive
\begin{eqnarray}
&&
\frac{m}{2}\left(\frac{\lambda_2-\lambda_1}{\bar\gamma_2-\bar\gamma_1}\right)\,\ddot\theta =
\nonumber\\
&&
\qquad\qquad
\gamma\,\dot\theta - 2\sin{\theta}\cos{\theta}(\bar\gamma_1-\bar\gamma_2) + \frac{1}{\sqrt{2}}\xi_1\sin{\theta}
- \frac{1}{\sqrt{2}}\xi_2\cos{\theta}
\; . 
\end{eqnarray}
(Note that the terms in $z_{\rm GNM}$ cancel out.)
Using the identity $2\sin{\theta}\cos{\theta}=\sin{2\theta}$ and defining 
\begin{equation}
\vartheta=2\theta
\end{equation}
we end up with
  \begin{eqnarray}
&& 
\ddot\vartheta + \underbrace{(-1) \frac{2}{m}\left(\frac{\bar\gamma_1-\bar\gamma_2}{\lambda_1-\lambda_2}\right) \gamma}_{{\rm damping \; constant} \; \eta} \;\; \dot\vartheta
\ =  \underbrace{- \frac{4}{m}\frac{(\bar\gamma_1-\bar\gamma_2)^2}{\lambda_1-\lambda_2}}_{{\rm natural \; frequency \;} \omega^2} \sin{\vartheta}
\nonumber\\
&&
\qquad\qquad\qquad\qquad
+ \underbrace{\frac{4}{m\sqrt{2}}\left(\frac{\bar\gamma_1-\bar\gamma_2}{\lambda_1-\lambda_2}\right)\left(\xi_1\sin{\frac{\vartheta}{2}}
- \xi_2\cos{\frac{\vartheta}{2}}\right)}_{\rm multiplicative \; stochastic \; forcing}
\; . 
\label{eq:N2-Langevin}
\end{eqnarray}
This is a non-linear second order stochastic differential equation for the forced motion of a damped angle
$\vartheta$. The meaning of the terms and parameters are written below the underbraces in the equation itself.
The force and potential are
\begin{equation}
F(\vartheta) = - \frac{4}{m} \frac{(\bar\gamma_1-\bar\gamma_2)^2}{\lambda_1-\lambda_2} \sin \vartheta
\qquad
 V(\vartheta) = \underbrace{\frac{4}{m} \frac{(\bar\gamma_1-\bar\gamma_2)^2}{\lambda_2-\lambda_1}}_{\omega^2>0} \cos \vartheta
\end{equation}
since $\lambda_2>\lambda_1$.

The potential minima are at $\vartheta=\pi$ and $\vartheta=3\pi$ that correspond to 
$\theta=\pi/2$ and $\theta=3\pi/2$, angles highlighted with filled stars in Fig.~\ref{fig:sketchN2}.
The maxima are at $\vartheta=0$ and $\vartheta=2\pi$ that correspond to 
$\theta=0$ and $\theta=\pi$.
The stability of these stationary points is decided by 
\begin{equation}
V''(\vartheta) = - \omega^2 \, \cos\vartheta
\end{equation}
which yields
\begin{eqnarray}
&& V''(\vartheta = \pi) = V''(\vartheta = 3\pi)  =  \omega^2 > 0
\qquad
{\rm stable \ minima}
\\
&& V''(\vartheta = 0) = V''(\vartheta = 2\pi)  =  - \omega^2 < 0
\quad\;
{\rm unstable \ maxima}
\end{eqnarray}

Consider $T=0$ and 
expand around the minimum of the potential $\vartheta = \vartheta_{\rm min} +\epsilon = \pi +\epsilon$. 
The force term is proportional to $\sin \vartheta = \sin(\vartheta_{\rm min} +\epsilon) = \sin \vartheta_{\rm min} \; \cos \epsilon + \cos \vartheta_{\rm min} \;  \sin \epsilon
=  -  \sin \epsilon \approx - \epsilon$. Then, 
 \begin{eqnarray}
&& 
\ddot\epsilon + \eta 
\dot\epsilon +  \omega^2 \epsilon \approx 0
\end{eqnarray}
with $\eta$ and $\omega^2$ defined above. The solutions are
\begin{equation}
\epsilon(t) =  \frac{[\epsilon(0) y_2 - \dot \epsilon(0)]}{y_2-y_1} \; e^{y_1t} +  \frac{[\dot \epsilon(0) - \epsilon(0) y_1]}{y_2-y_1} \;  e^{y_2t} 
\end{equation}
with 
\begin{equation}
y_{1,2} = \frac{1}{2} \left[ - \eta \pm \sqrt{\eta^2 - 4 \omega^2} \right]
\; . 
\end{equation}
Hence, 
\begin{eqnarray}
\epsilon(t) 
&=& 
 \frac{1}{y_2-y_1} \; e^{-\eta t/2}
\; 
\left\{ 
[\epsilon(0) y_2 - \dot \epsilon(0)] \; e^{\sqrt{\eta^2 - 4 \omega^2} t/2} 
\right.
\nonumber\\
&& \qquad\qquad\qquad\;\;
\left. 
+ [\dot \epsilon(0) - \epsilon(0) y_1] \;  e^{- \sqrt{\eta^2 - 4 \omega^2} t/2} 
\right\}
\; . 
\end{eqnarray}
So, going back to the damping constant $\eta$, the condition
\begin{equation}
\eta = - \frac{2}{m} \left(  \frac{\bar\gamma_1-\bar\gamma_2}{\lambda_1-\lambda_2} \right) \gamma = 
 \frac{2}{m} \left(  \frac{\bar\gamma_1-\bar\gamma_2}{\lambda_2-\lambda_1} \right) \gamma
 > 0 
 \; , 
\end{equation}
 implies
\begin{equation}
\eta> 0  \qquad \Rightarrow   \qquad
\bar\gamma_1-\bar\gamma_2> 0
\end{equation}
and the dynamics will be damped towards the minima only if the $\bar\gamma$ are so ordered.

Some particular cases are the following.
\begin{itemize}
\item
For $\bar\gamma_{\mu}=-\lambda_{\mu}/2$, 
\begin{equation}
\ddot\vartheta + \frac{\gamma}{m}\, \dot\vartheta
+ \frac{\lambda_1-\lambda_2}{m}\, \sin{\vartheta}
= \frac{\sqrt{2}}{m}\left(\xi_2\cos{\frac{\vartheta}{2}} - \xi_1\sin{\frac{\vartheta}{2}}\right) 
\; . 
\end{equation}

\item
Newtonian dynamics  is achieved by switching off the damping and noise,
$\gamma=\xi_1=\xi_2=0$ in Eq.~(\ref{eq:N2-Langevin}). The 
angular evolution is then ruled by 
\begin{equation}
\ddot\vartheta 
+ \frac{4}{m}\frac{(\bar\gamma_1-\bar\gamma_2)^2}{\lambda_1-\lambda_2} \sin{\vartheta}
= 0
\; .
\label{eq:N2-Newton}
\end{equation}
If we take $\bar\gamma_{\mu}=-\lambda_{\mu}/2$,  
\begin{equation}
\ddot \vartheta + \frac{(\lambda_1-\lambda_2)}{m}\sin{\vartheta} = 0
\; . 
\end{equation}
Thus, the  Newtonian motion of the particle on the circle, subject to the harmonic anisotropic potential,
is {equivalent to pendulum motion, with small-oscillation frequency $\omega^2=(\lambda_2-\lambda_1)/m$, 
about $\vartheta=\pi$ modulo $2\pi$}.
Note that, in the case of isotropic harmonic oscillators, with equal elastic constants $\lambda_1=\lambda_2=\lambda$,
the angular velocity of the particle is constant, i.e. the motion reduces to a free
particle on the circle, or uniform circular motion. This is consistent with a constant value of the potential
energy on the circle (centripetal force):
\begin{equation}
V(\vec x) = -\frac{\lambda_1}{2}x_1^2 - \frac{\lambda_2}{2}x_2^2 = -\frac{\lambda}{2} (x_1^2+x_2^2)=-\lambda
\; .
\end{equation}

\item
For $\bar\gamma_\mu = {(\pm |a|/J)} \lambda_\mu^2$ and zero temperature,  
 \begin{equation}
\ddot\vartheta + \underbrace{ (\mp  |a|) \frac{2}{mJ} \, (\lambda_1+\lambda_2) \, \gamma}_{\eta}  \, \dot\vartheta
- \underbrace{\frac{4 {a^2}}{m{J^2}}  \; |\lambda_1-\lambda_2| (\lambda_1+\lambda_2)^2}_{\omega^2>0} \, \sin{\vartheta} =0
\; . 
\end{equation}
and  $\eta$ can be negative. 

\end{itemize}

\subsection{The momentum}
\label{app:C22}

Complementarily, one can study the evolution of the modulus of the
momentum, $p=(p_1^2+p_2^2)^{1/2}$, close to the North and
South poles of the circle. At zero bath temperature, its linearized evolution
is coupled to the angular displacement, as shown in the next section.
For distinct eigenvalues, the general linearized solution has the form
\begin{equation}
p(t)=\left|R_+e^{\lambda_+t}+R_-e^{\lambda_-t}\right|
\; , \nonumber
\end{equation}
where $\lambda_\pm$ are the eigenvalues derived below and the constants
$R_\pm$ are fixed by the initial conditions so that the signed momentum is real.
For a repeated eigenvalue $\lambda$, the corresponding form is
$p(t)=|(R_1+R_2t)e^{\lambda t}|$.
Thus, stability and the possible oscillations are determined by the coupled
system, rather than by the friction term alone.

\subsection{Full linear stability analysis and transitions}
\label{app:C23}

We now present the full stability analysis around the poles $\theta = \pm \pi/2$ and 
$p=0$. Here $p$ denotes the signed local momentum coordinate; its absolute value 
gives the momentum modulus to linear order. Introducing $\Theta = \theta \mp \pi/2$, 
the linearized evolution equations read
\begin{eqnarray}
\dot{\Theta} = \pm\frac{\sqrt{2}(\bar\gamma_1 - \bar\gamma_2)}{m(\lambda_1 - \lambda_2)} p
\; , \nonumber\\
\dot p = \mp 2 \, \sqrt{2} \; (\bar\gamma_2 - \bar\gamma_1)\Theta + \frac{2\gamma(\bar\gamma_1 - \bar\gamma_2)}{m(\lambda_1 - \lambda_2)} p
 \label{sist_2D_lin}
\end{eqnarray}
which in matricial form become
\begin{eqnarray}
\frac{d}{dt} 
\left(
\begin{array}{c}
\Theta 
\\ 
p 
\end{array} 
\right)
= 
A 
\left(
\begin{array}{c}
\Theta 
\\ 
p 
\end{array}
\right)
\quad 
\mbox{where} 
\quad
A = 
\left(
\begin{array}{cc}
0 & a 
\\ b & c 
\end{array}
\right)
\end{eqnarray}
with 
\[
a = \pm \frac{\sqrt{2}(\bar\gamma_1 - \bar\gamma_2)}{m(\lambda_1 - \lambda_2)}
\; , \quad 
b = \mp 2 \, \sqrt{2} \; (\bar\gamma_2 - \bar\gamma_1)
\; , \quad 
c = \frac{2\gamma(\bar\gamma_1 - \bar\gamma_2)}{m(\lambda_1 - \lambda_2)}
\; .
\]
The characteristic equation is
\[
\lambda^2 - c\lambda - ab = 0,
\]
with solutions
\[
\lambda_{\pm} 
= \frac{\gamma \, (\bar\gamma_1 - \bar\gamma_2)}{m \, (\lambda_1 - \lambda_2)} \pm 
\frac{\gamma\, |\bar\gamma_1 - \bar\gamma_2|}{m \, |\lambda_1 - \lambda_2|} \sqrt{1+ \frac{4m}{\gamma^2} \, 
(\lambda_1 - \lambda_2)}
\; .
\]

Depending on the sign of $\bar\gamma_1 - \bar\gamma_2$ and the value of the friction
parameter $\gamma^2$ relative to $4m\, (\lambda_1 - \lambda_2)$, different regimes arise:
\begin{itemize}
\item 
$\bar\gamma_1 = \bar\gamma_2$. Then $\dot{\theta} = \dot p = 0$ and there is no dynamical evolution: the system remains trapped in its initial condition, which can be anywhere on the circle. 
This point marks a {\it phase transition} in the parameter space $(\bar\gamma_1,\bar\gamma_2)$.

\item 
Take $\bar\gamma_1 \neq \bar\gamma_2$. Then, two cases have to be distinguished.
\begin{itemize}
\item If $\gamma^2 + 4m (\lambda_1 - \lambda_2) > 0$, 
the eigenvalues are real. The modulus of the term after the $\pm$ depends on the 
sign of $\lambda_1-\lambda_2$. With our convention, $\lambda_2>\lambda_1$ and $\lambda_1-\lambda_2$ is 
negative. The modulus of the term after the $\pm$ is then smaller than the one 
of the first term; therefore, both eigenvalues have the same sign and it is 
determined by the sign of  the ratio $(\bar\gamma_1 - \bar\gamma_2)/(\lambda_1 - \lambda_2)$.
In cases in which this is negative, 
the trajectories decay toward the North/South pole 
({\it condensed states} in the large $N$ case) without oscillations ({\it attractive fixed points}).
In contrast, when the ratio is positive, there is an exponential divergence from the poles 
({\it repulsive fixed points}).

\item 
If $\gamma^2 + 4m (\lambda_1 - \lambda_2) < 0$
the eigenvalues are complex with, for the same reason as above, 
 negative real parts if $(\bar\gamma_1 - \bar\gamma_2)/(\lambda_1 - \lambda_2)<0$: 
the trajectories converge to the condensed state 
 with damped oscillations. Instead, for $(\bar\gamma_1 - \bar\gamma_2)/(\lambda_1 - \lambda_2)>0$
the trajectories exhibit oscillatory growth getting away from the pole.
\end{itemize}

\end{itemize}

\subsection{Constants of motion}
\label{app:C3}

The constants of motion under Newtonian dynamics,  $I_{\mu}$, also take simple forms:
\begin{eqnarray}
&& I_1  = x_1^2 + \frac{1}{2m}\frac{(x_1p_2-x_2p_1)^2}{\lambda_2-\lambda_1} \; , 
 \label{eq:i1xp}\\
&& I_2 = x_2^2 + \frac{1}{2m}\frac{(x_2p_1-x_1p_2)^2}{\lambda_1-\lambda_2}
\; . 
\end{eqnarray}
Note that $I_1+I_2=x_1^2+x_2^2=2$.
Equation~(\ref{eq:dottheta}) allows us to rewrite the last terms as functions of $\dot\theta$. Moreover, 
replacing $x_1$ and $x_2$ by their expressions in terms of $\theta$, 
\begin{eqnarray}
&& I_1 = 2\cos^2{\theta} + \frac{m}{2}\frac{\lambda_2-\lambda_1}{(\bar\gamma_2-\bar\gamma_1)^2}\dot \theta^2 
= 2\cos^2{\frac{\vartheta}{2}} + \frac{m}{8}\frac{\lambda_2-\lambda_1}{(\bar\gamma_2-\bar\gamma_1)^2}\dot \vartheta^2 
\; , 
\\
&& I_2 = 2\sin^2{\theta} - \frac{m}{2}\frac{\lambda_2-\lambda_1}{(\bar\gamma_2-\bar\gamma_1)^2}\dot \theta^2
=
2\sin^2{\frac{\vartheta}{2}} - \frac{m}{8}\frac{\lambda_2-\lambda_1}{(\bar\gamma_2-\bar\gamma_1)^2}\dot \vartheta^2 
\; . 
\end{eqnarray}

\section{The numerical code}
\label{app:numerics}
\setcounter{equation}{0}
\renewcommand{\theequation}{D.\arabic{equation}}
\setcounter{figure}{0}
\renewcommand{\thefigure}{D.\arabic{figure}}

\subsection{Random couplings}

The anisotropies $\lambda_\mu$ are generated from a Wigner semicircle distribution. Concretely, we draw a random Gaussian Orthogonal Ensemble (GOE) matrix, compute its eigenvalues, and sort them. These $\lambda_\mu$ 
thus ordered are used in the expressions of the $I_\mu$. Moreover, we use them  to build the $\bar\gamma_\mu$ in the quadratic model
or to evaluate the $\bar\gamma_\mu$ from the {solution to the} $N\to\infty$ equation~\cite{BaCuLoNe22} 
for quenches of the isolated model, 
{presented in App.~\ref{app:new-derivation}}.

\subsection{Stochastic dynamics and SRK2 integration}

The stochastic evolution follows a Langevin equation with Gaussian white noise and  friction coefficient $\gamma$. 
The noise term is drawn at each step as
\begin{equation}
\xi_\mu = \sigma\, \eta_\mu \; , \qquad 
\sigma = \sqrt{\frac{2 \gamma k_B T}{\Delta t}} \; ,
\end{equation}
with $\eta_\mu$ standard normal random variables.  
We integrate the dynamics using a second-order Stochastic Runge--Kutta (SRK2) scheme.

The core of the integrator is the function \texttt{computeDerivatives}, which evaluates the deterministic and stochastic contributions to the time derivatives of $x_\mu$ and $p_\mu$. The function first computes $\sum_\mu x_\mu^2$ for the Lagrange multiplier and then the Hamiltonian derivatives. The coupling between modes involves the precomputed matrix
\begin{equation}
\Lambda^{\mathrm{diff}}_{\mu\nu} = \lambda_\mu - \lambda_\nu
\; , \, \qquad \mu\neq\nu \; , 
\end{equation}
and the terms
\begin{equation}
\propto (x_\mu p_\nu - p_\mu x_\nu)\, \frac{\gamma_\mu - \gamma_\nu}{\Lambda^{\mathrm{diff}}_{\mu\nu}}
\; .
\end{equation}
The spherical constraint is enforced by computing the Lagrange multiplier
\begin{equation}
z = -2 \sum_\mu \gamma_\mu x_\mu^2 \Big/ \sum_\mu x_\mu^2
\; .
\end{equation}

The SRK2 algorithm proceeds as:
\begin{enumerate}
    \item Generate noise $\xi_\mu$.
    \item Compute first-stage increments $k^{(1)}_x$, $k^{(1)}_p$.
    \item Construct intermediate fields 
    \begin{equation}
    x_\mu^{1/2} = x_\mu + \Delta t\, k^{(1)}_{x,\mu}
    \; , 
    \qquad
    p_\mu^{1/2} = p_\mu + \Delta t\, k^{(1)}_{p,\mu}
    \; .
    \end{equation}
    \item Evaluate second-stage increments $k^{(2)}_x$, $k^{(2)}_p$ using $(x^{1/2},p^{1/2})$.
    \item Update:
    \begin{equation}
    x_\mu \leftarrow x_\mu + \frac{\Delta t}{2} \left( k^{(1)}_{x,\mu} + k^{(2)}_{x,\mu} \right)
    \; ,
    \qquad
    p_\mu \leftarrow p_\mu + \frac{\Delta t}{2} \left( k^{(1)}_{p,\mu} + k^{(2)}_{p,\mu} \right)
    \; .
    \end{equation}
\end{enumerate}

This algorithm ensures the 
correct handling of the spherical constraint and provides second-order accuracy in the weak sense for the stochastic dynamics.

\subsection{Numerical errors}

\begin{figure}[h!]
(a) \hspace{5cm} (b) \hspace{5cm} (c) 
\\
\centerline{
\includegraphics[scale=0.28]{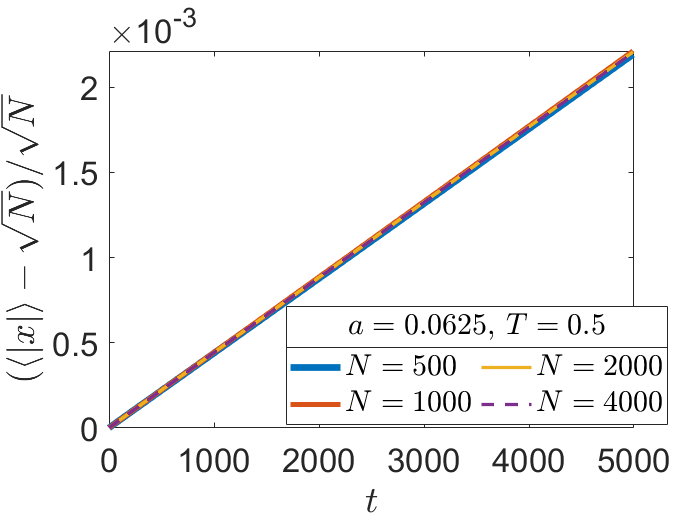}
\includegraphics[scale=0.28]{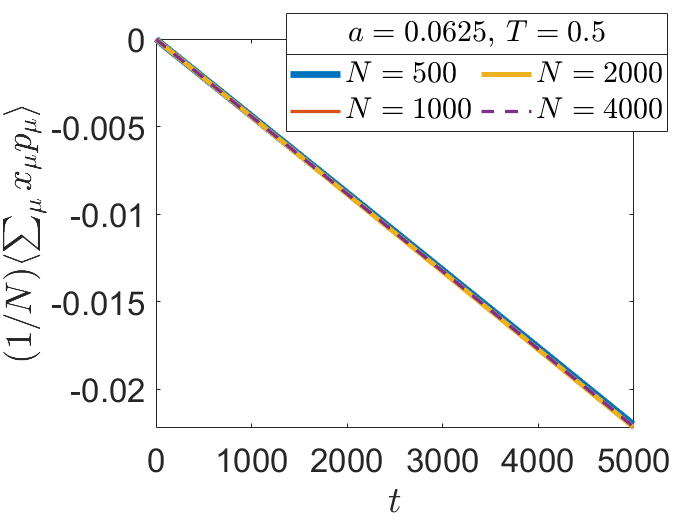}
\includegraphics[scale=0.28]{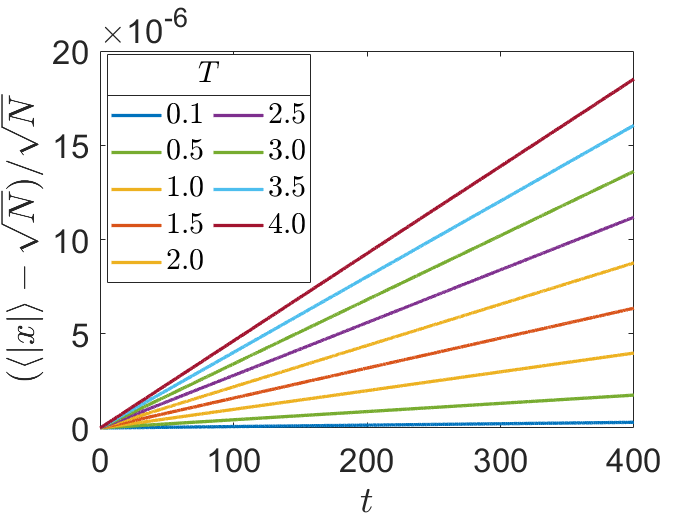}
}
 (d) \hspace{5cm} (e) \hspace{5cm} (f)
\\
\centerline{
\includegraphics[scale=0.28]{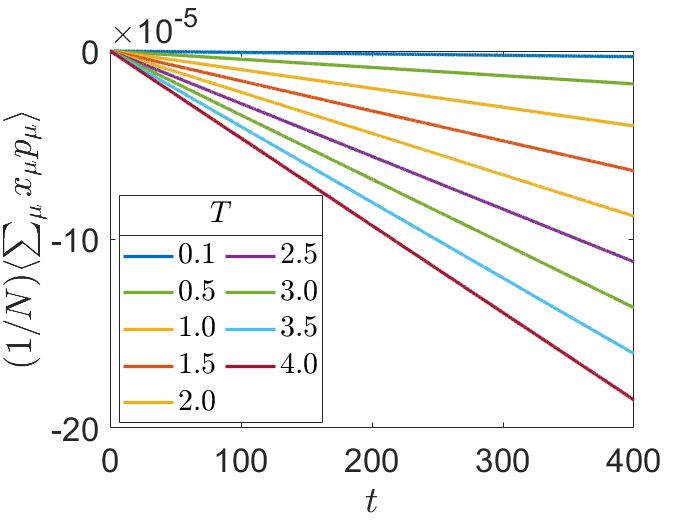}
\includegraphics[scale=0.28]{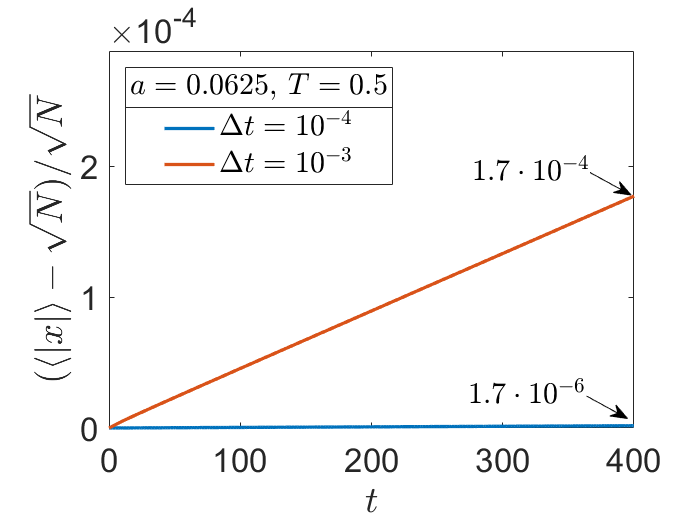}
\includegraphics[scale=0.28]{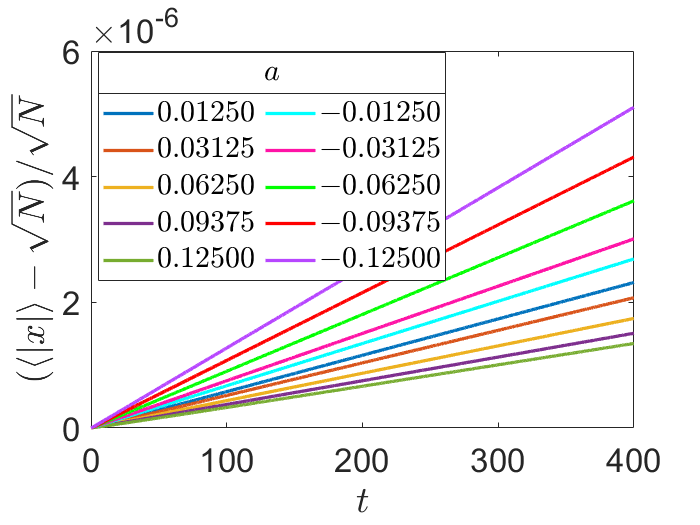}
}
\caption{\small 
Numerical errors in the quadratic model with $\bar\gamma_\mu = a \lambda_\mu^2 - 1/2 \, \lambda_\mu$.  
Time $t$ is real time, that is, iterations times $\Delta t$.
(a)-(b) $N$ independence of the error in the spherical constraint and orthogonality condition between 
momentum and position. $\Delta t=10^{-3}$. (c)-(d) $T$ dependence of the same errors. $\Delta t=10^{-4}$ and $N=500$. 
(e) $\Delta t$ dependence of the 
error of the spherical constraint. $T=0.5$ and $N=500$.
(f) $a$ dependence of the 
error of the spherical constraint. $T=0.5$ and $N=500$.
}
\label{fig:app-E1}
\end{figure}

The numerical integration has some errors which manifest in weak violations of the spherical and orthogonality conditions
$
N^{-1} (\sum_\mu x_\mu^2 - N) =0
$ 
and
$
N^{-1} \sum_\mu x_\mu p_\mu = 0
$, where we used the normalization factors which are natural  to make the left-hand-sides 
${\mathcal O}(1)$ (note that $x_\mu = {\mathcal O}(N^{1/2})$).
In practice, we obtain
\[
N^{-1/2} (\sum_\mu |x_\mu| - \sqrt{N}) = \epsilon_1
\; , 
\qquad 
N^{-1} \sum_\mu x_\mu p_\mu = \epsilon_2
\; ,
\]
where $\epsilon_1$ and $\epsilon_2$ are errors that grow approximately linearly in time:
\[
\epsilon_1 = c_1\, s_1(\Delta t, T)\, t
\; , 
\qquad 
\epsilon_2 = c_2\, s_2(\Delta t, T)\, t
\; ,
\]
with $c_1$ and $c_2$ constants. This claim is exemplified in Fig.~\ref{fig:app-E1}. In panel (a) we test the 
$N$ independence of $\epsilon_1$ and $\epsilon_2$. 
The slopes $s_1$ and $s_2$ depend both on the time step $\Delta t$ and on the temperature $T$. The smaller $\Delta t$ is, the smaller the slopes are (e); similarly, the errors are weaker at  lower temperatures (c)-(d). Therefore, reducing $\Delta t$ is the most effective way to keep the constraints satisfied over long times.  Good fits of the errors are obtained with
$s_1 = f_1(a) \, (\Delta t)^2 \, T$  
and
$s_2 = f_2(a) \, (\Delta t)^2 \, T$.

For large system sizes $N$, reaching equilibrium requires very long integration times due to slow dynamics. 
To obtain results within a feasible computational time, we cannot always reduce $\Delta t$ as much as we would like, 
and thus we must accept a trade-off between precision and efficiency.

\section{The momentum and the velocity}
\label{app:momentum-vs-velocity}
\setcounter{equation}{0}
\renewcommand{\theequation}{E.\arabic{equation}}
\setcounter{figure}{0}
\renewcommand{\thefigure}{E.\arabic{figure}}

\begin{center}
\begin{figure}[h!]
 \hspace{2.25cm} (a) \hspace{5.75cm} (b) 
\\
\centerline{
\includegraphics[scale=0.3]{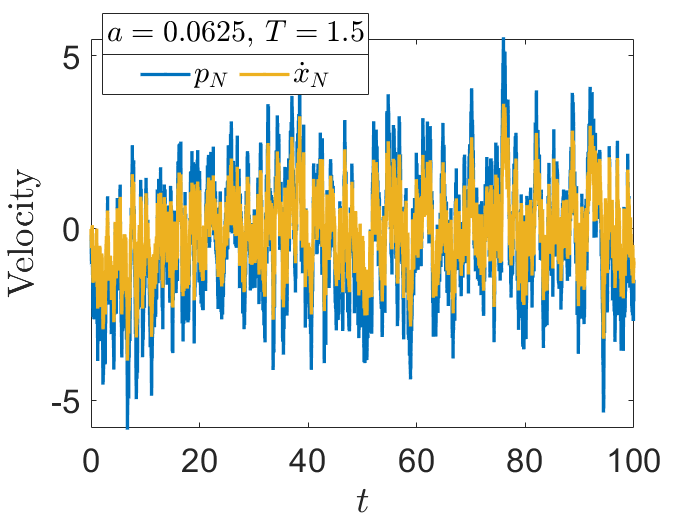}
\includegraphics[scale=0.3]{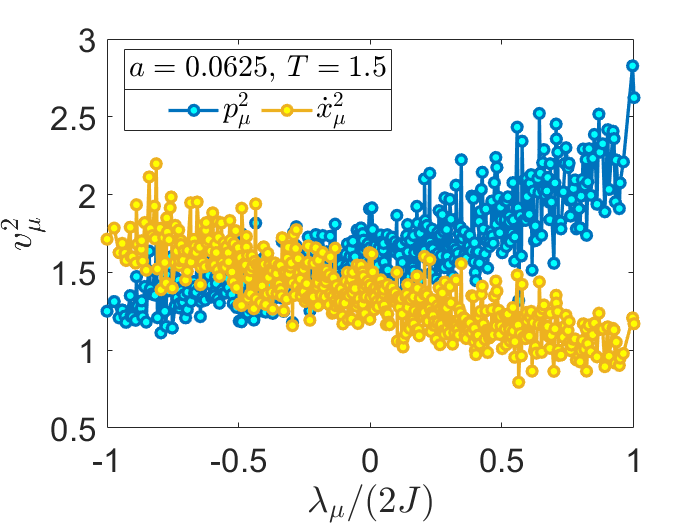}
}
\\
\; \hspace*{2.25cm} (c) \hspace{5.75cm} (d) 
 \\
 \centerline{
\includegraphics[scale=0.3]{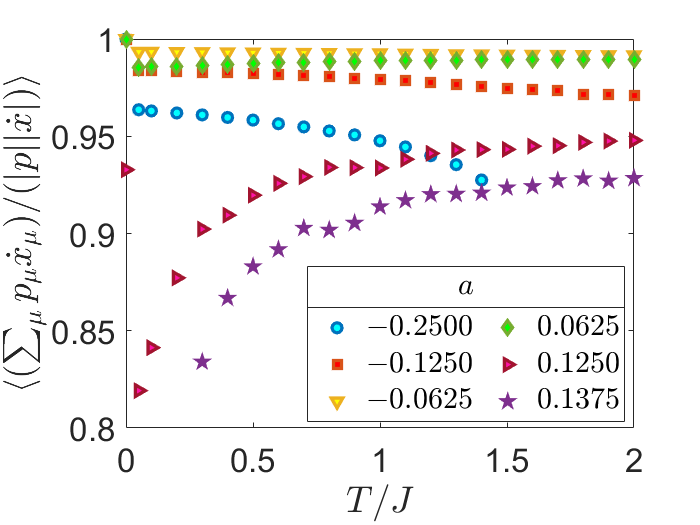}
\includegraphics[scale=0.3]{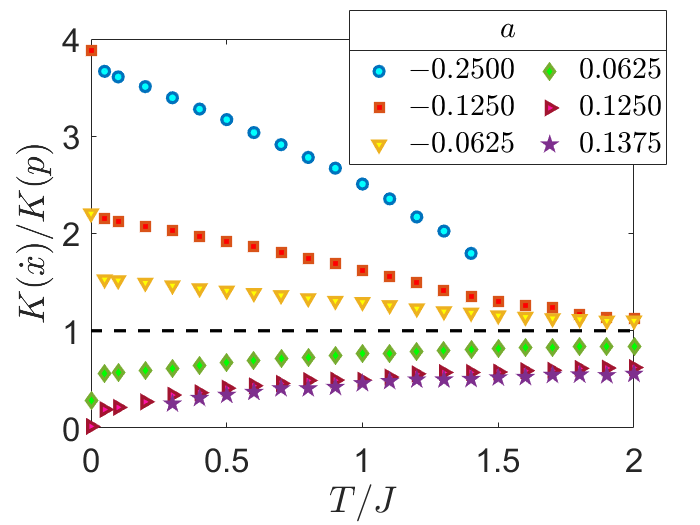}
}
\caption{\small Comparison between momentum and velocity.
(a) Time dependence of the largest mode of the momentum and velocity  in the extended phase. (b) Spectra of averaged square instantaneous velocities and momenta also in the extended phase. 
(c) Normalised scalar product between the momentum and 
velocity vectors for different values of $a$. (d) Ratio between kinetic energies defined using the 
momentum or the instantaneous velocity.}
\label{fig:velocity-vs-momentum}
\end{figure}
\end{center}

In the generalised model, the velocity and momentum are not equal. They tend to fluctuate in the same way, 
Fig.~\ref{fig:velocity-vs-momentum}(a), but their spectra are different (b).
The smaller the $|a|$, the closer is the model to the SSK/Neumann model, 
and the velocity and momentum tend to equalise
at all temperatures, green and yellow datapoints in (c). For the generalised model, they are not. all other 
data points in (c).

\section{Calculations for quadratic $\bar\gamma_\mu(\lambda_\mu)$}
\label{app:details-of-calculations}
\setcounter{equation}{0}
\renewcommand{\theequation}{F.\arabic{equation}}
\setcounter{figure}{0}
\renewcommand{\thefigure}{F.\arabic{figure}}

We here present some details of the manipulations of Eq.~(\ref{eq:Tmu-biss}) that lead to the parameter
dependencies of $e_{\rm kin}$, $z_{\rm GNM}$ and the various critical curves  that limit the extent of the 
extended and condensed phases.

\subsection{No condensation}

\vspace{0.25cm}

\noindent
{\it One equation linking $e_{\rm kin}$ and $z_{\rm GNM}$.}

\vspace{0.25cm}

Summing Eq.~(\ref{eq:Tmu-biss}) over $\mu$, dividing by $N$, and next taking $N\to\infty$, we derive
\begin{eqnarray}
     \frac{4a}{J}  \, \lim_{N\to \infty}  \frac{e_{\rm kin}}{T} 
    =  \int d\lambda \, \frac{ \rho(\lambda)}{A -  \lambda  } 
    \; . 
   \quad
\end{eqnarray}
In the following the limit $\lim_{N\to\infty}$ will be implicitly considered unless otherwise stated.
This integral can be explicitly performed and yields
\begin{eqnarray}
    8aJ \,  \frac{e_{\rm kin}}{T} 
    = 
     \left[ A
    - {{\rm sign}(A)} \sqrt{A^2 - 4J^2}
    \right]
    \; , 
\end{eqnarray}
assuming 
\begin{equation}
|A| = \Big| \frac{b}{a} J + 2e_{\rm kin} -z_{\rm GNM} \Big|>2J
\; .
\end{equation} 
Isolating the square root and squaring:
\begin{eqnarray}
   \left( 8a J  \frac{e_{\rm kin}}{T} -
   A 
   \right)^2
    =
   A^2 - 4J^2 
\end{eqnarray}
and then
\begin{eqnarray}
   (8a J)^2 \,  \frac{e^2_{\rm kin}}{T^2} - 2 (8a J)  \frac{e_{\rm kin}}{T}
   A
    =
   - 4J^2 
   \; . 
\end{eqnarray}
Replacing $A$ in terms of $e_{\rm kin}$, $z_{\rm GNM}$ and 
simplifying a bit
\begin{eqnarray}
 \frac{T}{2e_{\rm kin}} &=&
   - 8 a^2 \,  \frac{e_{\rm kin}}{T} 
   +   2b
   +   \frac{4a}{J} \, e_{\rm kin}
   -   \frac{2a}{J} \, z_{\rm GNM}
   \; .
   \label{eq:zGGE}
     \end{eqnarray}
\noindent
  This is the first equation linking $e_{\rm kin}(T,J,a,b)$ and $z_{\rm GNM}(T,J,a,b)$.
{Squaring removes the distinction between the two square-root branches. Candidate solutions must therefore also 
  satisfy the unsquared integral equations and the sign conditions.}
     
At high $T$ it turns out to be licit to assume the linear dependence, 
    $
     e_{\rm kin}(T,J,a,b) \sim \eta(J,a,b) T + \zeta(J,a,b)
 $.
      Then,  the Lagrange multiplier should also be linear in $T$, 
   $ z_{\rm GNM}(T,J,a,b) \sim 
       - J/(4a \eta(J,a,b)) \, 
  - 4 a J \, \eta(J,a,b)
   +  (b/a) J
     + 2 \, \zeta(J,a,b) 
     +  2 \, \eta(J,a,b) T
    $, with a slope which is twice the one of $e_{\rm kin}$.
     
\vspace{0.25cm}

\noindent
{\it Another equation linking $e_{\rm kin}$ and $z_{\rm GNM}$.}

\vspace{0.25cm}

Multiplying  Eq.~(\ref{eq:Tmu-biss}) by $[-z_{\rm GNM} +\lambda_\mu]^{-1}$, 
summing  over $\mu$, dividing by $N$, and next taking $N\to\infty$, we derive
\begin{eqnarray}
-   \frac{2a}{J}  \frac{1}{T} 
  & \to& 
   \int d\lambda \, \rho(\lambda) \, 
    \frac{1}{A -  \lambda  } 
    \; 
    \frac{1}{\lambda -z_{\rm GNM} }
   \nonumber\\
   &=&
     \frac{1}{A -z_{\rm GNM}}
         \int d\lambda \, \rho(\lambda) \, 
  \left(   \frac{1}{A -  \lambda  }  
   +
    \frac{1}{ \lambda -z_{\rm GNM}}
   \right)
   \end{eqnarray}
 which, after performing the integral assuming $z_{\rm GNM}>2J$, yields
  \begin{eqnarray}
  \frac{4aJ}{T}  &=&
      \frac{-1}{A-z_{\rm GNM}}      
   \left(\!
    A- 
     {{\rm sign}(A)} \sqrt{A^2 - 4J^2}
    -
      z_{\rm GNM}  +
      \sqrt{z^2_{\rm GNM}  - 4J^2}
      \right) \!.
      \qquad
      \label{eq:ekin}
\end{eqnarray}
Together with the constraint $2J < z_{\rm GNM} $ already used we have
 \begin{equation}
 2J < z_{\rm GNM} < \frac{b}{a} J + 2 e_{\rm kin} - 2J
 \end{equation}
 where we assumed that the larger bound is positive.
We can now manipulate Eq.~(\ref{eq:ekin}) and render it nicer with rather simple operations. We multiply the left-hand-side 
by the first factor in the right-hand-side, we take the square of both sides, we expand the 
powers and we apply other similar manipulations to arrive at the factorized form:
 \begin{eqnarray}
             && 0 = (A-z_{\rm GNM})^2  
             \qquad\qquad\qquad {\rm or}
             \\
             && 0 = 
                     \left(  \frac{4a J}{T} \right)^2  \left(  \frac{2a J}{T} + 1\right)^2
(A-z_{\rm GNM})^2
-    \frac{8a J}{T} \left(  \frac{2a J}{T} + 1\right) 
 A  \ z_{\rm GNM}
 \nonumber\\
 &&
 \qquad
 +  \frac{32a J^3}{T} \left(  \frac{2a J}{T} + 1\right) 
     +4J^2 
     \; . 
                 \end{eqnarray}    
                 The first equation yields 
$
                 2z_{\rm GNM} = 2 e_{\rm kin} + (b/a) J
                 $ which turns out to be a spurious solution with the wrong $a\to 0$ limit and we discard it.
The second equation simplifies to 
                    \begin{eqnarray}
   && 0 = 
                     \left(  \frac{2a}{T} \right)^2   \left(  \frac{2a J}{T} + 1\right)^2
 \left(  \frac{b}{a} J + 2e_{\rm kin} -2z_{\rm GNM} \right)^2
  +  \frac{8a J}{T} \left(  \frac{2a J}{T} + 1\right) 
     +1 
\nonumber\\
&&
\qquad
-    \frac{2a}{T}  \frac{1}{J} \left(  \frac{2a J}{T} + 1\right) 
 \left( \frac{b}{a} J + 2e_{\rm kin} -z_{\rm GNM} \right)  \ z_{\rm GNM}
 \label{eq:good-eq}
                 \end{eqnarray}  
                 and is a second equation relating $e_{\rm kin}$ to $z_{\rm GNM}$. 
Equation~(\ref{eq:zGGE}) allows us to replace $z_{\rm GNM}$ in terms of $e_{\rm kin}$ here, and 
obtain a closed equation for the latter. Once this is done, we can replace $e_{\rm kin}$ in Eq.~(\ref{eq:zGGE})
and determine the dependence of $z_{\rm GNM}$ on the parameters.
In the following subsections we give the explicit expressions of $e_{\rm kin}$ and  $z_{\rm GNM}$
in terms of $T$, $J$, $a$ and $b$.

\vspace{0.25cm}

\noindent
{\it The kinetic energy.}
    
\vspace{0.25cm}
    
  Replacing $z_{\rm GNM}$ in Eq.~(\ref{eq:good-eq}):
    \begin{eqnarray}
   && 0 = 
                     \left(  \frac{2aJ}{T} \right)^2   \left(  \frac{2a J}{T} + 1\right)^2
 \left( \frac{b}{a} 
   +   \frac{2e_{\rm kin}}{J} -   \frac{8ae_{\rm kin}}{T}  \,   -  \frac{T}{2ae_{\rm kin}} \right)^2
\nonumber\\
&&
\qquad
-    \frac{2aJ}{T} \left(  \frac{2a J}{T} + 1\right) 
 \left(  \frac{4 a e_{\rm kin}}{T} \,   +    \frac{T}{4ae_{\rm kin}} \right)  \ 
 \nonumber\\
 && 
 \qquad\qquad\quad
 \times \left(  -  \frac{4 a e_{\rm kin}}{T} \,  
   +    \frac{b}{a} 
   +   \frac{2e_{\rm kin}}{J}
   -     \frac{T}{4a e_{\rm kin}}
   \right)
 \nonumber\\
 &&
 \qquad
 +  \frac{8a J}{T} \left(  \frac{2a J}{T} + 1\right) 
     +1 
     \label{eq:eq-to-solve-ekin}
                 \end{eqnarray} 
                 which is now written in terms of the adimensional variable $e_{\rm kin}/J$ and the parameter $T/J$.
                   This expression can be written as the product of two quadratic factors:
                                \begin{eqnarray*}
&&
0 = 
\left(
4 J T^2 a 
- 8 J T a b\, e_{\rm kin} 
+ 64 J a^3\, e_{\rm kin}^2 
+ 32 a^2 T e_{\rm kin}^2
+ T^3 
- 4 T^2 b\, e_{\rm kin}
\right)
\nonumber\\
&&
\qquad \times \left(
8 J^2 T^2 a^2 
- 16 J^2 T a^2 b\, e_{\rm kin} 
+ 128 J^2 a^4\, e_{\rm kin}^2 
+ 6 J T^3 a 
\right.
\nonumber\\
&& 
\qquad\qquad\qquad
\left.
- 8 J T^2 a b\, e_{\rm kin} 
+ T^4 
- 16 T^2 a^2\, e_{\rm kin}^2
\right)
\; . 
\end{eqnarray*}
Either using Mathematica to solve Eq.~(\ref{eq:eq-to-solve-ekin}), or focusing on the  first factor of 
its factorized version, the physical solution  is ($b>0$)
                 \begin{eqnarray}
                 && e_{\rm kin} 
=   \frac{bT}{16 a^2} \left[ 1 - \sqrt{ 1 -  \frac{8  a^2}{b^2} \left( \frac{ 4 a J+T }{2 a J+T} \right) } \right] 
                 \label{eq:ekin-analytic-GGE}
                 \end{eqnarray}
                            restricting the  parameters such that $e_{\rm kin}$ is real and positive,  
                            \begin{equation}
0 \ < \ \frac{4aJ+T}{2aJ+T} \ <\  \frac{b^2}{8a^2}
\; , 
\label{eq:one-condition}
                            \end{equation}
                            The first inequality follows from requiring $e_{\rm kin}>0$.
              
              The analysis of this expression in terms of the parameters and the limitations it imposes to extended phase are discussed
              in the main text.               
                            
\vspace{0.25cm}

\noindent
{\it The Lagrange multiplier.}

\vspace{0.25cm} 

Replacing $e_{\rm kin}$  in Eq.~(\ref{eq:zGGE})
and simplifying a bit:
\begin{eqnarray}
&& 
z_{\rm GNM} =
 \frac{b [ 4(8aJ+7T) (aJ)^2 + (8aJ+T) T^2]}
{8 a^2 (2aJ+T) (4 a J + T)}
\nonumber\\
&& 
\qquad\quad
- \frac{bT }{8a^2} \frac{6aJ+T}{4aJ+T} 
\sqrt{1 - \frac{8a^2}{b^2} \left( \frac{4aJ+T}{2aJ+T} \right)}
\label{eq:zGGE-aboveTc}
\end{eqnarray}
(the condition in Eq.~(\ref{eq:one-condition}) ensures that the square root yields a real result)
      with limits
       \begin{eqnarray}
z_{\rm GNM}
 \to 
\left\{
\begin{array}{ll}
\displaystyle{ \frac{4 b^2 J^2+T^2}{2 b T}}
+  {\mathcal O}(a)  
\qquad\quad\qquad\quad  
a\sim 0  &
\\
[15pt]
\displaystyle{ \frac{T}{8a^2} \left(
b-\sqrt{b^2-8 a^2}\right)}
+{\mathcal O}\left(1\right)
  \quad\;\;\;
    T\to \infty \;\; \& \;\;  a \neq 0 &
\end{array}
\right.
\end{eqnarray}
We confirm that the slope of $z_{\rm GNM}$ at large $T$
 is twice the one of $e_{\rm kin}$.

\vspace{0.25cm}
 
  \noindent
 {\it Critical properties. }

\vspace{0.25cm}

We fix the critical temperature  $T_c$ by requiring
$z^c_{\rm GNM}=2J$, as the numerics suggest (see Sec.~\ref{sec:numerics}).
Inserting this value in
 Eqs.~(\ref{eq:zGGE}) and (\ref{eq:ekin}) they become
\begin{eqnarray}
 &&  0 = - \frac{T_c}{2e^c_{\rm kin}}  
   - 8 a^2 \,  \frac{e^c_{\rm kin}}{T_c} 
   +   2b
   +  \frac{4a}{J} \, e^c_{\rm kin}
   -  4a
   \label{eq:zGGE-crit}
\\
&&
0
=
 \left(  \frac{4a J}{T_c} +1\right) 
\left(  \frac{b}{a} J + 2e^c_{\rm kin}\!  -4J \right)
  - 
      \sqrt{\left( \frac{b}{a} J + 2e^c_{\rm kin} - 2J \right)^2 - 4J^2}
         \label{eq:ekin-crit}
          \end{eqnarray}
          These equations are compatible with $e^c_{\rm kin} = T_c/2$ and $T_c=J$
           for 
          $a\to 0$. 
                    
      Rearranging and squaring, Eq.~(\ref{eq:ekin-crit})  becomes a quadratic equation
      on $e_{\rm kin}^c$, 
      with solution
       \begin{equation}
                  2e^c_{\rm kin} =  
                  2J -  \frac{b}{a} J + 2J \left\{ v+1 \pm 1 \right\}  \frac{1}{v}
                  \end{equation}
                  and
                                    \begin{equation}
                  v =  \left(  \frac{4a J}{T_c} \right)^2 + \frac{8a J}{T_c}  
                  \; . 
                  \end{equation}
                                   Finally, taking the positive sign, since the negative one would yield a divergent $e^c_{\rm kin}$
                  for $a\to 0$,              
                                     \begin{equation}
                  2e^c_{\rm kin} = 
                           2J -  \frac{b}{a} J + \frac{1}{2a} \;  \frac{ 8 \left(a J \right)^2 
               + 4a J T_c  + T_c^2}{ 2 a J  + T_c  }   
               \; . 
                \label{eq:first-eq-ekinTc}
                               \end{equation}
                               We already see from here that $T_c(a\to 0) = 2bJ$, otherwise the kinetic energy at criticality
                                would diverge.

Concerning the critical temperature, it follows from the replacement of $e_{\rm kin}^c$ in Eq.~(\ref{eq:zGGE-crit}) or 
(\ref{eq:ekin-crit}):
\begin{equation}
T_c = b  J \Big(1 -  \frac{3a}{b}  \pm \sqrt{1- \frac{2a}{b} - \frac{7a^2}{b^2}} \Big)
\; . 
\label{eq:Tc}
\end{equation}

\subsection{Condensed phase}

Following the same steps as in the previous section, with the necessary 
modifications, 
\begin{eqnarray}
- \frac{T_\mu}{T} &=&
 \Big\{ \frac{2a}{J} (1-q) \lambda_\mu - 2b (1-q) 
-   \frac{4a}{J} \, e_{\rm kin}
+ 
 4a (1-q)
 \nonumber\\
 &&
 \quad
 + \; 2q 
 \left[ \frac{a}{J}  (\lambda_\mu+2J) - b \right]
\Big\}^{-1}
 \; . 
 \label{eq:Tmu-condensed}
\end{eqnarray}
In the SSK case $a=0$ and $b=1/2$ and
\begin{eqnarray}
T_\mu &=& T
 \left\{   2b (1-q) 
 + \; 2b \, q \right\}^{-1}  = T/(2b) 
 = T
 \; , 
 \label{eq:Tmu-condensed-a0}
\end{eqnarray}
where we used $q=1-T/J$, which is self-consistently derived from the spherical constraint
setting $T_\mu=T$.

\vspace{0.25cm}

\noindent{\it The kinetic energy.}

\vspace{0.25cm}

One can manipulate Eq.~(\ref{eq:Tmu-condensed}) as we have already done in the case without condensation,
only that the parameters in the integral over $\lambda$ are different.
We can sum over $\mu$ and divide by $N$, to obtain $e_{\rm kin}$ on the left-hand-side. 
Then, 
\begin{eqnarray}
- \frac{2e_{\rm kin}}{T} 
&=&
\frac{1}{C} \int d\lambda \; \rho(\lambda) \;  \frac{1}{ \lambda - B/C}
\end{eqnarray}
with 
\begin{eqnarray}
C &=& \frac{2a}{J} (1-q) + \frac{2a}{J} \, q=  \frac{2a}{J} 
\; , \\
B &=& -2 (2a-b) (1-q) + \frac{4a}{J} e_{\rm kin} - 4a  q + 2bq
\nonumber\\
&=& 
- 2 (2a-b) 
+ \frac{4a}{J} e_{\rm kin} 
\; , 
\end{eqnarray}
and both are independent of $q$. 
We assume  that $B/C$ is larger than $2J$; then
\begin{eqnarray}
\frac{2e_{\rm kin}}{T} 
&=&
 \frac{1}{C}\frac{1}{2J^2}  \left[ \frac{B}{C} - \sqrt{\frac{B^2}{C^2} - 4 J^2} \right]
\end{eqnarray}
from where 
\begin{eqnarray}
&& 
\left( \frac{2e_{\rm kin}}{T} C \, 2J^2 - \frac{B}{C} \right)^2 
=
\frac{B^2}{C^2} - 4 J^2
\nonumber\\
&&\qquad \Rightarrow
\left( \frac{2e_{\rm kin}}{T}\right)^2  (2CJ^2)^2 - 2 B \, 2J^2 \,  \left( \frac{2e_{\rm kin}}{T}\right) 
+ 4 J^2 = 0
\; . 
\end{eqnarray}
Simplifying 
\begin{eqnarray}
\left( \frac{2e_{\rm kin}}{T}\right)^2  (CJ)^2 -  B  \,  \left( \frac{2e_{\rm kin}}{T}\right) 
+  1 = 0
\end{eqnarray}
and replacing $B$ and $C$:
\begin{eqnarray*}
&& 
\left( \frac{2e_{\rm kin}}{T}\right)^2 
 \left[ (2a)^2 - \frac{2aT}{J} \right]
+2(2a-b)  \ \left( \frac{2e_{\rm kin}}{T}\right) 
+ 1 = 0
\end{eqnarray*}
which is just a quadratic equation on $2e_{\rm kin}/T$, 
with solutions
\begin{eqnarray}
&& 
\frac{2e_{\rm kin}}{T}  
=
\frac{b-2a 
-
\sqrt{(b-2a)^2 - \frac{2a}{J} ( 2a J- T ) }
}
{\frac{2a}{J}  (2a J - T )}
\; . 
\label{eq:ekin-belowTc}
\end{eqnarray}
We chose the sign in front of the square root to get the correct $a\to 0$, $b=1/2$ limit, 
$2e_{\rm kin} = T$ at all $T$. 
At $T\to 0$:
\begin{eqnarray}
&& e_{\rm kin} =\frac{T}{2}  \left[
\frac{b-2a - \sqrt{b^2-4ab}}{(2a)^2}
\right]
\; . 
\end{eqnarray}
We note that the functional 
form that we got for $e_{\rm kin}$ below $T_c$ is not the same as the one we had above 
$T_c$. In Sec.~\ref{sec:numerics} we show that the full expression, with Eq.~(\ref{eq:ekin-analytic-GGE}) above $T_c$ and 
Eq.~(\ref{eq:ekin-belowTc}) below $T_c$, describes the asymptotic kinetic energy at all temperatures.
 
 \vspace{0.25cm}
 
 \noindent
{\it The $N$th mode temperature.} 

\vspace{0.25cm}

Evaluating Eq.~(\ref{eq:Tmu-condensed}) at $\mu=N$, and setting $\lambda_N=2J$, one has
\begin{eqnarray}
- \frac{T_N}{T} 
&=&
  \Big( 8a  -2b 
-   \frac{4a}{J} \, e_{\rm kin}
\Big)^{-1}
\end{eqnarray}
which does not explicitly depend on $q$.
Using now the expression for $e_{\rm kin}$ in Eq.~(\ref{eq:ekin-belowTc})
\begin{eqnarray}
-\frac{T_N}{T} 
&=& 
 \left( 8a  -2b 
-   \, T
\frac{b-2a 
-
\sqrt{(b-2a)^2 - \frac{2a}{J} ( 2a J- T ) }
}
{ (2a J - T )}
\right)^{-1}
 \label{eq:TN-lowT}
\end{eqnarray}
which for $T\to 0$ approaches
\begin{equation}
T_N \ \  \mathop{\longrightarrow}\limits_{T\to 0} \  \
\frac{T}{2}  (b-4a)^{-1}
\; . 
\end{equation}
If we set $4a=b$, before taking $T\to 0$, 
\begin{equation}
T_N
= 
\left( 
\frac{2a 
-
\sqrt{ \frac{2a}{J} T }
}
{ 2a J - T }
\right)^{-1} 
 = 
 \frac{2aJ-T}{2a - \sqrt{ \frac{2a}{J} T} }  \ \ \mathop{\longrightarrow}\limits_{T\to 0} \  \ J
\end{equation}

\vspace{0.25cm}

\noindent{\it The Edwards-Anderson parameter $q$.}

\vspace{0.25cm}

We can get the equation for $q$ multiplying Eq.~(\ref{eq:Tmu-condensed}) 
by $1/(2J-\lambda_\mu)$, summing over $\mu$, dividing by $N$, taking $N\to\infty$, 
and proceeding as for the non condensed case:
\begin{eqnarray}
- \frac{(1-q)}{T} 
&=& \frac{1}{C} \; \frac{1}{2J-B/C} \; \frac{1}{2J^2} 
\left[ - \frac{B}{C} + \sqrt{\left(\frac{B}{C} \right)^2 -4J^2}  + 2J \right]
\end{eqnarray}
where we assumed, again, that $B-2J\, C>0$.
Then, after some simple steps, 
\begin{eqnarray}
&& \left[ \frac{(1-q)}{T} \, C \,  \left(2J-\frac{B}{C} \right) \, J + 1\right]^2 2J
\nonumber\\
&&
\qquad 
- \frac{2B}{C}  \left[ \frac{(1-q)}{T} \, C \,  \left(2J-\frac{B}{C} \right) \, J + 1\right] 
= 
 -2J
\end{eqnarray}
which is a quadratic equation for 
$y=(1-q)/T$:
\begin{equation}
y^2 \alpha_1 + y \alpha_2 + \alpha_3 =0
\qquad
\Rightarrow
\qquad
y = \frac{-\alpha_2 \pm \sqrt{ \alpha_2^2 - 4 \alpha_1 \alpha_3}}{2\alpha_1}
\end{equation}
with 
\begin{eqnarray}
&& 
\alpha_1  =
2J  \left( B-2J \, C \right)^2 J^2 
\; , 
\nonumber\\
&& 
\alpha_2
=
- 2J \; 2J \left( B - 2J \, C \right)  + 2J \, \frac{B}{C} \left( B - 2J \, C \right) 
=
\frac{2J}{C} (B-2J \, C)^2
\; , 
\nonumber\\
&& 
\alpha_3 
=
- \frac{2}{C}  \left(B -  2J \, C \right)
\; . 
\end{eqnarray}
The solution is 
\begin{eqnarray}
&& 
\frac{1-q}{T} = \frac{1}{2J^2 \, C}
\Big[ -1 \pm \sqrt{ \displaystyle{\frac{B+2J \, C}{B-2J \, C}}} \Big]
\; . 
\end{eqnarray}
After replacing $B$ and $C$ using the expression above for $e_{\rm kin}$, we find
\begin{eqnarray}
\frac{1-q}{T} 
= 
\frac{1}{4Ja}
\Big[
-1 + \sqrt{
\frac{b +2a e_{\rm kin}/J}{b-4a +2a e_{\rm kin}/J}
}
\Big]
\; .   
\end{eqnarray} 
One has to replace $e_{\rm kin}$ by the expression in Eq.~(\ref{eq:ekin-belowTc}) to get the explicit dependence on 
$a$, $b$, $T$ and $J$:
\begin{eqnarray}
&&
q =
1+ 
\frac{ T - 
 T \sqrt{
 \displaystyle{
 \frac{-4 a b J + 2 a T + b T + T \sqrt{-4 a b + b^2 + \frac{2 a T}{J}}}{
  16 a^2 J - 4 a b J - 6 a T + b T + 
   T \sqrt{-4 a b + b^2 + \frac{2 a T}{J}}}}}}{4 a J}
   \;\;
   \label{eq:qEA}
\end{eqnarray}

\section{The role of temperature}
\label{app:temperature}
\setcounter{equation}{0}
\renewcommand{\theequation}{G.\arabic{equation}}
\setcounter{figure}{0}
\renewcommand{\thefigure}{G.\arabic{figure}}

\begin{center}
\begin{figure}[h!]
\centerline{
\includegraphics[scale=0.3]{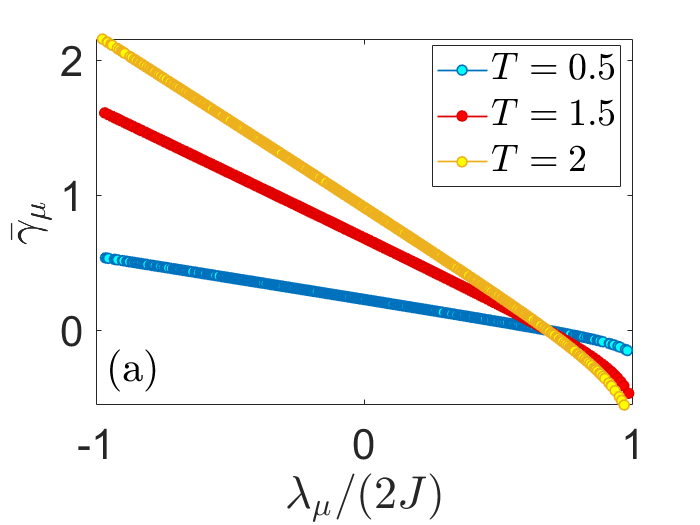}
\includegraphics[scale=0.3]{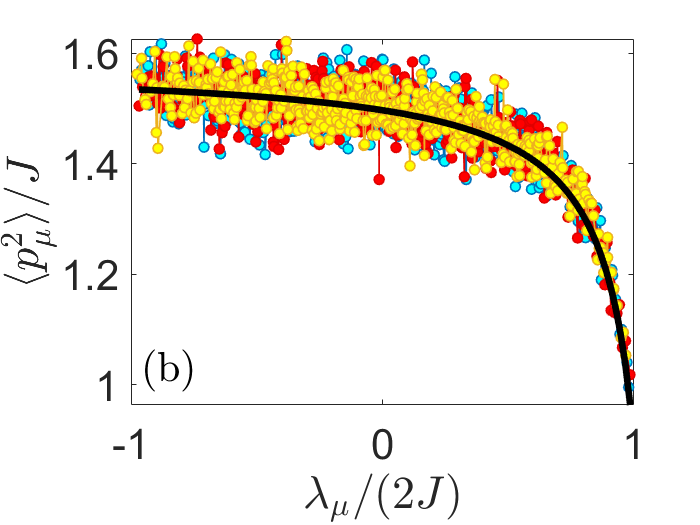}
}
 \centerline{
\includegraphics[scale=0.3]{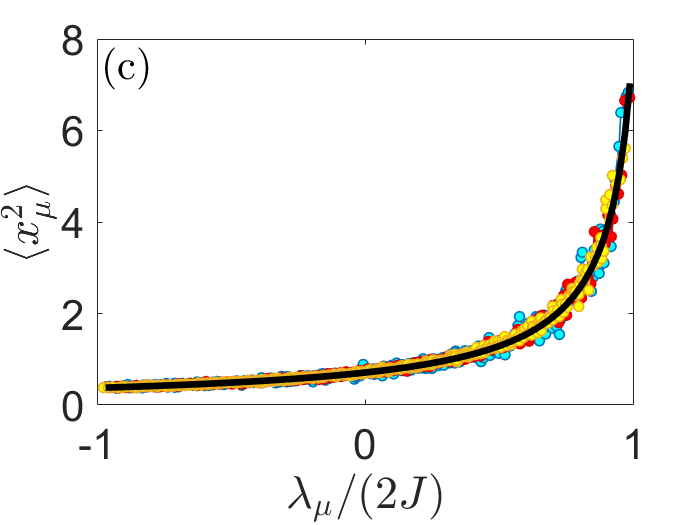}
\includegraphics[scale=0.3]{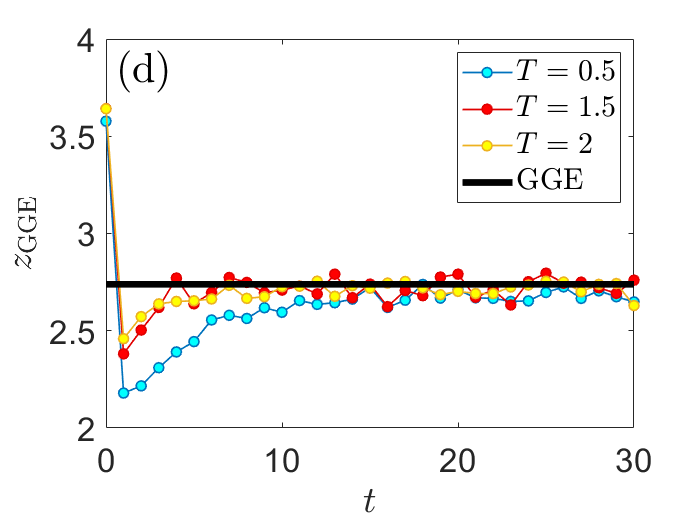}
}
\caption{\small (a) The coefficients $\bar\gamma_\mu = T \gamma_\mu$ used to build the Hamiltonian ${\mathcal H}_{\rm GGE}$, 
at three temperatures $T$, with $J/J_0$ and $T_0/J_0$ chosen so as to be in phase Ib of the phase diagram in 
Fig.~\ref{fig:fig14}. (b) The stationary spectrum of the square momentum components, (c) the 
stationary spectrum of the square
coordinates, and (d) $z_{\rm GGE}$ as a function of time. Data points are for the results of the Langevin simulations while 
the black lines represent the analytic predictions.}
\label{fig:T-indep}
\end{figure}
\end{center}

In the main text we argued that the Langevin dynamics applied to ${\mathcal H}_{\rm GGE}$ 
with  $\bar\gamma_\mu = T \gamma_\mu$ and the $\gamma_\mu$ selected {from the solution described 
in App.~\ref{app:new-derivation} to the equation derived in}~\cite{BaCuLoNe22}, 
takes to the measure $ \propto \exp(-\beta {\mathcal H}_{\rm GGE})$ for all $T$. In Fig.~\ref{fig:T-indep} we show
numerical confirmation of this statement. In (a) the coefficients  $\bar\gamma_\mu = T \gamma_\mu$ used to build the 
Hamiltonian ${\mathcal H}_{\rm GGE}$, at three temperatures $T$, with $J/J_0$ and $T_0/J_0$ chosen so as to 
be in phase Ib of the phase diagram in Fig.~\ref{fig:fig14} are shown. Note that different slopes induced by the 
factor $T$. In (b) and (c) we plot the spectra of $\langle p_\mu^2\rangle$ and $\langle x_\mu^2\rangle$ 
in the asymptotic limit, confirming the independence of $T$. Finally, in (d) we see that, for these 
parameters $\bar \gamma_\mu$, 
the approach to the asymptotic limit is only slightly influenced by $T$ in the very early stages of the evolution.

\section{The finite $N$ GNM}
\label{sec:stationary-states}
\setcounter{equation}{0}
\renewcommand{\theequation}{H.\arabic{equation}}
\setcounter{figure}{0}
\renewcommand{\thefigure}{H.\arabic{figure}}

In this Appendix we give some details on the stationary 
points and relaxational dynamics of the finite $N$ GNM.

\subsection{Stationary states}

To understand the relaxational (Langevin) dynamics of the SSK model (and even more complex cases) it is
useful to know the stationary points of the potential energy (including the constraint)~\cite{Ros23}.
Those are saddle points of different
stability, including  minima and maxima as extreme cases. Operationally, one gets these
configurations by assuming that the Lagrange multiplier  is an external parameter and
then finds which value it should take at each saddle-point.

In the case  we are interested in, 
${\mathcal H}^z_{\rm GNM} = {\mathcal H}_{\rm GNM}+ z_{\rm GNM} (x^2-N)/2=
 \sum_\mu \bar\gamma_\mu I_\mu + z_{\rm GNM} (x^2-N)/2$, 
 the stationary states are the configurations that satisfy
$\{ \dot x_\mu = 0, \dot p_\mu =0 \}$, 
that is
\begin{eqnarray}
0 &   =  & \frac{\partial {\mathcal H}^z_{\rm GNM}}{\partial p_\mu}= 
 \sum_{\nu (\neq \mu)}  \, x_\nu  \, (x_\mu p_\nu - p_\mu x_\nu) \, 
 \frac{\bar\gamma_\mu - \bar\gamma_\nu}{\lambda_\mu - \lambda_\nu}
 \; , 
 \label{eq:explicit11}
\\
0  &  =   & -\frac{\partial {\mathcal H}^z_{\rm GNM}}{\partial x_\mu}= 
-  2\bar\gamma_\mu x_\mu 
+ 
\! \! \frac{2}{N} \sum_{\nu (\neq \mu)}  \,  \frac{p_\nu}{m}  \, (x_\mu p_\nu - p_\mu x_\nu) \, 
 \frac{\bar\gamma_\mu - \bar\gamma_\nu}{\lambda_\mu - \lambda_\nu}
 -    
z_{\rm GNM} \, x_\mu
 \nonumber\\
 &  =  &  
 -  \left( 2\bar\gamma_\mu + z_{\rm GNM} \right) x_\mu 
+ 
\frac{2}{N} \sum_{\nu (\neq \mu)}  \,  \frac{p_\nu}{m}  \, (x_\mu p_\nu - p_\mu x_\nu) \, 
 \frac{\bar\gamma_\mu - \bar\gamma_\nu}{\lambda_\mu - \lambda_\nu}
 \; , 
 \label{eq:explicit12}
\end{eqnarray}
where we have already used $x^2=N$ to set to zero the terms proportional to 
$\partial z_{\rm GNM}/\partial x_\mu$ and $\partial z_{\rm GNM}/\partial p_\mu$.
Recall that $z_{\rm GNM}$ is a function of $\vec p$ and $\vec x$.

We note that multiplying Eq.~(\ref{eq:explicit12}) by $p_\mu$ and summing over $\mu$, several 
terms vanish identically and $\sum_\mu \bar\gamma_\mu x_\mu p_\mu = 0$. A solution with 
$p_\mu\neq 0$ should satisfy this condition, together with $\sum_\mu x_\mu p_\mu = 0$.

\subsubsection{The SSK/Neumann case.}
\label{subsubsec:SSK}

As a warm-up exercise, let us show which are the 
stationary states of the conventional SSK/Neumann model, recovered 
by setting $\bar\gamma_\mu = - \lambda_\mu/2$ 
(the proof is recalled in App.~\ref{app:GGE-to-SSK}). The stationarity equations reduce to 
\begin{eqnarray}
0 & = & 
 \sum_{\nu (\neq \mu)}  \, x_\nu  \, (x_\mu p_\nu - p_\mu x_\nu) 
=
  \sum_{\nu}  \, x_\nu  \, (x_\mu p_\nu - p_\mu x_\nu) 
  \nonumber\\
  &=& 
  x_\mu \sum_\nu x_\nu p_\nu  - p_\mu \sum_\nu x_\nu^2 = 
- p_\mu N \ \ \ \ \ \ \Rightarrow  \ \ \ \ \ \ p_\mu = 0
\end{eqnarray}
which fixes the momentum to zero, and 
\begin{eqnarray}
0 & = &  
  \left(  \lambda_\mu -  z_{\rm GNM} \right) x_\mu 
- 
 \frac{1}{N} \sum_{\nu (\neq \mu)}  \,  \frac{p_\nu}{m}  \, (x_\mu p_\nu - p_\mu x_\nu) 
  \nonumber\\
 & = &  
 \left(  \lambda_\mu -  z_{\rm GNM} - 2e_{\rm kin} \right) x_\mu 
 \; . 
\label{eq:explicit13}
\end{eqnarray}
The momentum has already been set to zero, so the kinetic energy vanishes
and  $z_{\rm GNM} = z_{\rm GNM}(\vec x, \vec p=0)$.
The  solution of the remaining equation is $x_\mu=0$ or 
\begin{equation}
 x_\mu \neq 0 \qquad  \mbox{and} \qquad  z_{\rm GNM}(\vec x, \vec p=0)  = \lambda_\mu 
\; ,
 \end{equation}
 which fixes the value of the Lagrange multiplier to be the one of the 
 spring constant along the direction of the position of the particle. 
 Thus, 
 {there are $2N$ axis-aligned stationary configurations, organized into $N$ antipodal pairs labelled by $\mu$, and given by}
\begin{eqnarray}
 &&
 \vec p =0 \; , \qquad\qquad\qquad
 z_{\rm GNM}(\vec x, \vec p=0)  = \lambda_\mu 
\; , 
 \nonumber\\
 &&
 x_\nu = 0 \;\; \forall \nu \neq \mu 
 \; , 
 \qquad \  x_\mu = \pm \sqrt{N} 
 \; . 
 \end{eqnarray}
 One can also write the position vector  as 
   \begin{equation}
  \vec x = \pm  \vec v_\mu 
 \label{eq:stat-points1}
\end{equation}
where  $\vec v_\mu$ are vectors such that $v_\mu^2 = N$ and point 
in the direction of any of the Cartesian axes. 

\subsubsection{The generic model.}

Going back to the interesting case, we note that Eq.~(\ref{eq:explicit11}) does not 
necessarily set $p_\mu =0$ and, depending on the choice of $\{\bar\gamma_\mu\}$,  
there may be solutions with non-vanishing momentum.

Setting $p_\mu=0$ to ask for configurations with vanishing 
momentum, we recover the condition $x_\mu=0$ or else
$x_\mu \neq 0$ with,  from Eq.~(\ref{eq:explicit12}), 
\begin{eqnarray}
z_{\rm GNM} = - 2\bar\gamma_\mu
\; . 
\end{eqnarray}
Thus, 
{this construction also gives $2N$ axis-aligned stationary configurations, 
organized into $N$ antipodal pairs labelled by $\mu$, and given by}
\begin{eqnarray}
 &&
 \vec p =0 \; , \qquad\qquad\qquad
 z_{\rm GNM}(\vec x, \vec p=0)  = -2\bar\gamma_\mu 
\; , 
 \nonumber\\
 &&
 x_\nu = 0 \;\; \forall \nu \neq \mu 
 \; , 
 \qquad \  x_\mu = \pm \sqrt{N} 
 \; . 
 \end{eqnarray}
This means that the stationary points are also the orthogonal vectors on the sphere
along the Cartesian axes, $\vec x = \pm \vec v_\mu$. 
Curiously enough, the parameters $\{\lambda_\mu\}$ do not 
play any role in the selection of the stationary points. They should do, though, in the stability and 
dynamics.

We now ask whether the stationary points with vanishing momentum are ordered as in the SSK/Neumann model, where 
$V_{\rm SSK}(\vec x = \pm {\vec v}_1, \vec p = 0, z_{\rm GNM}=\lambda_1) > V_{\rm SSK}(\vec x = \pm{\vec v}_2, \vec p = 0, z_{\rm GNM}=\lambda_2) > \dots > 
V_{\rm SSK}(\vec x = \pm {\vec v}_N, \vec p = 0, z_{\rm GNM}=\lambda_N) $
with the harmonic parameters organized as $\lambda_1 < \dots < \lambda_N$. 
After translation to the GNM model, the 
question is how are the  
${\mathcal H}^z(\vec x = \pm {\vec v}_\mu, \vec p = 0, z_{\rm GNM}=-2\bar\gamma_\mu) $ ordered.
First, we notice that $I_\mu(\pm \vec v_\rho, \vec 0)=N \delta_{\mu\rho} $, which implies
\begin{equation}
{\mathcal H}^z_{\rm GNM}(\pm \vec v_\rho, \vec 0)
\! = \!  
\sum_\mu \bar\gamma_\mu I_\mu(\pm \vec v_\rho, \vec 0)  + \frac{1}{2} \, z_{\rm GNM}(\pm \vec v_\rho, \vec 0) 
\left( v_\rho^2 -N \right) 
= \bar\gamma_\rho N
, 
\label{eq:energies-metastable-states}
\end{equation}
and we conclude that these stationary points are ordered according to the values of the $\{\bar\gamma_\mu\}$. If these are monotonically 
decreasing with the $\mu$ index, then the order of the energy of the stationary states in the GNM model is the same as the 
one in the SSK one, with the North/South  poles or $\pm N$th directions corresponding to two degenerate minima.
Interestingly enough, the hierarchy 
of stationary states can be reshuffled for other orderings of the $\bar\gamma_\mu$.
Indeed, the energies of the metastable states are organized in increasing order following the order of 
increasing $\bar\gamma_\mu$, as follows immediately from Eq.~(\ref{eq:energies-metastable-states}). 

To know about the stability of these stationary points, one 
needs to calculate the Hessian of  ${\mathcal H}^z_{\rm GNM}$ and evaluate it at the concerned 
stationary point. After some straightforward calculations, 
 one finds that the Hessian of ${\mathcal H}_{\rm GNM}^z$ 
at any of the stationary points is a diagonal matrix with elements
\begin{equation}
\underbrace{2\bar\gamma_\mu + z_{\rm GNM}(\vec v_\rho, \vec 0)}_{\partial^2 {\mathcal H}^z_{\rm GNM}/\partial x_\mu^2}, \; 
\underbrace{ - \frac{2}{m} \; 
      \, \frac{\bar\gamma_\mu - \bar\gamma_\rho}{\lambda_\mu - \lambda_\rho} \; (1-\delta_{\mu\rho})}_{\partial^2 {\mathcal H}^z_{\rm GNM}/\partial p_\mu^2}
      \; . 
      \label{diagonal-Hessian-1}
     \end{equation}

In the conventional SSK model there is no kinetic energy, and only the first elements 
in (\ref{diagonal-Hessian-1}) are relevant. They 
are $z_{\rm GNM}(\vec v_\rho, \vec 0)+2\bar\gamma_\mu=\lambda_\rho-\lambda_\mu$. 
The stationary states $\pm \vec v_\rho= \pm \vec v_N$ are degenerate minima related by the 
symmetry $\vec x \leftrightarrow -\vec x$, 
since $\lambda_N-\lambda_\mu>0$ for all $\mu<N$, but they are marginally stable, 
since they are flat in the $N$th direction, $\lambda_N-\lambda_N=0$. All other stationary
states are metastable with as many positive and  negative eigenvalues of the Hessian
depending on $\mu$. If one were to add a conventional kinetic term, 
the components of the Hessian along the momentum directions equal $ 1/m$ for $\mu\neq \rho$
and tell us that the system will leave the metastable states but not the minimum.

For the generic model with $\bar\gamma_\mu \neq -\lambda_\mu/2$, a similar analysis applies.
$\vec x = \pm \vec v_N, \vec p =0$ with $z_{\rm GNM}(\pm \vec v_N, \vec 0) = -2 \bar\gamma_N$ are marginally stable absolute 
minima iff $\bar\gamma_N < \bar\gamma_\mu$ for all $\mu \neq N$. 
The energies of the other stationary states increase monotonically for decreasing 
label $\mu$ iff the $\bar\gamma_\mu$ are ordered so that $\bar\gamma_N < \dots < \bar\gamma_1$.

If one works with a non-monotonic, say quadratic $\bar\gamma_\mu(\lambda_\mu)$, with minimum at $-2J < \lambda_\rho < 2J$, 
according to Eq.~(\ref{eq:energies-metastable-states}) the configurations
$\pm \vec v_\rho$ with $\vec p =\vec 0$ are the ones taking the minimal energy among the ones 
with $\vec p=0$. 
However, the diagonal Hessian matrix components are
\begin{equation}
\underbrace{2\bar\gamma_\mu -2\bar\gamma_\rho}_{>0 \; \forall \mu \neq \rho, \;\; =0 \; \mu=\rho} \!\! ,  \;
\underbrace{ - \frac{2}{m} \; 
      \, \frac{\bar\gamma_\mu - \bar\gamma_\rho}{\lambda_\mu - \lambda_\rho} \; (1-\delta_{\mu\rho})}_{>0 \; \forall \mu < \rho}
       ,  \;  0 \;  , 
      \underbrace{ - \frac{2}{m} \; 
      \, \frac{\bar\gamma_\mu - \bar\gamma_\rho}{\lambda_\mu - \lambda_\rho} \; (1-\delta_{\mu\rho})}_{<0 \; \forall \mu >\rho}
      \label{diagonal-Hessian-2}
     \end{equation}
     So the modes with $\mu>\rho$, for which $\bar\gamma_\mu$ grows with $\lambda_\mu$ are unstable. Their
     momentum can detach from zero and let the energy further decrease.
This has consequences on the dynamics. In App.~\ref{app:N2} we 
considered in detail some cases for $N=2$ and in App.~\ref{app:C4} 
for $N>2$, where we expose this instability.

\begin{figure}[h!]
\includegraphics[scale=0.27]{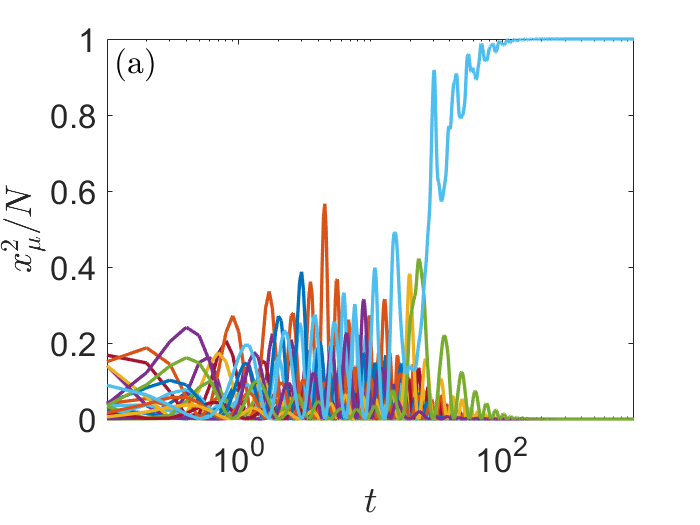}
\includegraphics[scale=0.27]{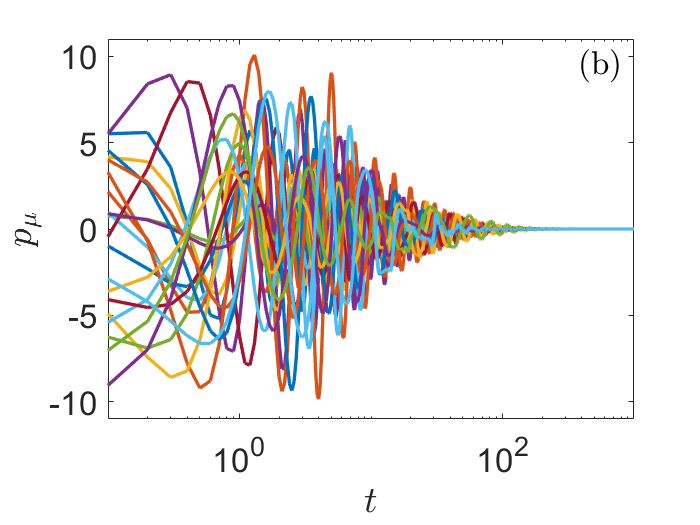}
\includegraphics[scale=0.27]{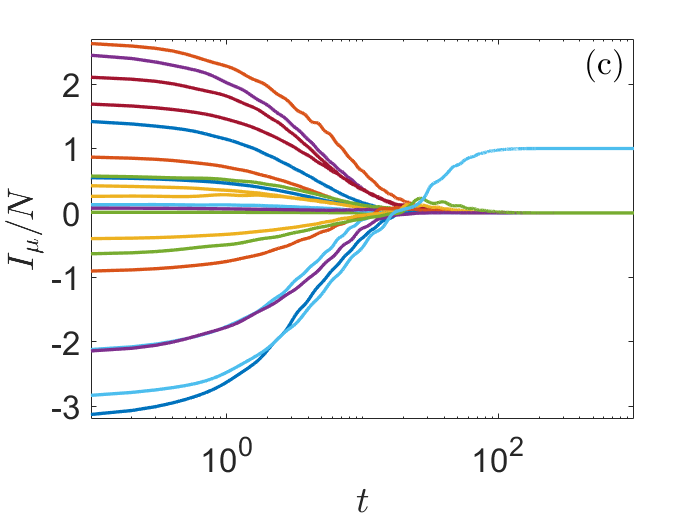}
\\
\includegraphics[scale=0.27]{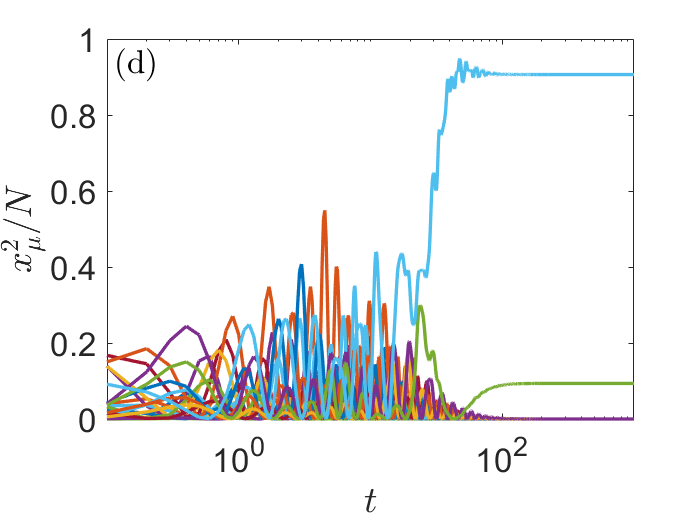}
\includegraphics[scale=0.27]{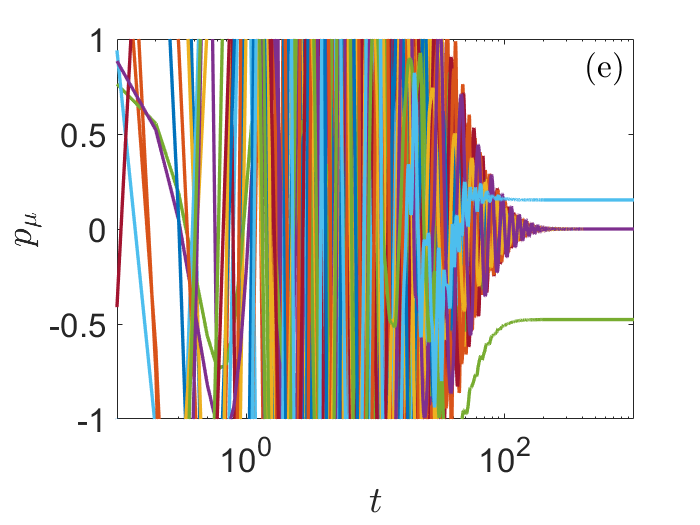}
\includegraphics[scale=0.27]{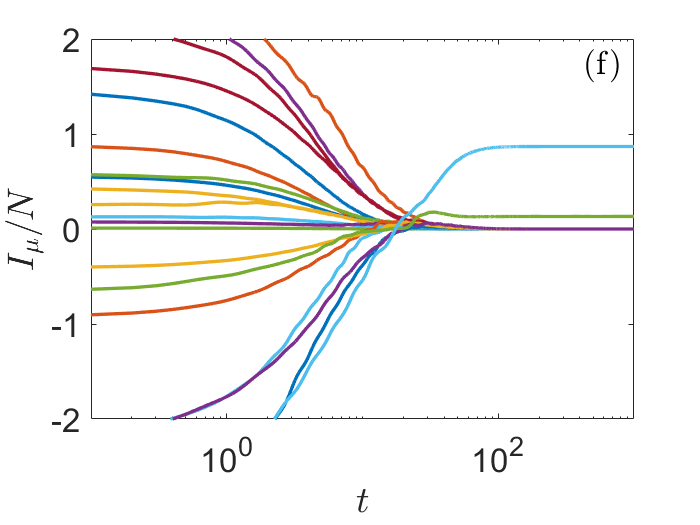}
\\
\includegraphics[scale=0.27]{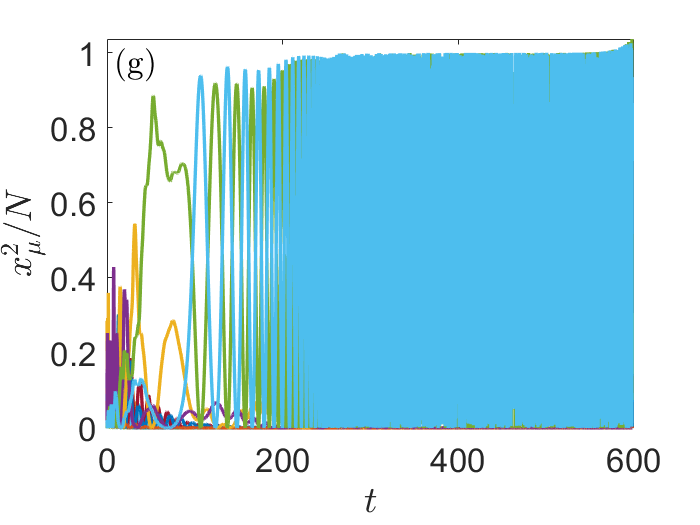}
\includegraphics[scale=0.27]{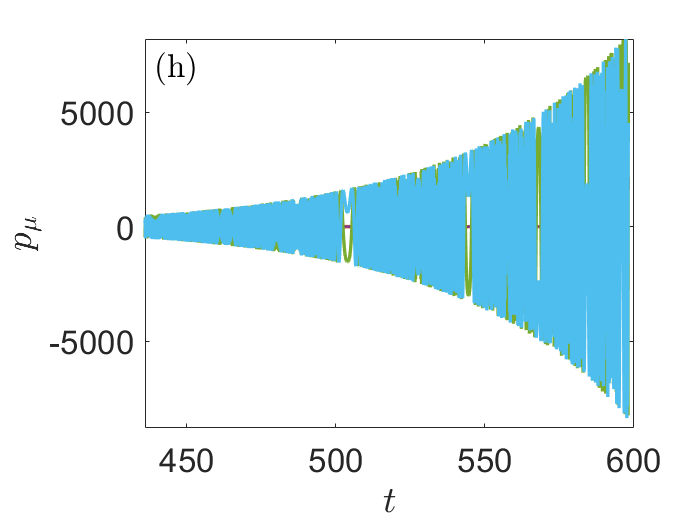}
\includegraphics[scale=0.27]{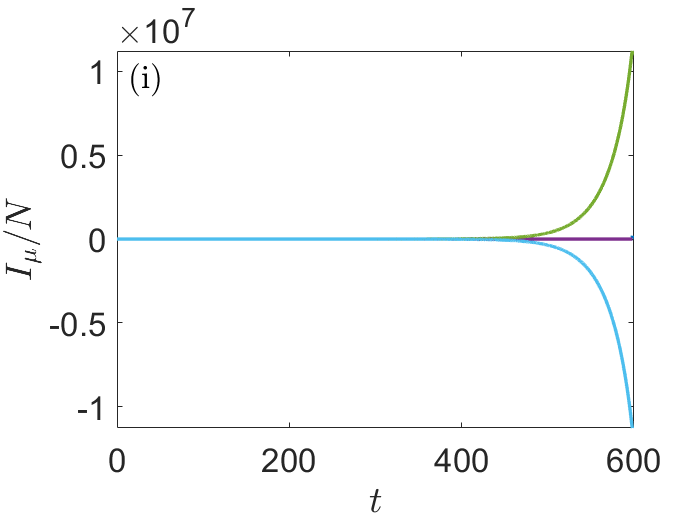}
\caption{\small Mode-resolved results for the Langevin dynamics of the 
${\mathcal H}_{\rm GNM}$ model with quadratic $\bar\gamma_\mu(\lambda_\mu)$, $a>0$ and $N=20$. Small friction coefficient $\gamma=0.1$ and zero temperature $T=0$.
One case per row: 
$\bar\gamma_{N-1} > \bar\gamma_{N}$ (first),
$\bar\gamma_{N-1} = \bar\gamma_{N}$ (second), 
$\bar\gamma_{N-1} < \bar\gamma_{N}$ (third). 
All other $\bar\gamma_\nu$ satisfy  $\bar\gamma_\nu< \bar\gamma_N$.
 Each curve corresponds to one 
mode, with the green one being the $N-1$th and the cyan one the $N$th.
In the first column the coordinates, in the second the components of the
momentum and in the third the conserved quantities of the Neumann model.
}
\label{fig:N20}
\end{figure}

Taking the previous observations into account and considering that the linear relation
$\bar\gamma_{\mu}= - \lambda_{\mu}/2$ recovers the SSK/Neumann model, we can use quadratic relations between the
$\bar\gamma_{\mu}$s and the $\lambda_{\mu}$s in order to explore other sectors of the parameter space.
Some examples of quadratic functions $\bar\gamma_\mu(\lambda_\mu)$, with $\lambda_\mu$ taking values in the 
interval $[-2J, 2J]$,  are given in the two small lower panels in Fig.~\ref{fig:phase-diagram}. This is what we 
do analytically in Sec.~\ref{subsec:quadratic-gamma}  and numerically in Sec.~\ref{sec:numerics}.

\subsection{Langevin dynamics}
\label{app:C4}

Consider a case with $N$ modes, $\lambda_1 < \dots < \lambda_N$, 
and focus on only two modes modes, the $\mu$th and the $N$th ones.
If $\bar\gamma_{\nu \neq (\mu,N)}> \bar\gamma_N$, 
all these modes vanish in the 
sense that all averages of $x_\nu$ and $p_\nu$ are negligible. 
The possible motion 
concentrates then on the plane $(\mu, N)$. The situation 
reduces to the one with $N=2$ discussed in App.~\ref{app:C2}:
If $\bar\gamma_\mu > \bar\gamma_N$, the $\mu$th mode also disappears and the
dynamics approaches one of the two stable states along the $N$th direction.
If $\bar\gamma_\mu=\bar\gamma_N$, the dynamics approach some stationary 
point on the circle  that depends on the initial conditions.
Finally, if $\bar\gamma_\mu<\bar\gamma_N$, the particle orbits on this plane. 

\begin{figure}[h!]
\centerline{
\includegraphics[scale=0.35]{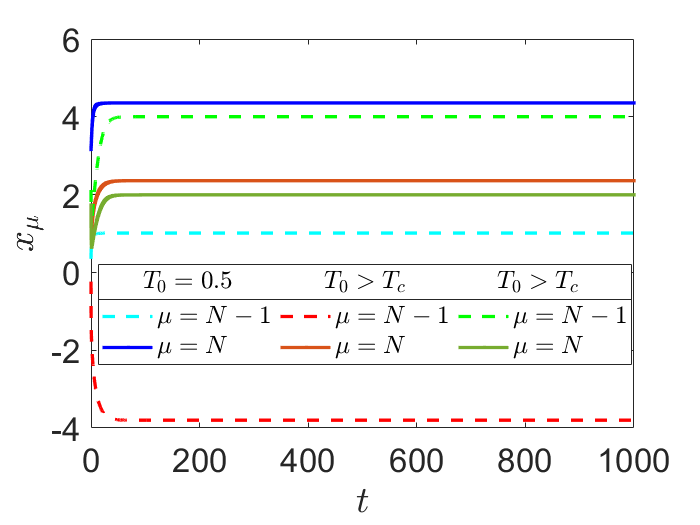}
\includegraphics[scale=0.35]{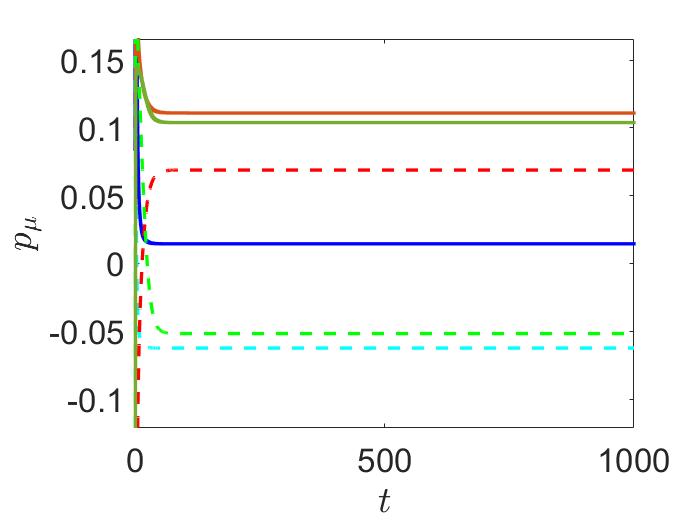}
}
\caption{\small 
Mode-resolved results for the Langevin dynamics of the 
${\mathcal H}_{\rm GNM}$ model
with quadratic $\bar\gamma_\mu(\lambda_\mu)$, $a>0$ and $N=20$.
Data for $\gamma =5$ and zero temperature,  having enforced $\bar\gamma_{N-1} =\bar\gamma_N$, 
generated for different initial conditions. 
The blue colour corresponds to the case with an initial condition in the condensed phase and 
$T_0 = 0.5$; red and green correspond to an initial condition in the extended phase but with a different random seed. Solid lines correspond to mode 
$\mu = N$, and dashed lines to $\mu = N - 1$. The solution depends on the initial condition.
}
\label{fig:app-E}
\end{figure}

This behaviour  is exemplified in Fig.~\ref{fig:N20}, where the 
results for the Langevin dynamics at $T=0$ with friction coefficient
$\gamma=0.1$  are shown for  three cases. $\bar \gamma_\nu$ are 
chosen so that $\mu=N-1$. In all panels, $\bar\gamma_{\nu \neq N-1}> \bar\gamma_N$ for all 
$\nu$ while different possibilities are considered for the relation between
$\bar\gamma_{N-1}$ and $ \bar\gamma_N$.
 (a)-(c) For $\bar\gamma_{N-1} > \bar\gamma_N$
the  North/South poles are stable. (d)-(f) For $\bar\gamma_{ N-1} = \bar\gamma_N$ all modes decay to zero apart from 
the $N-1$-th and $N$-th which reach asymptotic values that depend on the 
initial conditions, see Fig.~\ref{fig:app-E} (g)-(i) For $\bar\gamma_{ N-1} < \bar\gamma_N$ there is an instability, 
with ever more rapid oscillatory motion on the 
$(N-1, N)$ plane.

\subsection{Discussion}
Having developed a full stability analysis, and shown some runs of the Langevin evolution at zero temperature, we
stress that in the $N\to\infty$ limit, the dynamics can be 
stabilized at finite temperature, as shown in the main text both for the model 
with quadratic $\bar\gamma_\mu(\lambda_\mu)$ and for the model with $\bar\gamma_\mu(\lambda_\mu)$
extracted from the GGE analysis of the Newtonian dynamics of the SSK/Neumann model.
{This macroscopic dynamical stabilization is distinct from global canonical normalizability at finite $N$. A negative kinetic direction prevents the latter, but does not by itself determine the dynamics when the thermodynamic limit is taken first.}

\section{The Lagrange multipliers $\gamma(\lambda)$}
\label{app:new-derivation}
\setcounter{equation}{0}
\renewcommand{\theequation}{I.\arabic{equation}}

We present an alternative derivation of the Lagrange multipliers of the GGE measure, $\gamma(\lambda)$, describing the long-time 
limit of local observables the isolated Neumann model following a sudden quench. This derivation is simpler than the one presented in~\cite{BaCuLoNe22}. 
The functional forms obtained differ (except when the parameters satisfy $T^2_0 = J_0 J$, corresponding to what was termed the special line, or in conventional equilibrium $J=J_0$). The discrepancy arises because in~\cite{BaCuLoNe22} it was implicitly assumed
analyticity in the complex plane, based on explicit solutions along the 
special line and in equilibrium; 
however, this assumption fails in the remainder of phase diagram. We correct the derivation here, finding therefore
different expressions for $\gamma(\lambda)$ with respect to those presented in~\cite{BaCuLoNe22}.

In this Appendix, we start from Eq.~(130) in \cite{BaCuLoNe22} 
\begin{equation}
\frac1{2T(\lambda)}=\dashint d\lambda'\,\frac{\rho(\lambda')T(\lambda')}{z-\lambda'}\,\frac{\gamma(\lambda')-\gamma(\lambda)}{\lambda-\lambda'}+q\,\frac{\gamma(\lambda)-\gamma_N}{2J-\lambda}
\; .
\label{eq:gammas}
\end{equation}
and we show that it 
has one closed-form solution $\gamma(\lambda)$ that holds in all four phases, written as
a local term plus a single principal-value integral. The phase only enters through the 
inputs $T(\lambda)$, $z$, $q$ and $\langle I(\lambda)\rangle_{i.c.}$.

\subsection{Phase I and II}

In phases I and II, we do not expect macroscopic condensation: $q=0$. After multiplication by
 $2\pi \rho(\lambda) T(\lambda)/(z-\lambda)$, Eq.~(\ref{eq:gammas}) becomes
 \begin{eqnarray}
&
\frac{\pi \rho(\lambda)}{z-\lambda}
=
- \, 2 \pi \, \frac{\gamma(\lambda) \rho(\lambda) T(\lambda)}{z-\lambda} \;
\dashint d\lambda^{\prime}\ \frac{\rho(\lambda') \ T(\lambda')}{(z-\lambda')(\lambda-\lambda')}
\nonumber\\
&
\qquad\qquad
+ \, 2 \pi \, \frac{\rho(\lambda) T(\lambda)}{z-\lambda} \;
\dashint d\lambda^{\prime}\ \frac{\rho(\lambda') \ T(\lambda') \gamma(\lambda') }{(z-\lambda')(\lambda-\lambda')}
\; ,
\label{eq:inf_N_1-second}
\end{eqnarray}
and we recall that $z\geq 2J$ for $T_0\geq J_0$.
Let us extend the real $\lambda$ axis to the complex plane, by defining $\zeta =\lambda+ {\rm i} \eta \in {\mathbb C}$. 
Eventually we will take $\eta\to 0^+$ and 
$\zeta= \lambda + {\rm i} 0^+$ 
or $\eta\to 0^-$ and $\zeta= \lambda+ {\rm i} 0^- = \lambda - {\rm i} 0^+$ (we will use the notation $\pm {\rm i} \eta$ and 
take $\eta\to 0^+$). 
Consider now the function
\begin{equation}
    \phi(\zeta) = 
    \int_{-2J}^{2J} d\lambda' \, \frac{\rho(\lambda') T(\lambda')}{(z-\lambda') 
    (\zeta - \lambda')}
    \; .
    \label{eq:phi-zeta}
\end{equation}
Using the Sokhotski–Plemelj theorem
\begin{equation}
\lim_{\eta \to 0^+} \frac{1}{\lambda \pm {\rm i} \eta - \lambda'} = \mathcal{P}\left( \frac{1}{\lambda - \lambda'} \right) \mp i \pi \delta(\lambda - \lambda')
\end{equation}
with ${\mathcal P}$ the principal part, 
we have
\begin{equation}
 \phi_R(\lambda \pm {\rm i} 0^+) = 
\dashint_{-2J}^{2J} d\lambda' \, \frac{\rho(\lambda') T(\lambda')}{(z-\lambda') (\lambda - \lambda')}
\label{eq:phiR}
\end{equation}
and 
\begin{equation}
\phi_I(\lambda \pm {\rm i} 0^+) = \mp \pi \, \frac{\rho(\lambda) T(\lambda)}{(z-\lambda)}
\label{eq:phiI}
\end{equation}
Defining now
\begin{equation}
\Xi(\zeta) =  
\int_{-2J}^{2J} d\lambda' \, \frac{\rho(\lambda') T(\lambda') \gamma(\lambda')}{(z-\lambda') (\zeta - \lambda')}
\end{equation}
and proceeding in the same way, 
\begin{equation}
 \Xi_R(\lambda \pm {\rm i} 0^+) = 
\dashint_{-2J}^{2J} d\lambda' \, \frac{\rho(\lambda') T(\lambda')\gamma(\lambda')}{(z-\lambda') (\lambda - \lambda')}
\label{eq:XiR}
\end{equation}
and 
\begin{equation}
\Xi_I(\lambda \pm {\rm i} 0^+) = \mp \pi \, \frac{\rho(\lambda) T(\lambda) \gamma(\lambda)}{(z-\lambda)}
\; . 
\label{eq:XiI}
\end{equation}
With these expressions and taking $\zeta=\lambda+{\rm i} 0^+$ we can rewrite Eq.~(\ref{eq:inf_N_1-second}) as
\begin{equation}
\frac{\pi \rho(\lambda)}{2(z-\lambda)}
=
\phi_R(\lambda) \Xi_I(\lambda) -\phi_I(\lambda) \Xi_R(\lambda)
\label{eq:same-eq-as-ours}
\end{equation}
which is Eq.~(134) in Ref.~\cite{BaCuLoNe22}. Furthermore, from Eq.~(\ref{eq:phiI}),
\begin{equation}
\frac{\pi \rho(\lambda)}{2(z-\lambda)} = - \frac{\phi_I(\lambda)}{2T(\lambda)}
\label{eq:PhiI-final}
\end{equation}
so 
\begin{equation}
  \frac{\phi_I(\lambda)}{2T(\lambda)} =
  \phi_I(\lambda) \Xi_R(\lambda)
  -
 \phi_R(\lambda) \Xi_I(\lambda) 
 \;.
 \label{eq:new-eq}
\end{equation}

Now, comparing Eqs.~(\ref{eq:phiI}) and (\ref{eq:XiI}) we notice that 
\begin{equation}
\Xi_I(\lambda) = \gamma(\lambda) \phi_I(\lambda)
    \end{equation}
    at $\lambda \pm {\rm i} 0^+$. 
    Let us then define
    \begin{equation}
    R(\zeta) \equiv \frac{\Xi(\zeta)}{\phi(\zeta)}
    \; . 
    \end{equation}
    Its imaginary part, evaluated at $\lambda + {\rm i} 0^+$ is
    \begin{eqnarray}
    {\rm Im} R_+(\lambda) &=&  \frac{ {\rm Im} [\Xi(\lambda) \phi^*(\lambda) ] }{|\phi(\lambda)|^2}
    =
    \frac{\Xi_I(\lambda) \phi_R(\lambda) - \Xi_R(\lambda ) \phi_I(\lambda)}{\phi_R^2(\lambda)+\phi_I^2(\lambda)}
    \nonumber\\
    &=& - \frac{\phi_I(\lambda)}{2 T(\lambda) [\phi_R^2(\lambda)+\phi_I^2(\lambda)]} 
    \equiv \varphi(\lambda)
    \end{eqnarray}
    where we wrote $\lambda$ as arguments but we indicated that the limit is taken from 
    above with the $+$ subindex.
We saw above that the sign of the imaginary parts is the opposite one when
evaluating at $\lambda+{\rm i} 0^-$. Then, 
\begin{equation}
{\rm Im} R_+(\lambda) - {\rm Im} R_-(\lambda) = 2  \varphi(\lambda)
\end{equation}
and 
\begin{equation}
R_+(\lambda) - R_-(\lambda) = 2 {\rm i} \varphi(\lambda)
\; . 
\label{eq:prop-R}
\end{equation}

We can now deduce that the function $R_+$ is given by
\begin{equation}
R_+(\lambda) = C + \frac{1}{\pi} \; \dashint_{-2J}^{2J} d\lambda' \; 
\frac{\varphi(\lambda')}{\lambda'-\lambda}
+ {\rm i} \varphi(\lambda)
\label{eq:R+}
\end{equation}
with $C$ a constant. 

Using $[(z-\lambda')(\zeta-\lambda')]^{-1}
=
(z-\zeta)^{-1} \bigl[(\zeta-\lambda')^{-1}-(z-\lambda')^{-1}\bigr]$ and the spherical constraint,
the complex function $\phi(\zeta)$ defined in Eq.~(\ref{eq:phi-zeta}) can be rewritten as
\begin{equation}
\phi(\zeta)=\frac{D(\zeta)}{\zeta-z}
\;, 
\qquad
{\rm with} 
\qquad 
D(\zeta)=1-\int d\lambda'\,\frac{\rho(\lambda')T(\lambda')}{\zeta-\lambda'} 
\; .
\end{equation}
$D$ is the same function appearing in the right-hand side of Eq.~(\ref{eq:T-spectrum}), which fixes the temperature spectrum 
$T(\lambda)$.
For $T(\lambda')>0$, ${\rm Im} D>0$ in the upper half plane. On the real axis outside the cut, 
$D$ is real and monotonic with its only zero at $\zeta=z$ (thanks to the spherical constraint), 
which cancels against $1/(\zeta-z)$. So $\phi$ has no zeros in $\mathbb C\setminus[-2J,2J]$, and $R=\Xi/\phi$ has no poles there. Together with boundedness at $\pm 2J$ and the limit $C$ at infinity, this makes the solution~(\ref{eq:prop-R}) unique, and 
Eq.~(\ref{eq:R+}) follows.

Using now $\Xi = R \phi$ and $\Xi_I = \gamma \phi_I$:
\begin{eqnarray}
\gamma(\lambda) 
&=& 
C 
-
\frac{1}{\pi} \; \dashint_{-2J}^{2J} d\lambda' \; 
\frac{1}{\lambda'-\lambda}  \; \frac{\phi_I(\lambda')}{2T(\lambda') [\phi^2_R(\lambda') 
+ \phi_I^2(\lambda')]}
\nonumber\\
&& 
- \frac{\phi_R(\lambda)}{2T(\lambda) [\phi^2_R(\lambda) + \phi_I^2(\lambda)]}
\; . 
\label{eq:final-gamma-I-II}
\end{eqnarray}
The $\gamma(\lambda)$ are defined up to an additive  constant. One can 
select $C$ so that $\gamma(2J)$ equals zero, which is the choice adopted
in the analysis of phase II, in Fig.~\ref{fig:phaseIIa}.

The expression~(\ref{eq:final-gamma-I-II}) for parameters in 
equilibrium, $J=J_0$, and on the special line, $T_0^2 = J_0J$, 
coincides with the form given in~\cite{BaCuLoNe22}. However, 
for more generic parameters it does not. 

Fortunately, this error does not invalidate the results for 
$T(\lambda)$, and consequently for $\langle p^2(\lambda)\rangle$ and $\langle x^2(\lambda)\rangle $, given in~\cite{BaCuLoNe22},
since those are obtained from Eq.~(\ref{eq:T-spectrum}) which is independent of the $\gamma(\lambda)$ 
which are deduced later. 
For the numerical tests shown here it was crucial, though, to have the correct 
spectrum of $\gamma(\lambda)$, especially in phase II, 
since these are inputs in the deterministic force of the Langevin process.

\subsection{Phase III}

In phase III, $q\neq 0$. Proceeding as above, in the $q\neq 0$ case one adds 
\begin{equation}
2K_I(\lambda)\equiv\frac{2\pi\rho(\lambda)T(\lambda)}{z-\lambda}\,\frac{q\,[\gamma(\lambda)-\gamma_N]}{2J-\lambda}
\; .
\label{eq:KI}
\end{equation}
to the right-hand side Eq.~(\ref{eq:inf_N_1-second}) (using the same notation as in~\cite{BaCuLoNe22}).

The condensate is a point mass of weight $q$ at $\lambda_N=2J$ in the measure $\rho T/(z-\lambda)$
and is needed to satisfy the spherical constraint in phase III. Its effect is included by adding a pole to both 
complex functions:
\begin{equation}
\phi_q(\zeta)=\phi(\zeta)+\frac{q}{\zeta-2J}
\; ,\qquad
\Xi_q(\zeta)=\Xi(\zeta)+\frac{q\,\gamma_N}{\zeta-2J}
\; .
\end{equation}
On the cut, $|\lambda|<2J$, the added terms are real and only the real parts of $\phi$ and $\Xi$ change, 
\begin{equation}
\phi_{q,I}=\phi_I \; ,\qquad\qquad\qquad\  \Xi_{q,I}=\Xi_I=\gamma\,\phi_I \; ,
\end{equation}
\begin{equation}
\phi_{q,R}=\phi_R-\frac{q}{2J-\lambda},\qquad \Xi_{q,R}=\Xi_R-\frac{q\,\gamma_N}{2J-\lambda}.
\end{equation}
Replacing now in Eq.~(\ref{eq:inf_N_1-second}) and taking into account 
the added term~(\ref{eq:KI}), one has 
\begin{equation}
\frac{\pi\rho}{2(z-\lambda)}=\phi_{q,R}\,\Xi_I-\phi_I\,\Xi_{q,R}
\; ,\qquad
\frac{\pi\rho}{2(z-\lambda)}=-\frac{\phi_I}{2T}
\; ,
\end{equation}
with all functions evaluated at $\zeta = \lambda+{\rm i}0^+$. These equations have the same form as 
Eqs.~(\ref{eq:same-eq-as-ours}) and (\ref{eq:PhiI-final}), and the 
extra term has been absorbed.

Defining now $R_q = \Xi_q/\phi_q$, 
\begin{equation}
{\rm Im} R_{q,+}(\lambda)= -\frac{\phi_I}{2T(\lambda)\,[\phi_{q,R}^2+\phi_I^2]}\equiv \varphi_q(\lambda)
\end{equation}
and
$R_{q,+}-R_{q,-}=2 {\rm i}\varphi_q$. One can check that for real $\zeta$, 
outside the interval $[-2J,2J]$
$\phi_q$ has no zeroes and $R_q$ has no poles. Using $z=2J$, at $\zeta=2J$, 
$\phi_q$ and $\Xi_q$ have simple poles with residues $q$ and $q\gamma_N$, so
that
\begin{equation}
R_q(2J)=\gamma_N 
\; .
\end{equation}
Putting all these features together, 
\begin{equation}
R_q(\zeta)=C+\frac1\pi\int_{-2J}^{2J}d\lambda' \; \frac{\varphi_q(\lambda')}{\lambda'-\zeta}
\; .
\end{equation}

From $\Xi_I=\gamma\phi_I$ and $\Xi_q=R_q\phi_q$, $\gamma={\rm Re} R_{q,+} + 
{\rm Im} R_{q,+}\,\phi_{q,R}/\phi_I$, one finds
\begin{equation}
\gamma(\lambda)
=
C-\frac1\pi
\dashint_{-2J}^{2J} d\lambda' \; 
\frac{\varphi_q(\lambda')}{\lambda-\lambda'}
-
\frac{\phi_{q,R}(\lambda)}{2T(\lambda)
\,
\big[\phi_{q,R}(\lambda)^2+\phi_I(\lambda)^2\big]
}
\end{equation}
with
\begin{equation}
\phi_{q,R}(\lambda)=\phi_R(\lambda)-\frac{q}{2J-\lambda},\qquad
\varphi_q(\lambda)=\frac{\pi\rho(\lambda)}{2(2J-\lambda)\,[\phi_{q,R}^2+\phi_I^2]}.
\end{equation}
The $\gamma_N$ is
\begin{equation}
\gamma_N=C-\frac1\pi \int_{-2J}^{2J}d\lambda'\,\frac{\varphi_q(\lambda')}{2J-\lambda'}=\gamma(2J^-).
\end{equation}
The only differences from~(\ref{eq:final-gamma-I-II}) 
are the substitutions $\phi_R\to\phi_R-q/(2J-\lambda)$ and $\varphi\to\varphi_q$. $C$ is still an arbitrary additive constant which can be chosen to set 
$\gamma_N$ to zero.

\section*{References}
\bibliographystyle{iopart-num}
\bibliography{stochastic-Neumann2}

\providecommand{\newblock}{}
\begin{thebibliography}{10}
\expandafter\ifx\csname url\endcsname\relax
  \def\url#1{{\tt #1}}\fi
\expandafter\ifx\csname urlprefix\endcsname\relax\def\urlprefix{URL }\fi
\providecommand{\eprint}[2][]{\url{#2}}
% Bibliography created with iopart-num v2.1
% /biblio/bibtex/contrib/iopart-num

\bibitem{DeLuca16}
{De Luca} A and Mussardo G 2016 {\em J. Stat. Mech.\/} {\bf 2016} 064011 \urlprefix\url{https://doi.org/10.1088/1742-5468/2016/06/064011}

\bibitem{Doyon17}
Doyon B, Spohn H and Yoshimura T 2017 {\em Nucl. Phys. B\/} {\bf 926} 570 \urlprefix\url{https://doi.org/10.1016/j.nuclphysb.2017.12.002}

\bibitem{Doyon18}
Doyon B, Yoshimura T and Caux J~S 2018 {\em Phys. Rev. Lett.\/} {\bf 120} 045301 \urlprefix\url{https://doi.org/10.1103/PhysRevLett.120.045301}

\bibitem{Bastianello18}
Bastianello A, Doyon B, Watts G and Yoshimura T 2018 {\em SciPost Physics\/} {\bf 4} \urlprefix\url{https://doi.org/10.21468/SciPostPhys.4.6.045}

\bibitem{Goldfriend18}
Goldfriend T and Kurchan J 2018 {\em EPL\/} {\bf 124} 10002 \urlprefix\url{https://doi.org/10.1209/0295-5075/124/10002}

\bibitem{CuLoNePiTa18}
Cugliandolo L~F, Lozano G~S, Nessi N, Picco M and Tartaglia A 2018 {\em J. Stat. Mech.\/} {\bf 2018} P063206 \urlprefix\url{https://doi.org/10.1088/1742-5468/aac46f}

\bibitem{Cao19}
Cao X, Bulchandani V~B and Spohn H 2019 {\em J. Phys. A\/} {\bf 52} 495003 \urlprefix\url{https://doi.org/10.1088/1751-8121/ab5019}

\bibitem{Goldfriend19}
Goldfriend T and Kurchan J 2019 {\em Phys. Rev. E\/} {\bf 99} 022146 \urlprefix\url{https://doi.org/10.1103/PhysRevE.99.022146}

\bibitem{Doyon19}
Doyon B 2019 {\em J. Math. Phys.\/} {\bf 60} 073302 \urlprefix\url{https://doi.org/10.1063/1.5096892}

\bibitem{Spohn19}
Spohn H 2020 {\em J. Stat. Phys.\/} {\bf 180} 4 arXiv:1902.07751 \urlprefix\url{https://doi.org/10.1007/s10955-019-02320-5}

\bibitem{BaCuLoNePiTa19}
Barbier D, Cugliandolo L~F, Lozano G~S, Nessi N, Picco M and Tartaglia A 2019 {\em J. Phys. A\/} {\bf 52} 454002 \urlprefix\url{https://doi.org/10.1088/1751-8121/ab3ff1}

\bibitem{BaCuLoNe20}
Barbier D, Cugliandolo L~F, Lozano G~S and Nessi N 2020 {\em EPL\/} {\bf 132} 50002 \urlprefix\url{https://doi.org/10.1209/0295-5075/132/50002}

\bibitem{Myers20}
Myers J, Bhaseen J, Harris R~J and Doyon B 2020 {\em SciPost Physics\/} {\bf 8} \urlprefix\url{https://doi.org/10.21468/SciPostPhys.8.1.007}

\bibitem{Koch22}
Koch R, Caux J~S and Bastianello A 2022 {\em J. Phys. A: Math. Theor.\/} {\bf 55} 134001 \urlprefix\url{https://doi.org/10.1088/1751-8121/ac53c3}

\bibitem{BaCuLoNe22}
Barbier D, Cugliandolo L~F, Lozano G~S and Nessi N 2022 {\em SciPost Phys.\/} {\bf 13} 048 \urlprefix\url{https://doi.org/10.21468/SciPostPhys.13.3.048}

\bibitem{Baldovin21}
Baldovin M, Vulpiani A and Gradenigo G 2021 {\em J. Stat. Phys.\/} {\bf 183} 41 \urlprefix\url{http://dx.doi.org/10.1007/s10955-021-02781-7}

\bibitem{Kethepalli23}
Kethepalli J, Bagchi D, Dhar A, Kulkarni M and Kundu A 2023 {\em Phys. Rev. E\/} {\bf 107} 044101 \urlprefix\url{https://doi.org/10.1103/PhysRevE.107.044101}

\bibitem{Roy23}
Roy D, Dhar A, Spohn H and Kulkarni M 2023 {\em Phys. Rev. B\/} {\bf 107} L100413 \urlprefix\url{https://doi.org/10.1103/PhysRevB.107.L100413}

\bibitem{Pandey24}
Pandey S, Dhar A and Kundu A 2024 {\em J. Stat. Mech.\/} {\bf 2024} 103202 \urlprefix\url{https://doi.org/10.1088/1742-5468/ad7dfa}

\bibitem{Singh24}
Singh S, Dhar A, Spohn H and Kundu A 2024 {\em J. Stat. Phys.\/} {\bf 191} 66 \urlprefix\url{https://doi.org/10.1007/s10955-024-03282-z}

\bibitem{Baldovin25}
Baldovin M, Gradenigo G, Vulpiani A and Zanghì N 2025 {\em Phys. Rep.\/} {\bf 1132} 1--79 \urlprefix\url{http://dx.doi.org/10.1016/j.physrep.2025.05.003}

\bibitem{Gardiner2009}
Gardiner C~W 2009 {\em Stochastic Methods: A Handbook for the Natural and Social Sciences\/} 4th ed (Springer)

\bibitem{Coffey2004}
Coffey W~T, Kalmykov Y~P and Waldron J~T 2004 {\em The Langevin Equation\/} 2nd ed (World Scientific)

\bibitem{Kloeden2011}
Kloeden P~E and Platen E 2011 {\em Numerical Solution of Stochastic Differential Equations\/} 3rd ed (Springer)

\bibitem{LFC25}
Cugliandolo L~F 2024 {\em Adv. Phys.\/} {\bf 73} 3 \urlprefix\url{https://doi.org/10.1080/00018732.2025.2580840}

\bibitem{Neumann}
Neumann C 1859 {\em Journal für die reine und angewandte Mathematik\/} {\bf 56} 46--63 \urlprefix\url{https://doi.org/10.1515/crll.1859.56.46}

\bibitem{KoThJo76}
Kosterlitz J~M, Thouless D~J and Jones R~C 1976 {\em Phys. Rev. Lett.\/} {\bf 36} 1217 \urlprefix\url{https://doi.org/10.1103/PhysRevLett.36.1217}

\bibitem{Lund80}
Deift P, Lund F and Trubowitz E 1980 {\em Proc. Nat. Acad. Sci. U.S.A.\/} {\bf 77} 716--719 \urlprefix\url{https://doi.org/10.1073/pnas.77.2.716}

\bibitem{Lund80b}
Deift P, Lund F and Trubowitz E 1980 {\em Comm. Math. Phys.\/} {\bf 74} 141--188

\bibitem{Alba15}
Alba V 2015 Simulating the generalized gibbs ensemble (gge): a hilbert space monte carlo approach (\textit{Preprint} \eprint{1507.06994}) \urlprefix\url{https://arxiv.org/abs/1507.06994}

\bibitem{CuDe95a}
Cugliandolo L~F and Dean D~S 1995 {\em J. Phys. A: Math.Gen.\/} {\bf 28} 4213 \urlprefix\url{http://dx.doi.org/10.1088/0305-4470/28/15/003}

\bibitem{CuDe95b}
Cugliandolo L~F and Dean D~S 1995 {\em J. Phys. A: Math. Gen.\/} {\bf 28} L453 \urlprefix\url{http://dx.doi.org/10.1088/0305-4470/28/17/001}

\bibitem{BenAr01}
{Ben~Arous} G, Dembo A and Guionnet A 2001 {\em Probab. Th. Relat. Fields\/} {\bf 120} 1 \urlprefix\url{https://doi.org/10.1007/s00440000011}

\bibitem{ChCuYo06}
Chamon C, Cugliandolo L~F and Yoshino H 2006 {\em J. Stat. Mech\/} {\bf 2006} P01006 \urlprefix\url{https://doi.org/10.1088/1742-5468/2006/01/P01006}

\bibitem{CuDeYo07}
Cugliandolo L~F, Dean D~S and Yoshino H 2007 {\em J. Phys. A\/} {\bf 40} 4285 \urlprefix\url{https://doi.org/10.1088/1751-8113/40/16/003}

\bibitem{vanDu10}
van Duijvendijk K, Jack R~L and van Wijland F 2010 {\em Phys. Rev. E\/} {\bf 81} 011110 \urlprefix\url{https://doi.org/10.1103/PhysRevE.81.011110}

\bibitem{FyPeSc15}
Fyodorov Y~V, Perret A and Schehr G 2015 {\em J. Stat. Mech.\/} {\bf 2015} P11017 \urlprefix\url{https://doi.org/10.1088/1742-5468/2015/11/P11017}

\bibitem{BaFrCuSt21}
Barbier D, de~Freitas~Pimenta P~H, Cugliandolo L~F and Stariolo D~A 2021 {\em J. Stat. Mech.\/} {\bf 2021} 073301 \urlprefix\url{https://doi.org/10.1088/1742-5468/ac0900}

\bibitem{Lahoche22}
Lahoche V, Ousmane~Samary D and Ouerfelli M 2022 {\em J. Phys. Comm.\/} {\bf 6} 055002 \urlprefix\url{https://doi.org/10.1088/2399-6528/ac61b3}

\bibitem{Freitas23}
{de Freitas Pimenta} P~H and Stariolo D~A 2023 {\em Entropy\/} {\bf 25} 957 \urlprefix\url{https://doi.org/10.3390/e25060957}

\bibitem{DiTa24}
Wang D and Tang L~H 2024 {\em SciPost Phys.\/} {\bf 17} 106 \urlprefix\url{https://doi.org/10.21468/SciPostPhys.17.4.106}

\bibitem{Lahoche24a}
Lahoche V and {Ousmane Samary} D 2024 {\em Nuclear Physics B\/} {\bf 1005} 116582 \urlprefix\url{https://doi.org/10.1016/j.nuclphysb.2024.116582}

\bibitem{Lahoche24b}
Lahoche V and {Ousmane Samary} D 2024 {\em Nuclear Physics B\/} {\bf 1006} 116656 \urlprefix\url{https://doi.org/10.1016/j.nuclphysb.2024.116656}

\bibitem{Niedda24}
Niedda J, Tonolo T and Gradenigo G 2024 {\em J. Stat. Mech.\/} {\bf 2024} 113301 \urlprefix\url{https://doi.org/10.1088/1742-5468/ad890c}

\bibitem{Tarabolo25}
Tarabolo M and Dall'Asta L 2025 {\em SciPost Phys.\/} {\bf 19} 019 \urlprefix\url{https://scipost.org/10.21468/SciPostPhys.19.1.019}

\bibitem{Zhenpeng26}
Zhenpeng L 2026 Optimal persistence reveals hidden topology in complex energy landscapes arXiv:2605.19785 \urlprefix\url{https://arxiv.org/abs/2605.19785}

\bibitem{dAscoli2022}
d'Ascoli S, Refinetti M and Biroli G 2023 {\em Phys. Rev. Lett.\/} {\bf 130} 167401 \urlprefix\url{https://doi.org/10.1103/PhysRevLett.130.167401}

\bibitem{Fournier23}
Fournier S~J and Urbani P 2023 {\em J. Stat. Mech.\/} {\bf 2023} 113301 \urlprefix\url{https://10.1088/1742-5468/ad082d}

\bibitem{montanari25}
Montanari A and Urbani P 2025 Dynamical decoupling of generalization and overfitting in large two-layer networks (\textit{Preprint} \eprint{2502.21269}) \urlprefix\url{https://arxiv.org/abs/2502.21269}

\bibitem{Janssen25}
{Kerr Winter} M and Janssen L~M~C 2025 {\em Phys. Rev. Research\/} {\bf 7} 023010 \urlprefix\url{https://journals.aps.org/prresearch/abstract/10.1103/PhysRevResearch.7.023010}

\bibitem{Fournier26b}
Fournier S~J and Urbani P 2026 {\em J. Stat. Mech.\/} {\bf 2026} 043302 \urlprefix\url{https://iopscience.iop.org/article/10.1088/1742-5468/ae4f7e/pdf}

\bibitem{Lorenzana25}
{Garc\'{\i}a Lorenzana} G, Altieri A, Biroli G, Fruchart M and Vitelli V 2025 {\em Phys. Rev. Lett.\/} {\bf 135} 187402 \urlprefix\url{https://journals.aps.org/prl/abstract/10.1103/PhysRevLett.135.187402}

\bibitem{StarioloMetz2026}
Stariolo D~A and Metz F~L 2026 {\em J. Stat. Mech.\/} {\bf 2026} 033301 \urlprefix\url{https://doi.org/10.1088/1742-5468/ae4589}

\bibitem{Ikeda23}
Ikeda H 2023 {\em J. Stat. Mech.\/} {\bf 2023} 023302 \urlprefix\url{https://doi.org/10.1088/1742-5468/acb7d6}

\bibitem{Rubio}
Sidler D, Ruggenthaler M and Rubio A 2026 {\em Chem. Rev.\/} {\bf 126} 4 \urlprefix\url{https://doi.org/10.1021/acs.chemrev.5c00789}

\bibitem{CL1}
Cugliandolo L~F and Lozano G 1999 {\em Phys. Rev. B\/} {\bf 59}(2) 915--942 \urlprefix\url{https://link.aps.org/doi/10.1103/PhysRevB.59.915}

\bibitem{CL2}
Cugliandolo L~F and Lozano G 1998 {\em Phys. Rev. Lett.\/} {\bf 80}(22) 4979--4982 \urlprefix\url{https://link.aps.org/doi/10.1103/PhysRevLett.80.4979}

\bibitem{Moser81}
Moser J 1981 {\em Integrable Hamiltonian Systems and Spectral Theory\/} ({\em Lezioni Fermiane\/} vol~5) (Scuola Normal Superiore)

\bibitem{Uhlenbeck}
Uhlenbeck K~K 1982 Equivariant harmonic maps into spheres {\em Harmonic Maps\/} ({\em Lecture Notes in Mathematics\/} vol 949) (Springer-Verlag) p 146 \urlprefix\url{https://doi.org/10.1007/BFb0066266}

\bibitem{AvTa90}
Avan J and Talon M 1990 {\em Int. J. Mod. Phys. A\/} {\bf 05} 4477

\bibitem{BaTa92}
Babelon O and Talon M 1992 {\em Nucl. Phys. B\/} {\bf 379} 321 \urlprefix\url{https://doi.org/10.1016/0550-3213(92)90599-7}

\bibitem{Dullin01}
Dullin H~R, Richter P~H, Veselov A~P and Waalkens H 2001 {\em Physica D\/} {\bf 155} 159--183 \urlprefix\url{https://doi.org/10.1016/S0167-2789(01)00282-3}

\bibitem{Dullin12}
Dullin H~R and Hanssmann H 2012 {\em Cent. Eur. J. Math.\/} {\bf 10} 1627

\bibitem{BaBeTa09}
Babelon O, Bernard D and Talon M 2003 {\em Introduction to {C}lassical {I}ntegrable {S}ystems\/} (Cambridge University Press)

\bibitem{Baron24}
Baron P 2024 The {N}eumann-{M}oser dynamical system and the {K}orteweg-de {V}ries hierarchy (\textit{Preprint} \eprint{2402.18079}) \urlprefix\url{https://arxiv.org/abs/2402.18079}

\bibitem{Moser80}
Moser J 1980  (New York, NY: Springer-Verlag) chap Geometry of Quadrics and Spectral Theory

\bibitem{Russo04}
Arutyunov G, Russo J and Tseytlin A~A 2004 {\em Phys. Rev. D\/} {\bf 69} 086009 \urlprefix\url{https://doi.org/10.1103/PhysRevD.69.086009}

\bibitem{Schwartz}
Schwartz M and Navot Y 1997 {\em Physica A\/} {\bf 245} 517 \urlprefix\url{https://doi.org/10.1016/S0378-4371(97)00357-9}

\bibitem{Arnoulx23}
{Arnoulx de Pirey} T, Cugliandolo L~F, Lecomte V and van Wijland F 2023 {\em Adv. Phys.\/} {\bf 71} 1 \urlprefix\url{http://dx.doi.org/10.1088/1751-8121/ab3ad5}

\bibitem{Bray94}
Bray A~J 1994 {\em Adv. Phys.\/} {\bf 43} 357 \urlprefix\url{https://doi.org/10.1080/00018739400101505}

\bibitem{Onuki04}
Onuki A 2002 {\em Phase Transition Dynamics\/} (Cambridge University Press)

\bibitem{Puri09}
Puri S and Wadhawan V (eds) 2009 {\em Kinetics of Phase Transitions\/} (CRC Press)

\bibitem{KrReBe10}
Krapivsky P~L, Redner S and Ben-Naim E 2010 {\em A Kinetic View of Statistical Physics\/} (Cambridge University Press)

\bibitem{HePl10}
Henkel M and Pleimling M 2010 {\em Non-Equilibrium Phase Transitions: Ageing and Dynamical Scaling Far from Equilibrium\/} (Springer-Verlag)

\bibitem{LesHouches}
Cugliandolo L~F 2003 Dynamics of glassy systems {\em Slow Relaxations and Nonequilibrium Dynamics in Condensed Matter\/} ed Barrat J~L, Kurchan J, Feigel'man M~V and Dalibard J (Springer-Verlag) pp 367--521

\bibitem{Cugliandolo15}
Cugliandolo L~F 2015 {\em C. R. Physique\/} {\bf 16} 257 \urlprefix\url{https://doi.org/10.1016/j.crhy.2015.03.008}

\bibitem{Cugliandolo24}
Cugliandolo L~F 2024 {\em Annu. Rev. Condens. Matter Phys.\/} {\bf 15} 177--213 \urlprefix\url{https://doi.org/10.1146/annurev-conmatphys-040721-022848}

\bibitem{Zannetti15}
Zannetti M 2015 {\em EPL\/} {\bf 111} 20004 \urlprefix\url{https://doi.org/10.1209/0295-5075/111/20004}

\bibitem{Crisanti-etal20}
Crisanti A, Sarracino A and Zannetti M 2019 {\em Phys. Rev. Research\/} {\bf 1} 023022 \urlprefix\url{https://doi.org/10.1103/PhysRevResearch.1.023022}

\bibitem{Defenu25b}
Giachetti G and Defenu N 2026 Large-$n$ $o(n)$ with long-range interactions: integrability and resonance dynamics (\textit{Preprint} \eprint{2512.14868}) \urlprefix\url{https://arxiv.org/abs/2512.14868}

\bibitem{GoBaTs14}
{Gonz\'alez~Arenas} Z, Barci D~G and Tsallis C 2014 {\em Phys. Rev. E\/} {\bf 90} 032118 \urlprefix\url{https://doi.org/10.1103/PhysRevE.90.032118}

\bibitem{ArBaCuGoLo14}
Aron C, Barci D~G, Cugliandolo L~F, {Gonz\'alez~Arenas} Z and Lozano G~S 2014 {\em J. Stat. Mech.\/} {\bf 2014} P09008 \urlprefix\url{https://doi.org/10.1088/1742-5468/2014/09/P09008}

\bibitem{ArBaCuGoLo16}
Aron C, Barci D~G, Cugliandolo L~F, {Gonz\'alez~Arenas} Z and Lozano G~S 2016 {\em J. Stat. Mech.\/} {\bf 2016} 053207 \urlprefix\url{https://doi.org/10.1088/1742-5468/2016/05/053207}

\bibitem{vanKampen2007}
van Kampen N~G 2007 {\em Stochastic Processes in Physics and Chemistry\/} 3rd ed (Elsevier)

\bibitem{Ros23}
Ros V and Fyodorov Y 2023 {\em The high-d landscapes paradigm: spin-glasses, and beyond\/} (World Scientific) \urlprefix\url{https://arxiv.org/abs/2209.07975}

\end{thebibliography}

\end{document}